\documentclass[nopubldata]{lpr2}
\usepackage{color}
\graphicspath{{figs/}}
\begin{document}
%
\titlefigure{titlefigure}

\abstract{ Microwave photonics (MWP) is an emerging field in which radio frequency (RF) signals are generated, distributed, processed and analyzed using the strength of photonic techniques. It is a technology that enables various functionalities which are not feasible to achieve only in the microwave domain. A particular aspect that recently gains significant interests is the use of photonic integrated circuit (PIC) technology in the MWP field for enhanced functionalities and robustness as well as the reduction of size, weight, cost and power consumption. This article reviews the recent advances in this emerging field which is dubbed as integrated microwave photonics. Key integrated MWP technologies are reviewed and the prospective of the field is discussed.}

\title{Integrated microwave photonics}
%
\titlerunning{Integrated microwave photonics}
\author{David~Marpaung\inst{1,4,*}, Chris~Roeloffzen\inst{1}, Ren\'{e}~Heideman\inst{2}, Arne~Leinse\inst{2}, Salvador~Sales\inst{3}, and Jos\'{e}~Capmany\inst{3,5}}%
%
\authorrunning{D. Marpaung et al.}
%
\institute{%
Telecommunication Engineering group, University of Twente, the Netherlands
\and
LioniX BV, the Netherlands
\and
ITEAM Research Institute, Universidad Polit\'{e}cnica de Valencia, Spain
\and
Current affiliation: Centre for Ultrahigh Bandwidth Devices for Optical Systems (CUDOS), School of Physics University of Sydney, Australia 
\and
VLC Photonics S.L.,Spain}%
\mail{\textsuperscript{*}\,Corresponding author: e-mail:d.marpaung@physics.usyd.edu.au}
%
%
\maketitle

\section{Introduction}
\label{sec:intro}

Microwave photonics (MWP) \cite{CapmanyNatPhoton2007,SeedsMWP2002,SeedsMWP2006,YaoMWP2009}, a discipline which brings together the worlds of radio-frequency engineering and optoelectronics, has attracted great interest from both the research community and the commercial sector over the past 30 years and is set to have a bright future. The added value that this area of research brings stems from the fact that, on the one hand, it enables the realization of key functionalities in microwave systems that either are complex or even not directly possible in the radio-frequency domain and, on the another hand, that it creates new opportunities for information and communication (ICT) systems and networks.

While initially, the research activity in this field was focused towards defense applications, MWP has recently expanded to address a considerable number of civil applications \cite{CapmanyNatPhoton2007,SeedsMWP2002,SeedsMWP2006,YaoMWP2009}, including cellular, wireless, and satellite communications, cable television, distributed antenna systems, optical signal processing and medical imaging. Many of these novel application areas demand ever-increasing values for speed, bandwidth and dynamic range while at the same time require devices that are small, lightweight and low-power, exhibiting large tunability and strong immunity to electromagnetic interference. Despite the fact that digital electronics is widely used nowadays in these applications, the speed of digital signal processors (DSPs) is normally less than several gigahertz (a limit established primarily by the electronic sampling rate). In order to preserve the flexibility brought by these devices, there is a need for equally flexible front-end analog solutions to precede the DSP. Thus, there is a need of a wideband and highly flexible \textsl{analog signal processing engine}. Microwave photonics offer this functionality, by exploiting the unique capabilities of photonics, to bring advantages in terms of size, weight and power (SWAP) budgets in radio-frequency signal processing.

As an emerging technology, One of the main driving forces for MWP in the near future is expected to come from broadband wireless access networks installed in shopping malls, airports, hospitals, stadiums, and other large buildings. The market for microwave photonic equipment is likely to grow with consumer demand for wireless gigabit services. For instance, the IEEE standard WiMAX (the Worldwide Interoperability for Microwave Access) has recently upgraded to handle data rates of 1~Gbit/s, and it is envisaged that many small, WiMAX-based stations or picocells will soon start to spring up. In fact, with the proliferation of tablet devices such as the iPads, more wireless infrastructure will be required. Furthermore, it is also expected that the demand for microwave photonics will be driven by the growth of fiber links directly to the home and the proliferation of converged and in-home networks. To cope with this growth scenario, future networks will be expected to support wireless communications at data rates reaching multiple gigabits per second. In addition, the extremely low power consumption of an access network comprised of pico- or femtocells would make it much greener than current macrocell networks, which require high-power base stations.

For the last 25 years, MWP systems and links have relied almost exclusively on discrete optoelectronic devices and standard optical fibers and fiber-based components which have been employed to support several functionalities like RF signal generation, distribution, processing and analysis. These configurations are bulky, expensive and power-consuming while lacking in flexibility. We believe that a second generation, termed as \textsl{Integrated Microwave Photonics} (IMWP) and which aims at the incorporation of MWP components/subsystems in photonic circuits, is crucial for the implementation of both low-cost and advanced analog optical front-ends and, thus, instrumental to achieve the aforementioned evolution objectives. This paper reviews the salient advances reported during the last years in this emergent field.

\section{Fundamentals of microwave photonics} 
\label{sec:fundMWP}
\begin{figure}
  \includegraphics*[width=\linewidth]{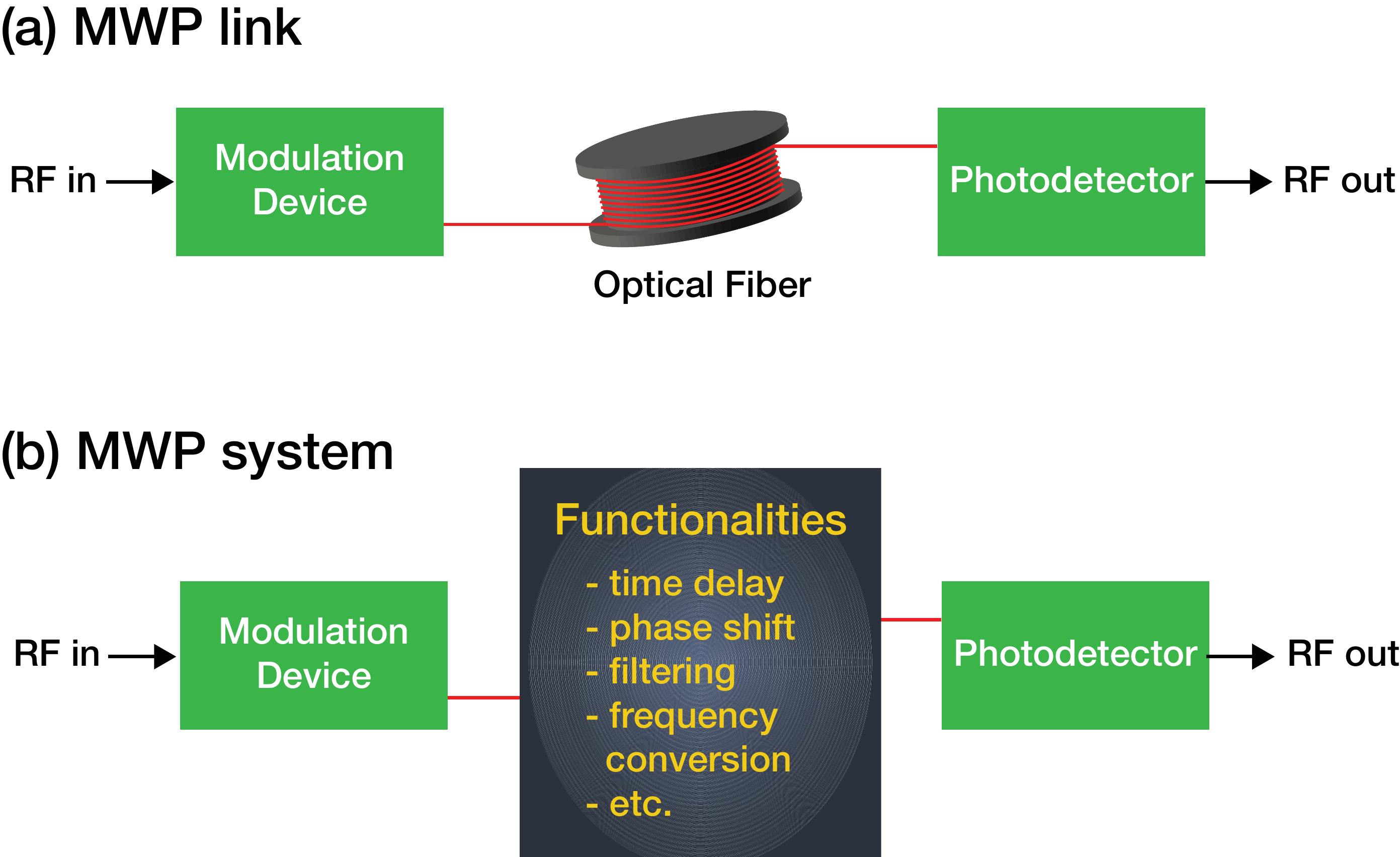}
  \caption{Schematics of (a) an MWP link and (b) a simple MWP system. The MWP link basically consists of a modulation device for E/O conversion and a photodetector for O/E conversion. Such an MWP link with added functionalities between the conversion will make an MWP system.}
  \label{MWPlink}
\end{figure}

The heart of any MWP system is an \textsl{MWP link}. As depicted in Figure~\ref{MWPlink}~(a), the link consists of a modulation device for electrical-to-optical (E/O) conversion connected by an optical fiber to a photodetector that does the O/E conversion. Most of the MWP links used today employ the intensity modulation-direct detection (IMDD) although as will be discussed in Section~\ref{sec:APL}, phase or frequency modulation schemes in combination with either direct detection or coherent detection are also gaining popularity. From the point of view of the modulation device, the modulation scheme employed in MWP links can be divided into two broad categories: direct modulation or external modulation. In the former, the modulation device is a directly modulated laser (DML) that acts as both  a light source and the modulator, while in external modulation the modulation device consists of a continuous wave (CW) laser and an external electro-optic modulator (EOM). Virtually all MWP links use p-i-n photodetectors for the O/E conversion. An excellent review of the range of devices that have been used in MWP links can be found in Chapter~2 of Ref~\cite{CoxBook2004}.

An \textsl{MWP system} is established by means of adding functionalities between the two conversions, i.e. processing in the optical domain (Figure~\ref{MWPlink}~(b)). The advantage of optical processing includes the large bandwidth, constant attenuation over the entire microwave frequency range, small size, lightweight, immunity to electromagnetic interference and the potential of large tunability and low power consumption. The capabilities of such MWP systems include the generation, distribution, control and processing of microwave signals. Some of the key functionalities in this case are high fidelity microwave signal transport, true time delay and phase shifting of microwave signals, frequency tunable and high selectivity microwave filtering, frequency up and down conversions and microwave carrier and waveform generations.     

In order to obtain full functionalities from the MWP systems, the MWP link needs to reach sufficient performance. The main hurdle for this is the fact that the E/O and O/E conversions add loss, noise and distortion to the RF signal being processed. Moreover, the relation between the RF loss and the optical loss in the MWP link is quadratic. This means that it is imperative to minimize the optical losses. For these reasons, we have seen that the best part of MWP activities of the 80's and 90's has been dedicated to design and to optimize the performance of MWP links. In the next section the figures of merit of MWP links is described. Comprehensive reviews of the progress in MWP links performance are reported in \cite{CoxMTT1997,CoxMTT2006}. 

\subsection{Figures of merit}
\label{subsec:FOM} 

The important figures of merit for MWP links are link gain, noise figure, input/output intercept points and spurious free dynamic range (SFDR). These metrics show the impact of losses, noise and nonlinearities in the link. 

The \textbf{link gain} describes the RF to RF power signal transfer in the MWP link or system. Due to the limited conversion efficiencies in the modulation device and photodetector, it is common that the MWP link shows negative link gain in the decibel scale, i.e. a net loss. This is especially true for the case of direct modulation since the link gain depends only on three parameters, namely the laser slope efficiency, the photodetector responsivity and the optical loss in the system. In the case of lossy impedance matching\footnote{In this case the impedances of both the modulation device and the photodetector are regarded as purely resistive, and resistors are added in series or in parallel to match the input and output impedances to the $50\,\Omega$ source and load resistances. For in-depth discussion of impedance matching in MWP links the reader is referred to Chapter~2 of Ref~\cite{CoxBook2004}.} employed at the laser and the photodetector , the link gain of of direct modulation link can be expressed as 

\begin{equation}
  \label{eq:DMLgain}
  g_{\mathrm{link,DM}} = \frac{1}{4}\left(\frac{s_{\mathrm{LD}}r_{\mathrm{PD}}}{L}\right)^2
\end{equation}
where $s_{\mathrm{LD}}$ is the laser slope efficiency in W/A, $r_{\mathrm{PD}}$ is the photodiode responsivity in A/W and $L$ is the optical loss in the link defined as $L=\left[1..\infty\right]$. In the case of external modulation, the link gain is a function of relatively more parameters, involving the laser, the EOM and the photodetector. For example, for a link employing a CW laser with output optical power $P_{\mathrm{o}}$ and a Mach-Zehnder modulator (MZM) with an insertion loss of $L_{\mathrm{MZ}}$, an RF half-wave voltage of $V_{\pi,\mathrm{RF}}$ and a biased at the angle of $\phi_{\mathrm{b}}=\pi V_{\mathrm{b}}/V_{\pi,\mathrm{DC}}$, the link gain for the case of lossy impedance matching can be expressed as   
      
\begin{equation}
  \label{eq:MZgain}
  g_{\mathrm{link,MZ}}=\left(\frac{\pi\,r_{\mathrm{PD}}\,R_{\mathrm{L}}\,P_{\mathrm{o}}\,\sin{\phi_{\mathrm{B}}}}{4\,L_{\mathrm{MZ}}\,V_{\pi,\mathrm{RF}}}\right)^2
\end{equation}
where $R_{\mathrm{L}}$ is the load resistance and $V_{\pi,\mathrm{RF}}$ is the \textcolor{black}{(frequency-dependent) RF} half-wave voltage. A careful look at  Eq.\eqref{eq:MZgain} will reveal that the link gain scales up quadratically with the input optical power from the laser. This means that the link gain can be increased by pumping more optical power to the system. This technique has been effectively used to demonstrate MWP links with net gain (i.e. positive link gain) instead of loss, where the value as high as +44~dB has been demonstrated using an ultra-low half-wave voltage MZM \cite{UrickElectLett2006}. Comparing Eq.\eqref{eq:DMLgain} and Eq.\eqref{eq:MZgain} one can identify that the gain of a direct modulation MWP link is relatively more difficult to increase since the most of the time the laser and photodetector have fixed range of slope efficiency and responsivity.  

The E/O and O/E conversions also add \textbf{noise} to the MWP system. The dominant noise sources are thermal noise, shot noise and relative intensity noise (RIN). In case of systems with optical amplifiers, the amplified spontaneous emission (ASE) noise of the amplifier will often dominate over the other sources. The total noise power in the link (with a receiver electrical bandwidth $B$) comprises of the electrical powers delivered to \textsl{a matched load} for the three sources considered above. Hence: 

\begin{equation}
\label{eq:noisepower}
p_{\mathrm {N}}= \left(1+g_{\mathrm{link}}\right)\,p_{\mathrm {th}}+\frac{1}{4}\,p_{\mathrm {shot}}+\frac{1}{4}\,p_{\mathrm {rin}}
\end{equation}
where $p_{\mathrm {th}}$, $p_{\mathrm {th}}$ and $p_{\mathrm {th}}$ are the thermal noise, shot noise and RIN powers defined as

\begin{equation}
p_{\mathrm {th}}=kTB\,
\end{equation} 

\begin{equation}
p_{\mathrm {shot}}=2q\,{I_{\mathrm {D}}}BR_{\rm L}\,
\end{equation} 

\begin{equation}
p_{\mathrm {RIN}}=\mathrm{RIN}\,{I_{\mathrm {D}}}^2BR_{\mathrm {L}}\,.
\end{equation} 
\nolinebreak 
The quantity $I_{\mathrm{D}}$ in the equations above is the average photocurrent. For an MWP link with MZM, the photocurrent can be expressed as  

\begin{equation}
\label{eq:IavMZ}
I_{\mathrm {D,MZ}}=\frac{r_{\mathrm {PD}}\,P_{\rm o}}{2L_{\mathrm{MZ}}}\left(1-\cos\phi_{\mathrm B}\right)\,.
\end{equation}   

The \textbf{noise figure} is a useful metric that measures the signal-to-noise ratio (SNR) degradation in the system, expressed in decibels. It is determined by the noise power and the link gain and can be written as

\begin{equation}
\label{eq:noisefigure}
{\mathrm {NF}}= 10\log_{10}\left(\frac{p_{\mathrm N}}{g_{\mathrm{link}}kTB}\right)\,.
\end{equation}
\nolinebreak
The noise figure is often used as an important measure of the usefulness of MWP links and systems. In recent years, efforts are directed towards realizing MWP links with sub-10~dB noise figure. Several groups  have been successful achieving this feature using the low biasing of a low $V_{\pi}$ MZM in conjunction with a very high power optical source and a customized high power-handling photodetector \cite{AckermanIMS2007,KarimPTL2007}. Low biasing the MZM away from the quadrature $\left(\phi_{\mathrm{b}}=\pi/2\right)$ towards the minimum transmission point $\left(\phi_{\mathrm{b}}=0\right)$ is advantageous for the noise figure because the noise powers reduces faster with the bias compared to the reduction of the link gain. This can be seen from Eq.~\eqref{eq:MZgain} and Eq.~\eqref{eq:IavMZ}. Sub-10~dB noise figure has also been achieved by using a high power source, a dual-output MZM and a balanced detection scheme \cite{AckermanIMS2007,McKinneyPTL2007}. By carefully matching the path length of the fibers going to the balanced photodetector (BPD) the RIN which is common mode noise in the paths can be canceled. The most recent review of progress in achieving MWP link gain with $G > 0~\mathrm{dB}$ and $\mathrm{NF} < 20~\mathrm{dB}$  can be found in \cite{UrickIMS2011}. 

\begin{figure}
  \includegraphics*[width=\linewidth]{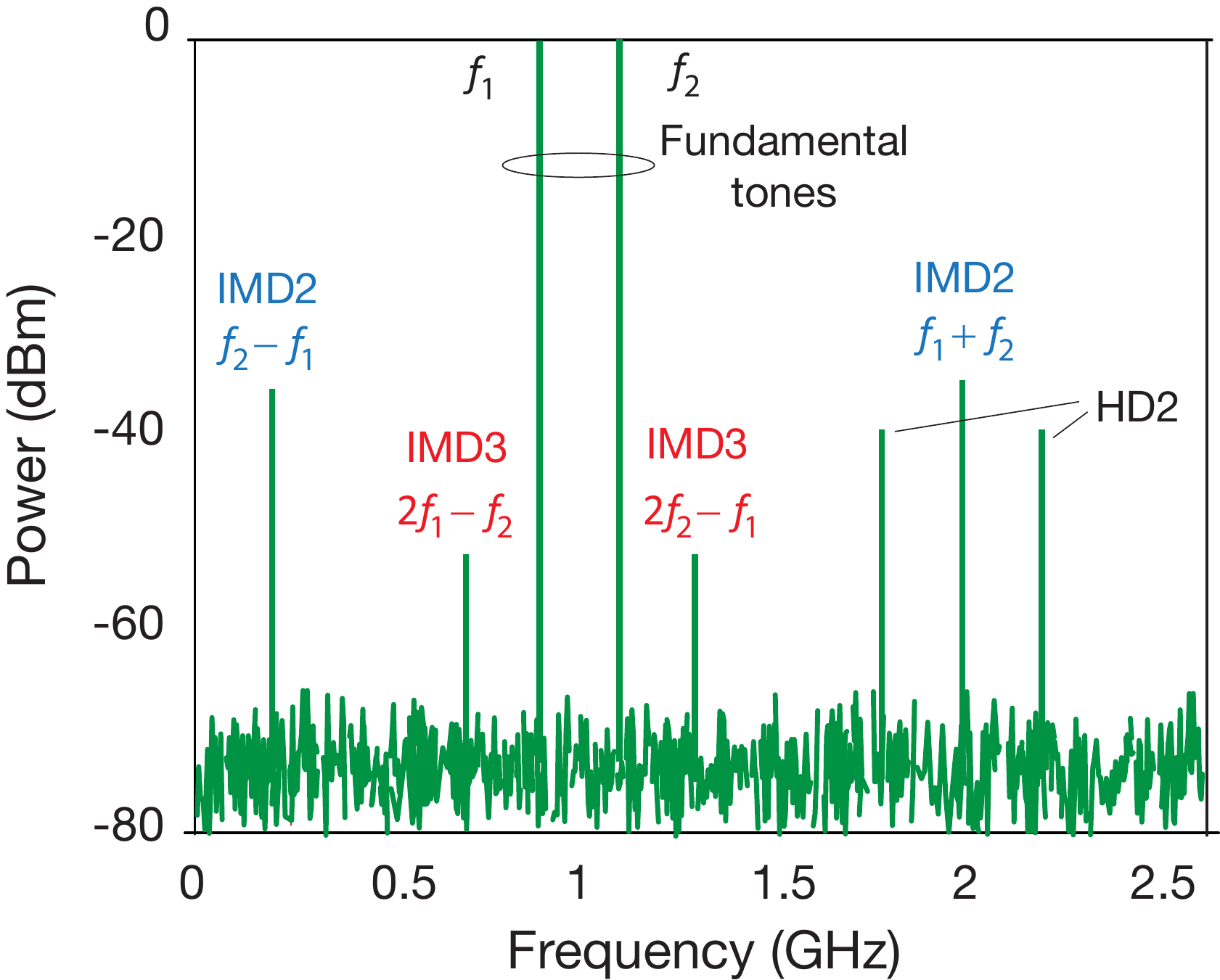}
  \caption{An illustration of a typical two-tone test output RF spectrum of an MWP link. IMD: intermodulation distortion, HD: harmonic distortion. }
  \label{twotone}
\end{figure}

The E/O and O/E conversions in the MWP link also add \textbf{nonlinear distortions} to the output RF signal. The most common way to probe these nonlinearities is to use the so-called two tone test. In such a test, the input to the link is a pair of two closely spaced tones, for example at frequencies $f_{1}$ and $f_{2}$. Due to the nonlinear response of the link (i.e. components like the EOM or the photodetector), these tones will generate new frequency components called the intermodulation distortions (IMDs). The second-order intermodulation (IMD2) is generated due to the quadratic nonlinearity in the link and the frequency components appear at the sum and the difference of the modulating frequencies $\left(f_{1} \pm f_{2}\right)$. The third-order intermodulation (IMD3) is generated by cubic nonlinearity in the link and appear at the sum and the difference of twice of one frequency with the other frequency $\left(2f_{1} \pm f_{2}, 2f_{2} \pm f_{1}\right)$. An illustration of the output two tone test spectrum of an MWP link depicting the fundamental tones and the IMDs is shown in Figure~\ref{twotone}. The spectrum in this figure reveals that the distortion component that fall closest to the fundamental signals are the IMD3 terms at $2f_{1}-f_{2}$ and $2f_{2}-f_{1}$, which most of the time cannot be filtered out. Thus, there is hardly any usable signal bandwidth that is free from these spurious signals. For this reason the IMD3 is regarded as the main limiting distortion factor in MWP links. As for the even order distortions, the IMD2 fall relatively far from the fundamental signals. But as the signal bandwidth increases, the separation between the signals and these distortion terms reduces. For a wideband system with a multioctave signal bandwidth, i.e. the case where the highest frequency component of the signal, $f_{\mathrm{high}}$ is more that twice of the lowest frequency component, $f_{\mathrm{low}}$, IMD2 will interfere with the signal. This  is in contrast with a  narrowband system with sub-octave bandwidth $\left(f_{\mathrm{high}}< 2f_{\mathrm{low}}\right)$, where IMD2 can easily be filtered out.     

\begin{figure}
  \includegraphics*[width=\linewidth]{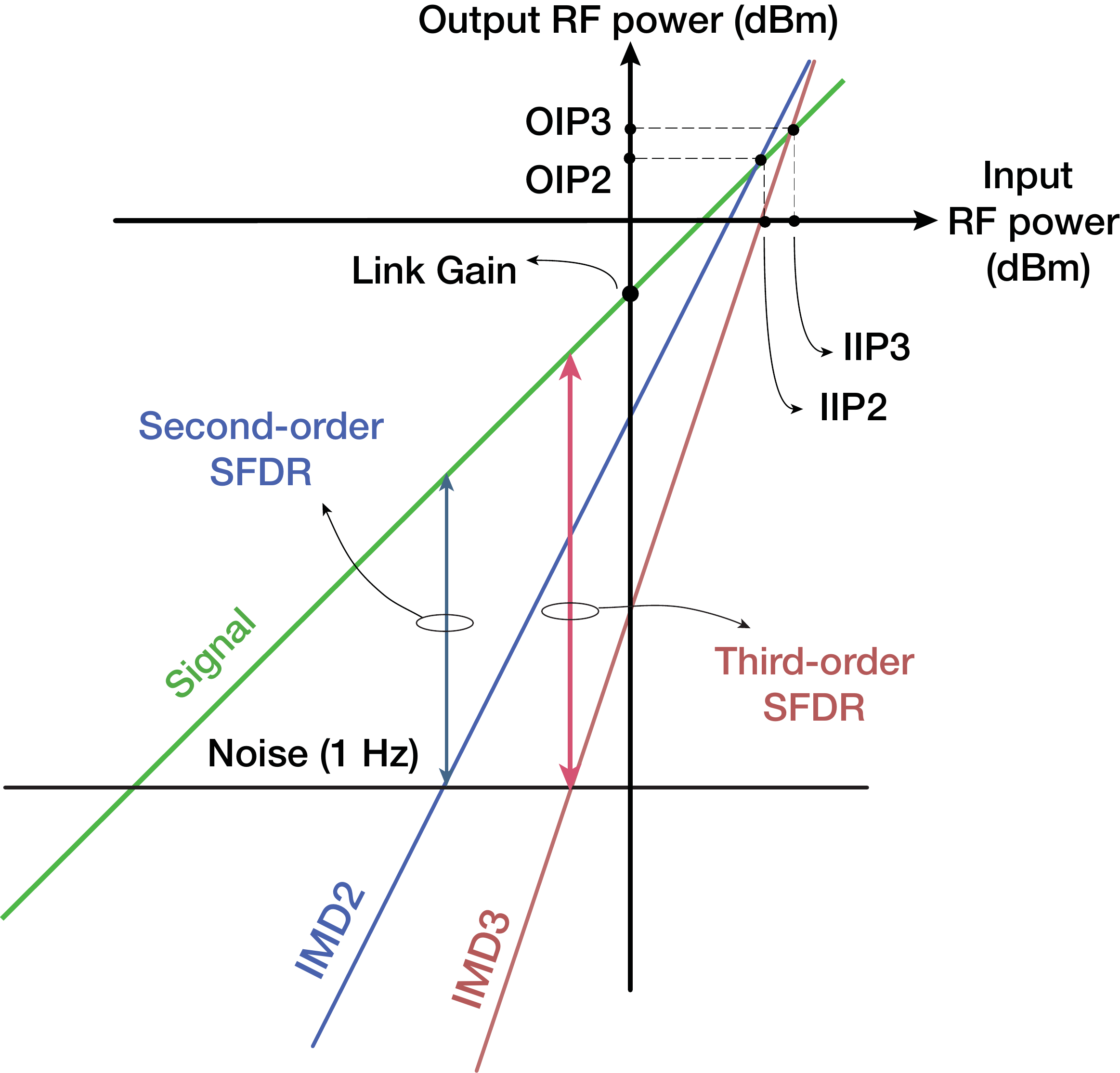}
  \caption{The relation of the input RF power to an MWP link with the output RF powers of the fundamental tone and the IMD products expressed in decibels. From such a graph, key link metrics such as gain, intercept points and SFDR can be deduced.}
  \label{SFDR}
\end{figure}

It is often useful to investigate how the power of each component in the output spectrum shown in Figure~\ref{twotone} varies with the input signal power. Such plot is shown in Figure~\ref{SFDR}. Here we have plotted the fundamental signal and the IMD2 and IMD3 powers decibels. The fundamental signal, being linearly dependent on the input signal, is plotted as line with the slope of one. The IMD$2$ power has a quadratic relation with the input RF power and thus in such a plot will appear as a line with a slope of 2 with respect to the input signal power. The IMD3, having a cubic dependence with the input power, is plotted as a line with the slope of 3. At some point, the extrapolated fundamental power and the $n^{\mathrm{th}}$ order IMD power will intersect. This intersection is known as the \textbf{intercept point}. Depending on which power this point is referred to, for each distortion order an input intercept point (IIP$n$) and output intercept point (OIP$n$) can be defined. These two intercept points are related to each other by the link gain via the relation $\mathrm{OIP}n\left(\mathrm{dBm}\right)=\mathrm{IIP}n\left(\mathrm{dBm}\right)+G_{\mathrm{link}}\left(\mathrm{dB}\right)$. It is important to mention however, that these intercept points cannot be directly measured since the fundamental powers will undergo a compression \cite{KolnerAO1987}. For this reason, the intercept points are deduced from the extrapolation of the measured fundamental and IMD powers.

It is useful to inspect the expressions of the intercept points of an MWP link with an MZM. The reason is that the nonlinearity profile is well known due to the well-defined sinusoidal transfer function of the MZM. Since the performance of such a link is well-explored, often its intercept points are used as the benchmarks for judging the performance of a novel type of MWP links and systems. We will see later on in Section~\ref{sec:APL} that it is indeed the case. The IIP2 and IIP3 of an MZM link can be written as 

\begin{equation}
\label{eq:IIP2MZ}
{\rm IIP2_{\mathrm {MZ}}}=\frac{2}{R_{\mathrm L}}\left({\frac{V_{\pi,\mathrm {RF}}}{\pi }}\tan{\phi_{\rm B}}\right)^2
\end{equation}
 
\begin{equation}
\label{eq:IIP3MZ}
{\rm IIP3_{\mathrm {MZ}}}=\frac{4\,\left(V_{\pi,\mathrm{RF}}\right)^2}{{\pi}^2\,R_{\mathrm{L}}}
\end{equation} 
\nolinebreak
The IIP2 is very sensitive  to the bias angle and ideally goes to infinity at quadrature because the even order distortion vanishes at this bias point. The IIP3 is, however, independent of the bias angle and virtually depends only to the modulator RF half-wave voltage. The simplicity of the IIP3 expression is very useful when comparing the performance of different types of links. The OIP3 of an MZM link on the other hand, is bias dependent. However, the expression for the OIP3 at quadrature bias is very simple, which is

\begin{equation}
\label{eq:OIP3MZMquad}
{\rm OIP3_{\rm Q}}={I^2_{\mathrm {Q}}}\,R_{\mathrm{L}}\,,
\end{equation}
where $I_{\mathrm {Q}}$ is the average (DC) photocurrent in the quadrature bias case, which can be obtained by substituting $\phi_{\mathrm{B}}=\pi/2$ into Eq.~\eqref{eq:IavMZ}. 

The figure of merit that incorporates the effect of noise and nonlinearity in the MWP link is the \textbf{spurious-free dynamic range (SFDR)}. The SFDR is defined as the ratio of input powers where, on the one hand, the fundamental signal power is equal to the noise power and, on the other hand, the $n^{\rm th}$-order intermodulation distortion (IMD$n$) power is equal to the noise power. In terms of output powers, this can be interpreted as the maximum output SNR that can be achieved while keeping the IMD$n$ power below the noise floor. \textcolor{black}{The latter definition} is illustrated in Figure~\ref{SFDR}. For link designers, it is desirable to express the $n^{\mathrm{th}}$-order SFDR (SFDR$_{n}$) in terms of other link measurable parameters such as the link gain, noise figure, and the intercept points. Such expressions can also be deduced from Figure~\ref{SFDR}. The SFDR$_{n}$ in terms of IIP$n$ can be written as : 
\nolinebreak
\begin{equation}
\label{eq:SFDRin}
{\rm SFDR}_{n}=\frac{n-1}{n}\left({\rm IIP}n-{\rm NF}+174\right)\,
\end{equation}
\nolinebreak
where the SFDR, IIP3 and NF are expressed in decibels. Alternatively we can express the SFDR in terms of OIP$n$, yielding
\nolinebreak
\begin{equation}
\label{eq:SFDRout}
{\rm SFDR}_{n}=\frac{n-1}{n}\left({\rm OIP}n-{\rm NF}-G_{\mathrm{link}}+174\right)\,.
\end{equation}
\nolinebreak
Again, here $G_\mathrm{link}$ is the link gain in decibels. SFDR$_{n}$ is usually expressed in ${\rm dB}\cdot{\rm Hz}^{\left(\frac{n-1}{n}\right)}$. This is essentially the same as saying that the SFDR is measured in dB in 1~Hz noise bandwidth. \textcolor{black}{However, in practice the noise bandwidth is larger than 1~Hz. In this case, it is useful to inspect how the SFDR scales with the bandwidth ($B$), as shown in the following equation:} 
\nolinebreak
\begin{equation}
\label{eq:SFDRscale}
{\rm SFDR}_{n}\left(B\,{\rm Hz}\right)={\rm SFDR}_{n}\left(1\,{\rm Hz}\right)-\left(\frac{n-1}{n}\right)10\log_{10}\left(B\right)\,.
\end{equation}
\nolinebreak
\textcolor{black}{As an example, consider a system with an SFDR$_{3}$ of 110~dB.Hz$^{2/3}$. To calculate the SFDR$_{3}$ in 1~MHz bandwidth, first we calculate the factor $10\log_{10}\left(B\right)$ which is $60\,{\rm dB}$. Thus, the dynamic range in 1~MHz is $70\,{\rm dB}$.}

Optimizing the SFDR of an MWP link in a wideband manner has been the holy-grail in microwave photonics. As mentioned earlier, using an MZM one can achieve noise figure reduction via low biasing. According to Eq.~\eqref{eq:IIP3MZ} and Eq.~\eqref{eq:SFDRin}, the third-order SFDR of such a link can be increased. However, the low biasing reduces the IIP2 dramatically, making the link only suitable for narrowband (i.e., less than one octave bandwidth) signals. Moreover, the SFDR of the MZM link is always bounded by the third-order nonlinearities of the MZM. For this reason in the past people have turned to linearization techniques to overcome the MZM nonlinearities. However, the distortion cancellation often works only in a relatively narrow operating bandwidth and is critically sensitive to the modulator parameters \cite{Cummings1998}. A type of link that has been theoretically predicted to show a very large dynamic range (over ${150~\rm dB}\cdot{\rm Hz}^{2/3}$) is called the Class-B MWP link \cite{DarcieJLT2007}. The realization of such a link, however, has been found challenging. The properties of Class-B MWP links will be discussed in more detail in Section~\ref{sec:APL}. Currently successful techniques to push the dynamic range of MWP links have shown the SFDR in the range of ${120-130~\rm dB}\cdot{\rm Hz}^{2/3}$. The most recent review of the techniques achieving a wideband high SFDR is reported in \cite{UrickSPIE2012}.

\subsection{Applications}
\label{subsec:function} 

The MWP concept has found widespread applications over the last 20 years. The earliest application for MWP technique was microwave signal distribution \cite{SeedsMWP2002}. Here the MWP link is used as a direct replacement of coaxial cables, exploiting the advantage in size, weight, flexibility and flat attenuation over the entire frequency range of interest. This concept is extended for antenna remoting purposes. Here, the MWP link is used to separate the highly sensitive and complex signal processing part of a radio receiver away from the antenna. In this way, the signal processing part can be protected in case of antennas deployed in harsh environments, for example in radar system \cite{Roman1998} or in radio astronomy \cite{Montebugnoli2005}. On the other hand, this concept is also very attractive for distributed or multiple antennas system, where a large number of antennas are needed to extend the coverage of service, for example in mobile communications. In this case, the MWP concept is used to centralize the signal processing chain (modulation, filtering etc.) and separate it away from the antennas. The advantage is that the antenna architecture can be simplified. This is the concept of radio over fiber \cite{SeedsMWP2006,LimJLT2010}, which is currently the main commercial driver\footnote{The commercial aspects of MWP were recently covered in the Nature Photonics Technology Focus, vol.~5, issue~12, December~2011.} for MWP. 

Although signal distribution is the main driver for MWP, applications like microwave signal generation and processing are catching up. The generation of high purity microwave signals and high complexity ultra-broadband waveforms are the latest development in MWP. The main attraction for signal generation using MWP technique is the large frequency tunability and the potential of reaching very high frequencies (up to the THz region) using relatively simpler technique compared to the traditional microwave/electronic approach. Moreover, the distribution of such high frequency signals using extremely low-loss optical fibers is also attractive, which would have been very lossy using coaxial cables. For waveform generations MWP techniques offer broad bandwidth and the full reconfigurabilities of the phase and amplitude of the the RF waveforms \cite{YaoNatPhoton2010,YaoOptComm2011}.   

For microwave signal processing, MWP techniques have enabled filtering, tunable true time delay and wideband phase shifting of microwave signals. The MWP concept added value are the operation bandwidth and the potential of fast and agile reconfigurabilities of these functionalities. Combining these basic functionalities lead to the realization of MWP processors for optical beamforming and phase array antenna systems.

\section{Integrated microwave photonics}
\label{sec:integratedMWP}

There are several factors that still distance the MWP concept to be widely implemented in real life applications and beyond the laboratory setups. The first factor is performance, particularly in terms of dynamic range. Typically MWP systems show prominent functionalities (for example true time delay or pulse shaping) over a large bandwidth but the performance in terms of dynamic range is not good enough to actually replace the traditional microwave solution. The other factors are reliability and cost. Most of MWP systems are composed of discrete components, i.e. lasers, modulators and detectors connected by fiber pigtails. This imposes several problems. Discrete components occupy larger size while interconnections with fiber pigtails reduce the system sturdiness. These lead to reduced reliability of the system. Second, the use of discrete components leads to a high system cost since each component will bear packaging cost. The use of discrete components might also lead to higher power consumption. These factors have counted against MWP solutions to replace traditional microwave solutions which have reached maturity over years of development.  

The promise of ultra-broad bandwidth and excellent reconfigurabilities of MWP systems are still very much tantalizing to be explored if the drawback factors mentioned earlier can be addressed. If MWP systems can be more viable in terms of cost, power consumptions and reliability, they will be able to replace microwave solutions for processing beyond only replacing the coaxial cables. Many believe that these challenges can be addressed by RF and photonics integration \cite{JacobsOFC2007,GasullaPhotonicsJournal2011,CapmanyNatPhoton2011, WoodwardMWP2011}. With photonic integration, one can achieve a reduction in footprint, inter-element coupling losses, packaging cost as well as power dissipation since a single cooler can be used for multiple functions \cite{ColdrenMWP2010}. Thus, the MWP functionalities can be brought a step closer towards real applications and subsequently commercial marketplace. 

Even though at first the concept of integrated MWP seems to be very much in line with the recent trend of large scale PIC technology \cite{JalaliMicrowaveMag2006,SorefJSTQE2006,JalaliJLT2006,LiangElectLett2009,ColdrenJLT2011,SmitLPR2012}, certain differences occur in terms of requirements and market/applications scale. Large scale photonic integration has heavily been driven by the so called "digital" applications, like high capacity optical communications and optical interconnects. This concept relies heavily on increasing speed, component counts as well as incorporating as much functionalities (active and passive) in a single chip/technology platform \cite{KishJSTQE2011,OrcuttOpex2011,OrcuttOpex2012,KoehlOPN2011}. Since currently there is no single technology platform that offers the best performance in all aspects, large scale integration often compromises the total photonic system performance. 

Due to the stringent requirements in handling analog signals, PIC technology for integrated MWP should show high performance, most of the time higher than the one expected from the digital applications. Moreover, at the present state, MWP is addressing lower-volume market, hence lower volume PIC productions. These are the aspects that we believe will force PIC technology players to take a different approach to integrated MWP.    

In the next section we will highlight a host of PIC technologies that have recently been demonstrated in MWP systems. Two key criteria that will be discussed from these technologies are the performance and the availability to MWP community. As have been discussed in the previous section, an important objective in MWP processing is to optimize the system link gain while maintaining a healthy noise figure. This objective thus dictates that the insertion loss of the PIC in the MWP system should be minimized. In most cases, this will lead to a stringent requirements in the propagation loss and the fiber-to-chip coupling losses in the PIC.     
 
As for the availability of PIC technology, integrated MWP will hugely benefit from initiatives like ePIXfab \cite{DumonEpixfab} and Jeppix \cite{Jeppix} in Europe and OpSIS \cite{HochbergNatPhotonics2010} in the USA, that allow users to access fabrication technologies (for silicon photonics or indium phosphide technologies) that would otherwise be too costly to bear by individual users. This is already reflected from the growing number of reported integrated MWP devices and systems that have enabled by these initiatives. In the next section some of the available platforms for integrated MWP are reviewed.

\section{Photonic integration technology}
\label{sec:technology} 

At this point, commercial wafer scale fabrication of photonic devices have crystallized into several major technologies: compound semiconductors (GaAs, InP), \textcolor{black}{also referred to as III-V, since the constituent elements come from columns 3 and 5 of the periodic table}, nonlinear crystals (LiNbO\textsubscript{3}), dielectrics (silica and silicon nitride based waveguides) and element semiconductor (silicon-on-insulator (SOI)). Each technology has boast specific strength like light generation and detection, modulation, passive routing with low propagation loss, electronic integration, ease in packaging, etc. Nevertheless, integration in single platform without sacrificing an overall system performance has not been achieved \cite{AurrionIPC2011}. In the past 20 years, four platforms have been frequently used to demonstrate integrated MWP functionalities. They are InP, silica planar lightwave circuits (PLCs) , silicon-on-insulator (SOI) and Si\textsubscript{3}N\textsubscript{4}/SiO\textsubscript{2}, known as the TriPleX\texttrademark\, waveguide technology. In this section we will focus on these technologies and briefly comment on other technologies such as LiNbO\textsubscript{3}, polymers and chalcogenides in Subsection~\ref{subsec:other}.

\subsection{Indium Phosphide (InP)}
\label{subsec:InP} 

The InP platform inherently supports light generation, amplification, modulation, detection, variable attenuation, and switching in addition to passive functionalities. For this reason, InP photonics is highly attractive for large scale photonic integration as have been consistently demonstrated by the company Infinera \cite{KishJSTQE2011}. In this case the PICs are highly complex with components (lasers, modulators, arrayed waveguide gratings) count of more than 400 integrated in a single chip. The PICs are developed for the application of the high-speed digital optical communications (100/500~Gb/s). As for MWP applications \cite{ColdrenMWP2010,ColdrenJLT2011}, InP PICs have been developed for a number of applications, like optical beamforming \cite{StulemeijerPTL1999}, fully programmable MWP filters using ring resonator structures \cite{NorbergPTL2010,NorbergJLT2011,GuzzonOpex2011,GuzzonJQE2012} and  monolithic integrated optical phase-locked loop for coherent detection scheme \cite{LiPTL2011,BhardwajEL2011,KrishnamachariMOTL2011}. 

It is well known that the propagation losses in InP passive optical waveguides can be an order of magnitude higher compared to waveguides based on silica or silicon \cite{LiangElectLett2009}. For example, in \cite{StulemeijerPTL1999} a waveguide propagation loss of 1.4~dB/cm was reported. For some applications this large propagation loss needs to be compensated by  optical gain from active components like semiconductor optical amplifiers (SOAs). This is especially important for cascade stages of resonator filters in the active MWP filters reported in \cite{NorbergPTL2010,NorbergJLT2011,GuzzonOpex2011} which are highly sensitive to losses. An issue arising with such active filters are the noise added from the SOAs that might limit the SFDR of the device. More detail about the SFDR of such filters can be found in \cite{GuzzonJQE2012}. This filter structure will be revisited in Section~\ref{subsec:coherent} where MWP filter is discussed in more details.    

The capability of InP PIC to support modulation and detection functionalities has been exploited for coherent receiver in phase modulated MWP links. The most critical element in this receiver is a linear OPLL for linear phase demodulator. Through feedback, the OPLL forces the phase of a local (tracking) phase modulator to mirror the phase of an incoming optical signal. Thus, the output from the photodetector is a scaled replica of the RF input \cite{LiPTL2011}. The tracking phase modulator has to nearly instantly track the phase deviation out of the photodetector, dictating that delay must be very small. This call for photonic and electric circuit integration. The most recent approaches in the realization of such coherent receiver will be discussed in more details in Section~\ref{subsec:OPLL}.


\begin{figure}
  \includegraphics*[width=\linewidth]{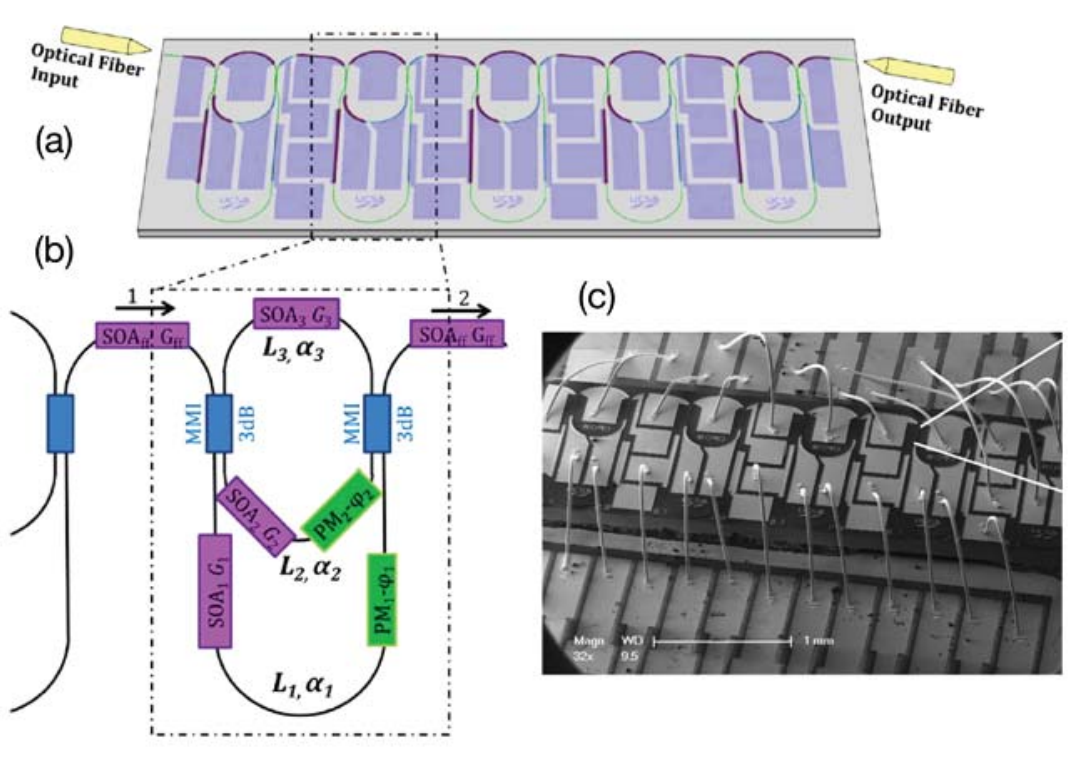}
  \caption{(a) Illustration of programmable filter array chip. (b) A single filter stage and its functional components (SOAs, PMs and 3 dB MMI couplers) shown
schematically. (c) Scanning electron microscopy (SEM) image of a programmable photonic filter device wire bonded to a carrier. (from \cite{NorbergJLT2011} courtesy of the IEEE).}
  \label{InPGuzzon}
\end{figure}

\subsection{Silica PLCs}

Silica glass planar lightwave circuits (PLCs) are widely used as key devices for wavelength division multiplexing (WDM) transmission and fiber-to-the-home (FTTH) systems because of their excellent optical properties and mass-producibility. In such applications, the PLCs have been used for wavelength multi/demultiplexers, optical add/drop or cross-connect switches and programmable filters \cite{HimenoJSTQE1998}. The silica-based waveguides are very popular due to the very low propagation loss characteristic. The lowest propagation loss in such a waveguide at $\lambda=1550$~nm has been demonstrated using a phosphorus-doped silica on silicon waveguide with a propagation loss of 0.85~dB/m \cite{AdarJLT1994}. But this has been shown on  a waveguide with a low refractive index contrast of 0.7\%.  Such a low contrast is less attractive for photonic chip integration since it only allows large bending radius and hence, larger chip size. 

Several MWP functionalities have been demonstrated in silica PLCs over the years. Horikawa et al. have demonstrated a true time-delay beamforming network based on silica waveguides with an index contrast of 1.5\%, a minimum bending radius of 2~mm and a propagation loss of 0.1~dB/cm \cite{HorikawaIMS1995,HorikawaOFC1996}. More recently Grosskopf et al. demonstrated a beamforming network on a lower index constrast silica waveguides ($\Delta n = 0.7$\%) and a minimum radius of 10~mm \cite{GrosskopfFIO2003}. In 2005 Rasras et al. \cite{RasrasPTL2005} proposed a wide-tuning-range optical delay line in a high ($\Delta n = 2$\%) index contrast waveguides. This device integrates four-stage ring resonator all-pass filters (APFs) with cascaded fixed spiral-type delay waveguides and enables continuous tuning ranges up to 2.56~ns. The minimum bending radius and the reported propagation loss are 1~mm and  0.07~dB/cm, respectively. More details on these functions can be found in Section~\ref{sec:delay}.

Beside delay lines and beamformer, the silica PLCs have also been used to demonstrate integrated frequency discriminator \cite{LaGassePTL1997,WyrwasMWP2011,WyrwasThesis2012} (more details in Section~\ref{subsec:PMIM}) and an arbitrary waveform generator \cite{SamadiOptComm2011} (see Section~\ref{subsec:arbitrary}). Recent investigations in silica PLCs for MWP applications are aimed at increasing the index contrast to 4\% (and above) and to reduce the footprint of the device \cite{CallenderSPIE2012}.

\subsection{Silicon photonics}
\label{subsec:SOI} 
Silicon photonics  is one of the most exciting and fastest growing photonic technologies in recent years. The initial pull of this technology is its compatibility with the mature silicon IC manufacturing. Another motivation is the availability of high-quality silicon-on-insulator (SOI) planar waveguide circuits that offer strong optical confinement due to the high index contrast between silicon $(n = 3.45)$ and SiO\textsubscript{2} $(n = 1.45)$. This opens up miniaturization and large scale integration of photonic devices. Moreover, it has also been shown that silicon has excellent material properties like high third-order optical nonlinearities which, together with the high optical confinement in the SOI waveguides, enable functionalities like amplification, modulation, lasing, and wavelength conversion \cite{JalaliJLT2006,RongNatPhoton2007}. Various review papers\footnote{See also Nature Photonics Technology Focus, vol.~4, issue~8, August~2010.} have been published highlighting recent breakthroughs and novel devices in this technology \cite{JalaliMicrowaveMag2006,SorefJSTQE2006,JalaliJLT2006} .  

The past 15 years have also seen significant increase of silicon photonics implementation in MWP systems. As highlighted earlier in this section, the progress showed in silicon photonics for MWP takes a slightly different direction compared to applications like photonic interconnect or high speed data communications. In these latter fields, the use of silicon photonics is focused on large scale monolithic integration combining passive waveguides, modulators, detectors and sometimes light sources. But silicon is not ideal for electro-optic modulators and detectors in 1550~nm operating wavelengths \cite{JalaliJLT2006,HochbergNatPhoton2012}. Thus, from the perspective of system performance, silicon modulators, detectors and lasers cannot yet provide the stringent requirement of MWP. For this reason, most of the advances in silicon photonics for MWP have been focused on passive reconfigurable devices and/or devices exploiting optical nonlinearities. 

\begin{figure}
  \includegraphics*[width=\linewidth]{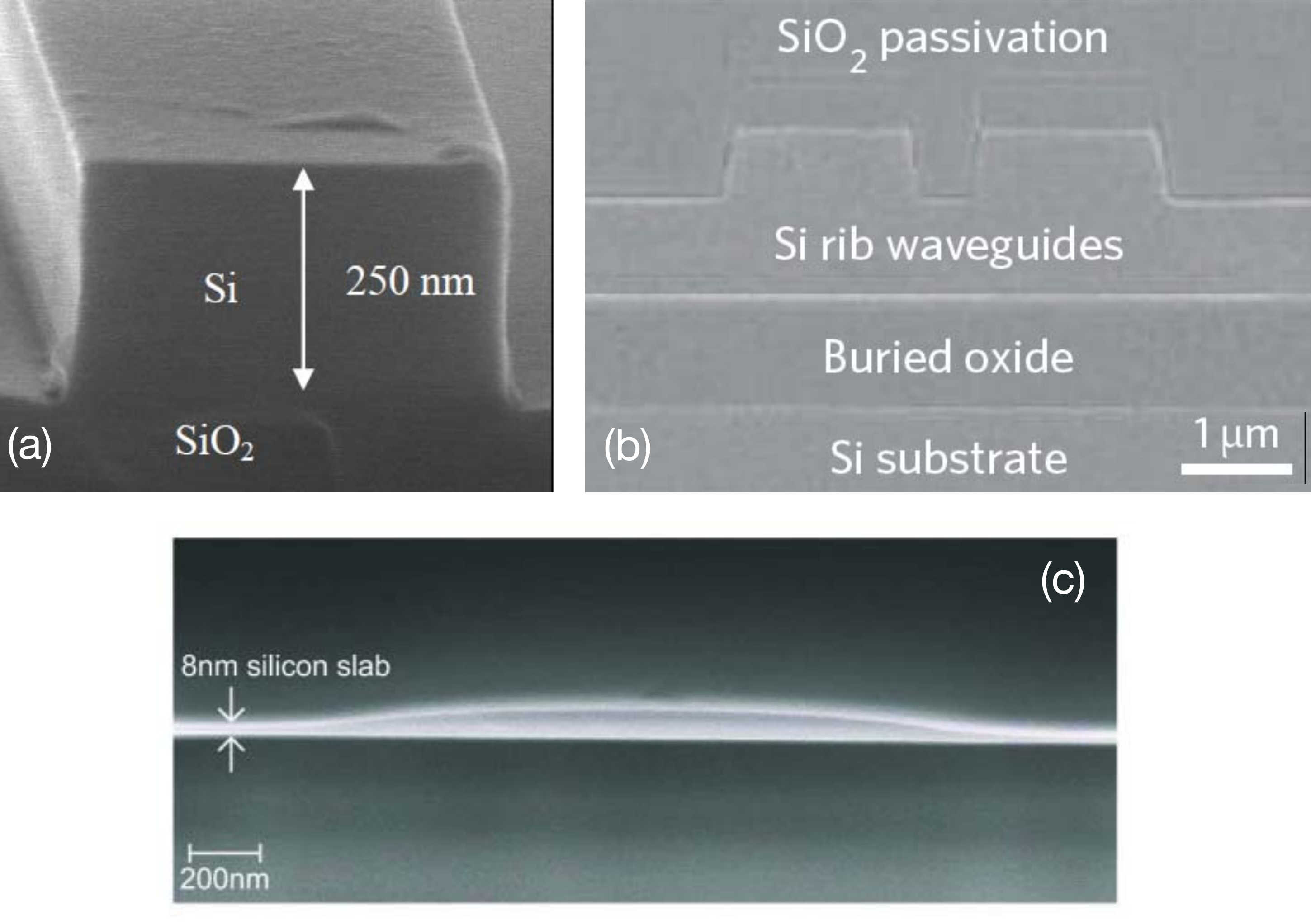}
  \caption{(a) Silicon strip waveguide/nanowire (from \cite{XiaoOpex2007}, courtesy of the OSA). (b) Rib waveguides (from \cite{RongNatPhoton2007}, courtesy of the Macmillan Publishers Ltd). (c) Etchless SOI strip (from \cite{CardenasOpex2009}, courtesy of the OSA).}
  \label{SOIwaveguide}
\end{figure}

The propagation loss in SOI waveguides has a large variation depending on the waveguide dimensions and processing conditions.  There are two types of waveguides commonly used in silicon photonics community: \textsl{shallow ridge or rib waveguides} with a width of 1-8 $\mu$m and \textsl{silicon strip waveguides (or nanowires)} with a dimension of approximately 500~nm wide by 250~nm thick. The rib waveguides exhibit relatively low losses down to 0.1-0.5~dB/cm, but limited in bending radius to hundred of micrometers \cite{FischerPTL1996,DongOpex2010_loss,RasrasJLT2009,IbrahimOpex2011,KhanOpex2011,GiuntoniOpex2012}. Strip waveguides on the other hand exhibit much higher losses with the lowest reported cases are in the order of 1-3~dB/cm \cite{XiaNatPhoton2007,XiaoOpex2007,GnanEL2008,BogaertsJLT2009} but they also allow ultra compact devices due to the tight minimum bending radius which is in the order of a few micrometers. Because of the high index contrast of the SOI waveguides surface roughness in due to imperfect etching will result in high scattering losses. But when an etchless process is used, the loss of SOI strip waveguides/nanowires could be as low as 0.3~dB/cm \cite{CardenasOpex2009}. A typical cross sections of nanowires, rib waveguides and the etchless strip waveguide are depicted in Figure~\ref{SOIwaveguide}~(a), (b) and (c), respectively.  

\begin{figure}
  \includegraphics*[width=\linewidth]{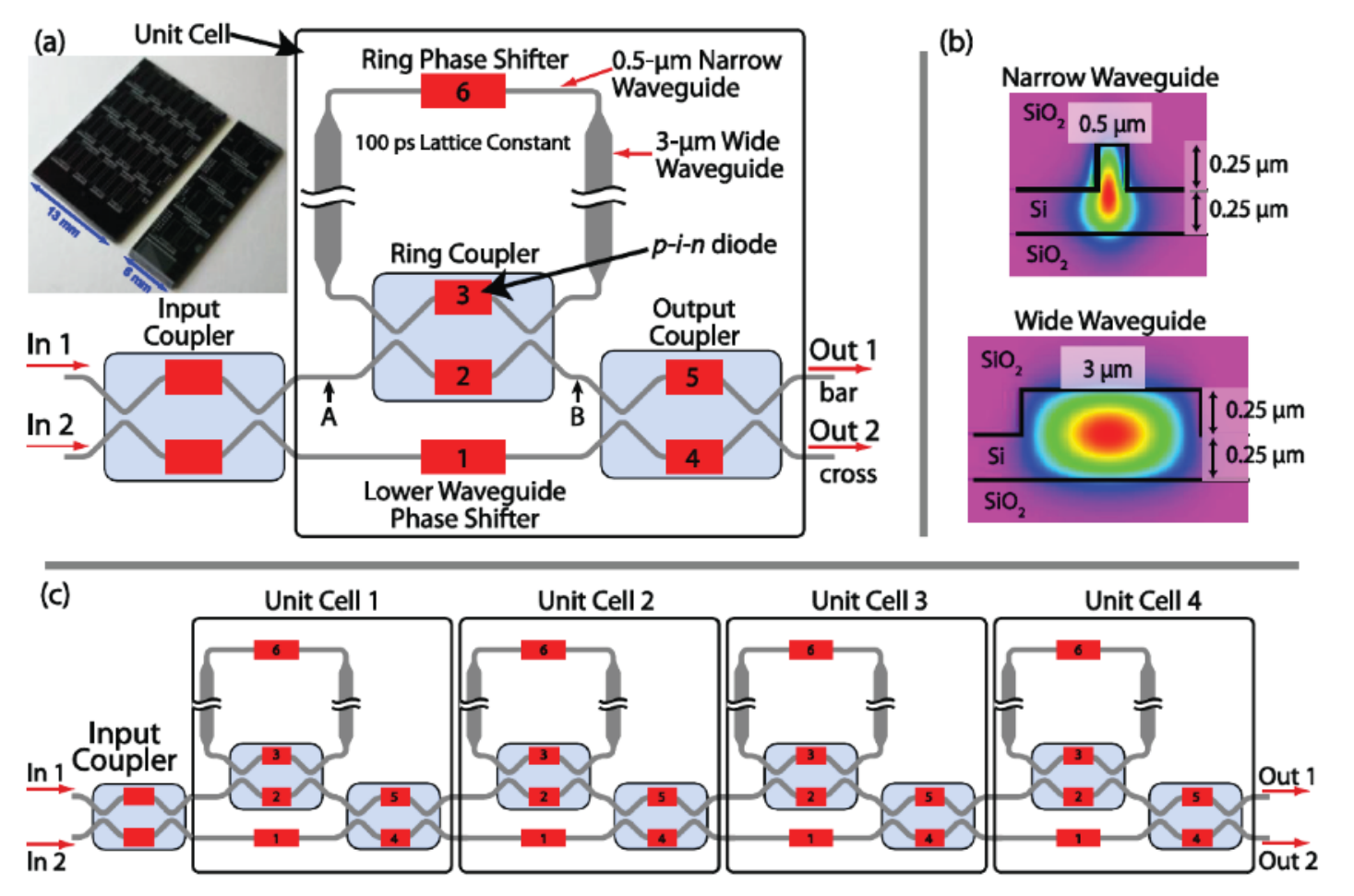}
  \caption{(a) Schematic of a single unit cell, red boxes indicate phase shifter electrodes. (b) Simulation showing the mode size in both the narrow (0.5 $\mu$m) and wide (3 $\mu$m) waveguides. (c) Schematic of a four-unit-cell filter (from \cite{IbrahimOpex2011}, courtesy of the OSA).}
  \label{Ibrahim}
\end{figure}

Most of the integrated MWP devices in SOI were demonstrated using the rib waveguides instead of the nanowires. This is expected since MWP systems have a strict requirement regarding losses. In 1997 Yegnanarayanan et al. \cite{YegnanarayananPTL1997} demonstrated the first optical delay lines in SOI for true time-delay phased array antenna. They used eight-channel 3~$\mu$m wide waveguides with an incremental time delay of 12.3~ps measured  over 2-20-GHz  frequency range. MWP filters have also been demonstrated in SOI waveguides. Rasras et al.\cite{RasrasJLT2007,RasrasJLT2009} demonstrated bandpass and notch MWP filters based on Mach-Zehnder interferometer (MZI) tunable couplers and ORRs fabricated in silicon-buried channel waveguides with a width of 2~$\mu$m and a propagation loss of 0.25~dB/cm. Dong et al.\cite{DongOpex2010_filter} demonstrated a bandpass filter with narrow passband using 5th-order ORR fabricated in shallow-ridge waveguides with a width of 1~$\mu$m and height of 0.25~$\mu$m. The waveguide propagation loss is 0.5~dB/cm and the ring radius is 248~$\mu$m. The filter characteristic will be discussed in more details in Section~\ref{subsec:coherent}. A similar waveguide structure with wider width (2~$\mu$m) was used in the delay lines in a programmable unit cell filter for signal RF processing reported in \cite{ToliverOFC2010,FengOpex2010}.  In another demonstration of MWP filter, two types of silicon rib waveguides, a narrow waveguide which is 0.5-$\mu$m wide and a wide waveguide which is 3-$\mu$m wide, were used in the lattice filter configuration shown in Figure~\ref{Ibrahim}. The narrow and wide waveguides are connected with a linear taper. The the narrow waveguides were used to obtain smaller bending radius and for fast and efficient reconfiguration of the filter response while the wide waveguides were used for the low propagation loss (measured value 0.5~dB/cm). Other demonstration of integrated MWP in wide SOI rib waveguides include optical delay lines \cite{KhanOpex2011,GiuntoniOpex2012} which will be discussed in more details in Section~\ref{sec:delay}.   
                      
Integrated MWP with SOI stripe waveguides has been demonstrated for optical delay lines \cite{CardenasOpex2010,MortonPTL2012}, arbitrary waveform generation \cite{ShenOpex2010,KhanNatPhotonics2010} and ultrawideband (UWB) signal generation \cite{YunhongIPC2011,YueOL2012,MirshafieiPTL2012}. In \cite{MortonPTL2012} SOI waveguides with dimensions of 250~nm-by-500~nm were used for fabricating 20~ORRs in a balanced side-coupled integrated spaced sequence of resonators (SCISSOR) structure. As reported in \cite{CardenasOpex2010}, the waveguides exhibit a propagation loss of 4.5~dB/cm.  In \cite{ShenOpex2010,KhanNatPhotonics2010} an eight-channel reconfigurable optical filter consisting of cascaded microring resonators and tunable MZI couplers is fabricated in SOI waveguides with a dimension of 500~nm-$\times$-250~nm. The rings have radius of 5~$\mu$m and the propagation loss is 3.5~dB/cm. In \cite{YueOL2012} the optical nonlinearity of SOI waveguides is exploited to generate UWB monocycles. The non-degenerate two-photon absorption (TPA) in a 4.046 cm long silicon waveguide with a 776-$\times$-300~nm$^{2}$ cross section is used to create a 143~ps Gaussian monocycle pulse. Details on the arbitrary waveform and UWB generation techniques are discussed in Section~\ref{sec:generation}. 
   

\subsection{TriPleX\texttrademark\,\,technology (Si\textsubscript{3}N\textsubscript{4}/SiO\textsubscript{2})}
\label{subsec:triplex} 

\begin{figure}
  \includegraphics*[width=\linewidth]{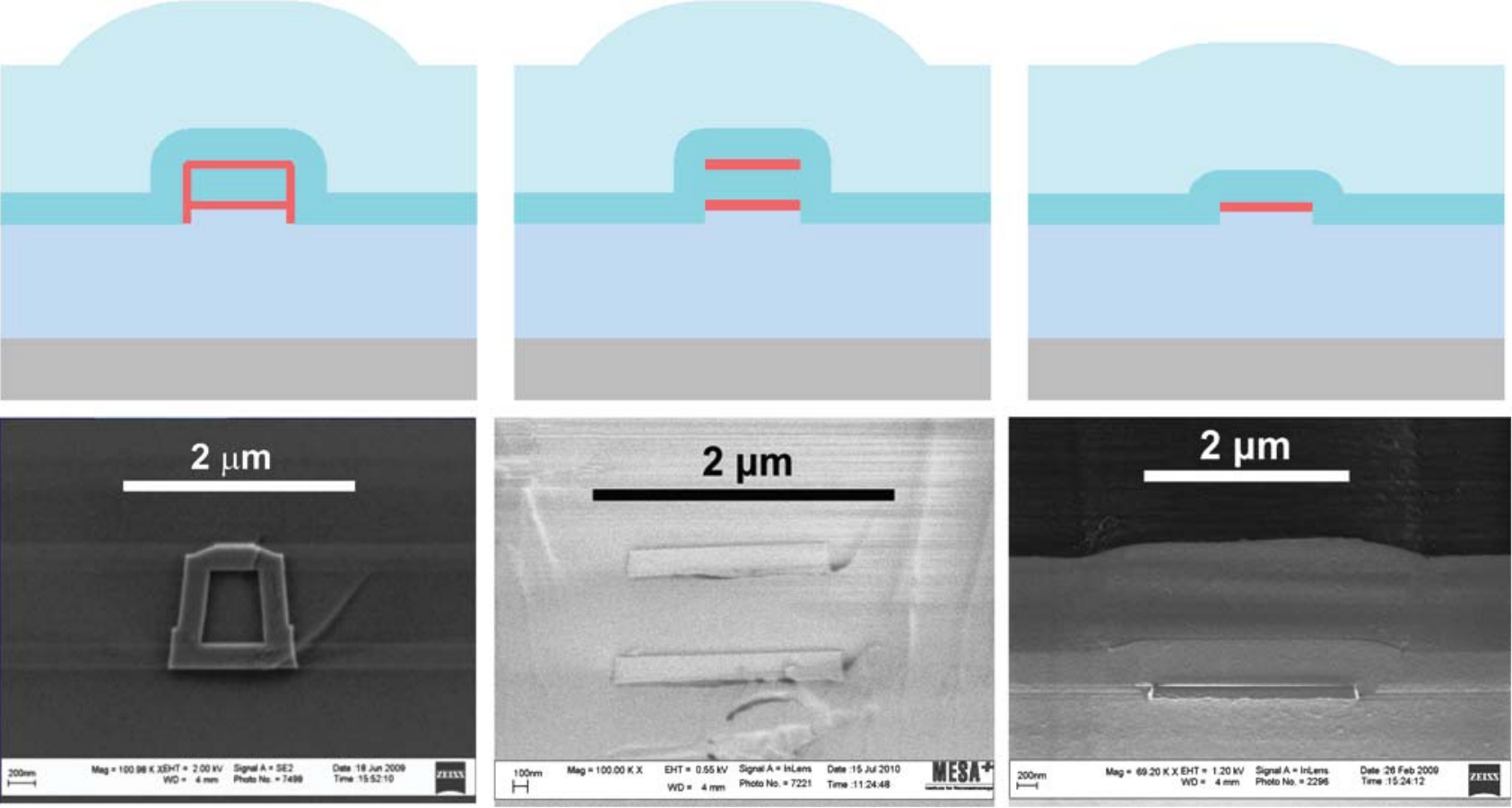}
  \caption{Schematics (top row) and corresponding SEM images of realized structures (bottom row) of three typical single-mode channel layouts: a symmetrical, box-shaped layout with minimal modal birefringence (left column), and two asymmetrical layouts with large modal birefringence: the double-stripe (=-shaped, center column), and a single-stripe layout (right column).}
  \label{TriplexWG}
\end{figure}

Recently, many MWP functionalities like beamforming \cite{ZhuangPTL2007,MeijerinkJLT2010,ZhuangJLT2010,MarpaungEuCAP2011,MarpaungMWP2011,BurlaAO2012}, optical frequency discriminator \cite{MarpaungMWP2010,MarpaungOpex2011}, UWB pulse shaping \cite{MarpaungOpex2011} and MWP filter \cite{ZhuangOpex2011,BurlaOpex2011} have been demonstrated in the TriPleX\texttrademark\, waveguide technology platform\footnote{The TriPleX\texttrademark\, waveguide technology is a proprietary technology of LioniX BV, Enschede, the Netherlands. See: http://www.lionixbv.nl}.  This waveguide technology is based on a combination of silicon nitride (Si\textsubscript{3}N\textsubscript{4}) as waveguide layer(s), filled by and encapsulated with by silica (SiO\textsubscript{2}) as cladding layers.  The consisting SiO\textsubscript{2} and Si\textsubscript{3}N\textsubscript{4} layers are fabricated with CMOS-compatible industrial standard low-pressure chemical vapor deposition (LPCVD) equipment which enables low cost volume production \cite{HeidemanSPIE2009}. TriPleX\texttrademark\, allows for extremely low loss integrated optical waveguides both on silicon and glass substrates for all wavelengths in between 405~nm (near UV) up to 2.35~$\mu$m, providing maximum flexibility from an integration standpoint. Several significantly different waveguide geometries (Figure~\ref{TriplexWG}) can be obtained by varying individual steps in the generic fabrication process. The details on the fabrication steps for these waveguide structures are give in \cite{HeidemanJSTQE2012}.

The three geometries shown in Figure~\ref{TriplexWG} are called \textsl{box-shape} (left), \textsl{double-stripe} ($=-$shape, middle) and \textsl{single-stripe} (right), respectively. All modal characteristics are controlled and tuned by the design of the geometry. While the values of the parameters of these geometries are quite similar each type of waveguide typically has core dimensions in the order of 1~$\mu\mathrm{m}^2$ their corresponding wavelength dependence, modal characteristics, birefringence and therefore desired application differ greatly.

\begin{figure}
  \includegraphics*[width=\linewidth]{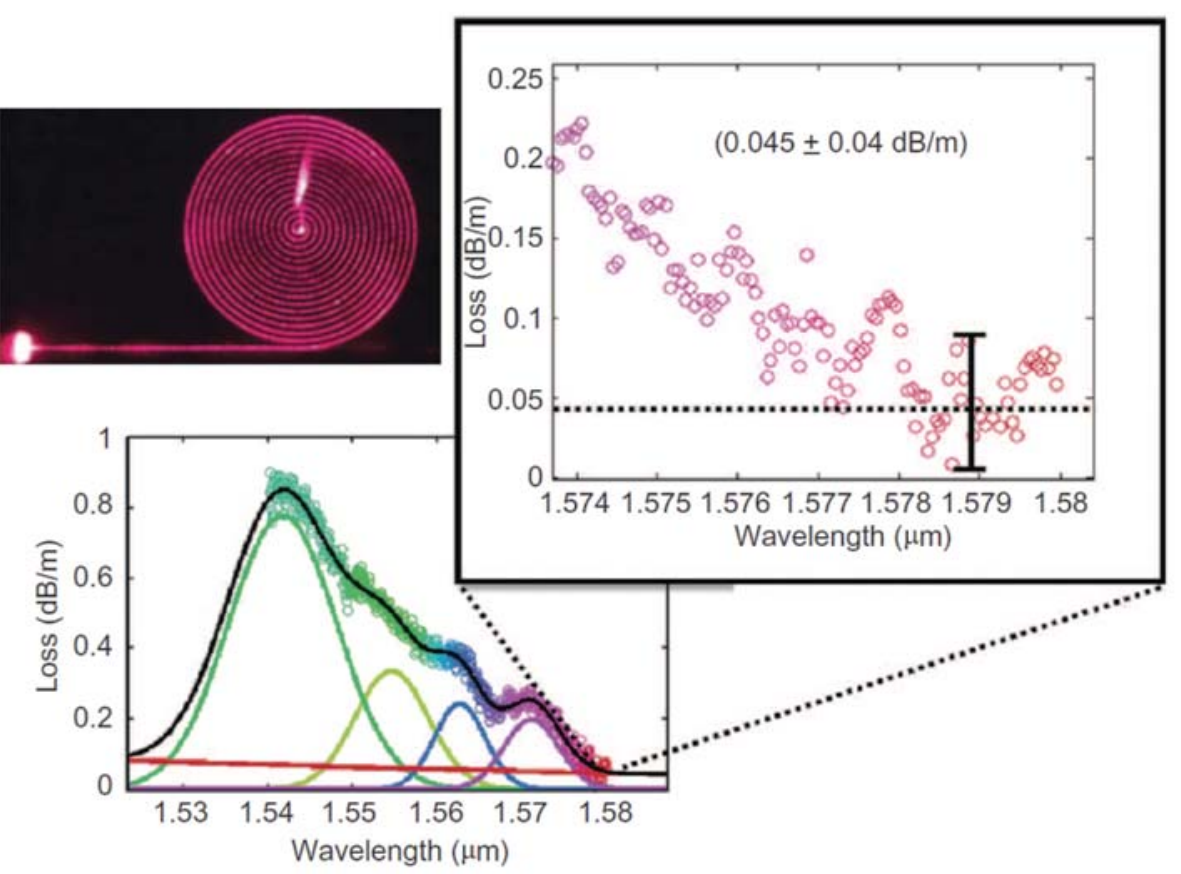}
  \caption{Propagation loss vs. wavelength for a Si\textsubscript{3}N\textsubscript{4}-on-SiO\textsubscript{2} (TriPleX\texttrademark\, single stripe) waveguide with a 40-nm-thick by 13-$\mu$m-wide core and bonded thermal oxide upper cladding, measured on a 1-m-long spiral waveguides (inset) (from \cite{DaiLSA2012}, courtesy of the Macmillan Publishers Ltd).}
  \label{TriplexstripeLoss}
\end{figure}

The \textsl{single-stripe} TriPleX\texttrademark\, structure (Figure~\ref{TriplexWG}, right) is well suited for (opto-fluidic) sensing applications. This layout goes with large modal birefringence, which is a prerequisite for most integrated optical interferometric sensing schemes to prevent signal fading. The single-stripe layout is also compatible with microfluidics \cite{HeidemanJSTQE2012}. But more importantly, this waveguide structure has been used to demonstrate ultra-low propagation loss \cite{TienOpex2010,BautersOpex2011_1,TienOpex2011,DaiOpex2011,BautersOpex2011_2,DaiLSA2012}. The Si\textsubscript{3}N\textsubscript{4} single stripe optical waveguide has a high aspect ratio Si\textsubscript{3}N\textsubscript{4} core (see Figure~\ref{TriplexWG}, right) to minimize the scattering loss at the sidewall, which is the dominant loss mechanism \cite{BautersOpex2011_1}. With an optimized fabrication process, the Si\textsubscript{3}N\textsubscript{4}-on-SiO\textsubscript{2} optical waveguide shows a loss as low as 0.045~dB/m \cite{BautersOpex2011_2}, which is presently a world record for planar waveguides. This value has been measured on a spiral waveguide with a core thickness of 40~nm and width of 13~$\mu$m and a bonded thermal oxide upper cladding. The measured propagation loss is depicted in Figure~\ref{TriplexstripeLoss}.  Several typical photonic integrated devices like arrayed waveguide gratings (AWGs) \cite{DaiOpex2011} and ultra-high Q ring resonators \cite{TienOpex2011} have also been demonstrated.

The \textsl{box-shaped} TriPleX\texttrademark\, (Figure~\ref{TriplexWG}, left) layout is best exploited for telecom applications: due to its symmetrical layout, the polarization birefringence is largely reduced \cite{MorichettiJLT2007}. For this geometry a library of standard optical components with predictable characteristics is available, and currently has been offered as multi project wafer (MPW) service runs\footnote{The MPW service was initially offered as a part of a Dutch National project MEMPHIS. See: http://www.smartmix-memphis.nl}. The box shape geometry has been applied for a variety of applications, for example in  ranging from polarization independent, thermally tunable ring resonators acting as mirrors to create narrow spectral bandwidth lasers tunable over the entire telecom C-band \cite{Oldenbeuving2012}. For MWP applications, the box-shaped waveguide has been used to fabricate a programmable optical beamformer \cite{MeijerinkJLT2010,ZhuangJLT2010,BurlaOpex2011,BurlaAO2012} and frequency discriminators for high SFDR phase modulated MWP link \cite{MarpaungMWP2010,MarpaungOpex2011}. 

The beamformer reported in \cite{MeijerinkJLT2010,ZhuangJLT2010} consists of 8~inputs, 2~balanced outputs, 8~ORRs for tunable true time delays, more than 23~tunable couplers and an optical sideband filter. The minimum bend-radius used in the chip is 700~$\mu$m and the waveguide propagation loss is 0.6~dB/cm. The details of this beamformer will be discussed in Section~\ref{subsec:OBFN}. The FM discriminator was fabricated using a higher index-contrast box-shaped waveguide.  It consists of five fully tunable race-track ORRs with a bend radius of 150~$\mu$m. The measured waveguide propagation loss in this chip amounts to 1.2~dB/cm, which is  dominated by the sidewall roughness in the waveguide. The details of the FM discriminator is discussed in Section~\ref{subsec:PMIM}.  

\begin{figure}
  \includegraphics*[width=\linewidth]{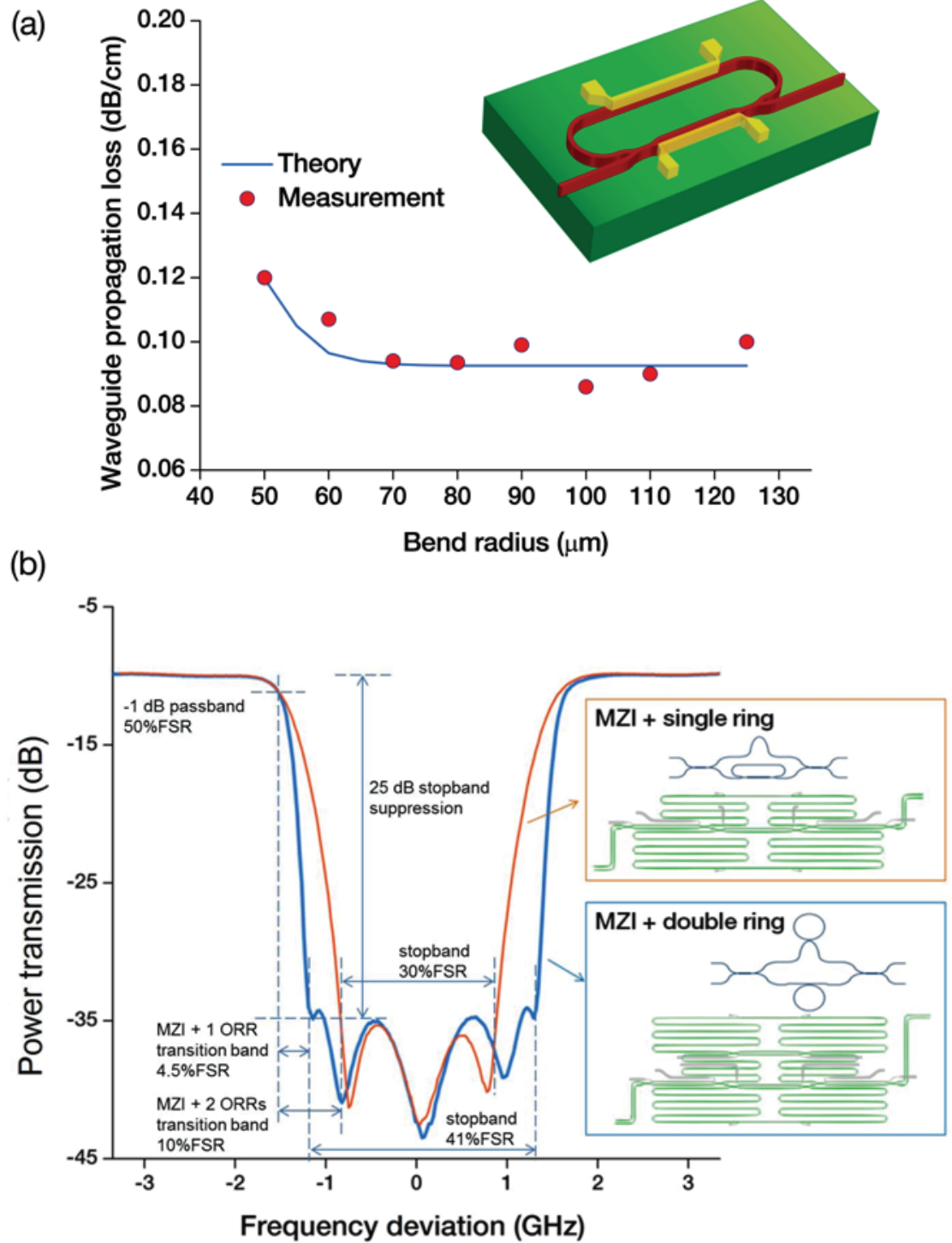} 
  \caption{(a) Propagation loss of the TriPleX double-stripe waveguide structure versus different bend radii in the race track-shaped ORR. (b) Measured filter shapes of the single and double ring-assisted MZI (Inset: schematic of the filter architecture and mask layout design) (from \cite{ZhuangOpex2011}, courtesy of the OSA).}
  \label{TriplexLoss}
\end{figure}

The \textsl{double-stripe} ($=$-shaped) TriPleX\texttrademark\, layout Figure~\ref{TriplexWG}, middle) is standardized especially for MWP applications. This geometry goes with large polarization birefringence, tight bending radii and low channel attenuation levels at 1.55 $\mu$m. The propagation loss of such waveguides measured in a ring resonator as a function of the bending radius is depicted in Figure~\ref{TriplexLoss}~(a). As shown in the result, the maximum waveguide propagation loss was measured to be 0.12~dB/cm in the ORR with a bend radius of 50~$\mu$m, and an average waveguide propagation loss as low as 0.095~dB/cm was achieved for the bend radii larger than 70~$\mu$m \cite{ZhuangOpex2011}. 

Using the double-stripe waveguides an MWP filtering function has been demonstrated \cite{MarpaungMWP2011,ZhuangOpex2011}. Two types of optical sideband filters (OSBFs) for optical single-sideband suppressed carrier modulation (OSSB-SC) scheme  have been fabricated and characterized. One filter consists of an asymmetric MZI with an ORR inserted in its shorter arm. The other is an upgraded version of the first one obtained by adding a second ORR to the longer arm of the asymmetric MZI. Both filters were designed with a full programmability by using the thermo-optic tuning mechanism. For the design of both filters, a waveguide bend radius of 125~$\mu$m was used, which results in footprints of 0.3$\times$1.5~cm (MZI~+~ORR) and 0.4$\times$1.5~cm cm (MZI~+~2~ORRs). The measured filter responses show high frequency selectivity as depicted in Figure~\ref{TriplexLoss}~(b).

All three standardized TriPleX\texttrademark\, geometries can be coupled very efficiently to the outside world  through the use of integrated spot size converters. These components are used to adiabatically transform the profile of the optical mode between two sections having different modal characteristics, and are commercially available with typical coupling efficiency (or corresponding coupling loss) between individual components better than 80\% (e.g. smaller than 1~dB). 

\textcolor{black}{For advanced on-chip MWP functionalities, it is desirable to have a photonic circuit that boasts transitions of different types of TriPleX\texttrademark\, waveguides according to the need. For example, one would simultaneously use the ultra-low loss single-stripe geometry for very long delay lines and the double stripe structure for sharp bends.  This is similar to the approach of using "narrow" and "wide" waveguides in SOI technology as demonstrated in \cite{RasrasJLT2009,IbrahimOpex2011}.} 

\subsection{Other technology}
\label{subsec:other} 

Besides the four platforms discussed above, integrated MWP has been demonstrated with a host of other materials, like GaAs \cite{NgSPIE1994,CombrieCLEO2010}, LiNbO\textsubscript{3} \cite{HorikawaMTT1995,MitchellLEOS2007,IlchenkoPTL2008,WangOpex2009,WangPTL2010,WijayantoElectLett2012}, polymers \cite{YeniayJLT2004,HowleyPTL2005,HowleyJLT2007,YeniayPTL2010} and chalcogenide glasses \cite{MaddenOpex2007,EggletonNatPhotonics2011,EggletonLPR2012,PantOpex2011,PantOL2012,PelusiNatPhotonics2009,ByrnesCLEO2012}. \textcolor{black}{Due to the limited scope of this paper, some of these technologies will only be discussed briefly.} 
\textcolor{black}{\textsl{Lithium niobate} has been the material of choice for developing linear, high performance electro-optic modulators. For MWP, the material has been used in applications like beamforming \cite{HorikawaMTT1995}, optoelectronic converters \cite{MitchellLEOS2007,IlchenkoPTL2008,WijayantoElectLett2012} as well as ultrawideband (UWB) signal generation \cite{WangOpex2009,WangPTL2010}.} 
\textcolor{black}{Optical waveguides based on \textsl{polymer} materials usually exhibit a low refractive index contrast as well as a low propagation loss. This is suitable for applications like long optical delay lines and beamforming \cite{YeniayJLT2004,HowleyPTL2005,HowleyJLT2007,YeniayPTL2010}, although the low index contrast often results in large chip footprint.} 
\textcolor{black}{\textsl{Chalcogenide glasses} have been used in ultrafast optical signal processing due to their strong Kerr-nonlinearity \cite{EggletonNatPhotonics2011,EggletonLPR2012}. For MWP, this platform has been used to demonstrate on-chip tunable optical delay lines \cite{PantOL2012} and MWP filters \cite{ByrnesCLEO2012}.}

 A summary of integrated MWP demonstrations since 1994 until mid-2012 is shown in Table~\ref{tlab}.  


\begin{table*}
  \centering
  \caption{Reported results in integrated MWP.}
  \label{tlab}
  \begin{tabular}{@{}llllccl@{}}
    \toprule
   Year & Functionality & Key component & PIC Technology & Loss &Bend radius  & $1^{\mathrm{st}}$ author [Ref]\\
   &  &  &  & (dB/cm) & $(\mu$m) & \\
    \midrule
    1994 & Beamforming & Rib waveguides  & GaAlAs/GaAs & 1  & 3000 & Ng \cite{NgSPIE1994,NgPTL1994}\\
    1995 & Beamforming & Switched delay  & Silica & 0.1 & 2000  & Horikawa \cite{HorikawaIMS1995,HorikawaOFC1996}\\
    1995 & Beamforming  & Phase shifter & LiNbO\textsubscript{3} & 0.3   & - & Horikawa \cite{HorikawaMTT1995}\\
    1997 & FM Discrim. & MZI filter & Silica & -  & - & Lagasse \cite{LaGassePTL1997}\\
    1997 & Delay  & Rib waveguide & SOI & -  & 5000 &  Yegnanarayanan \cite{YegnanarayananPTL1997}\\
    1999 & Beamforming & Phase shifter & InP & 1.4   & 250 & Stulemeijer \cite{StulemeijerPTL1999}\\
    2000 & Delay & Channel waveguide & Polymer & 0.02  & - & Tang \cite{TangOptEng2000}\\
    2003 & Beamforming &Switched delay  & Silica & -  & 10000 & Grosskopf  \cite{GrosskopfFIO2003}\\
    2005 & Delay &Switched delay, ORR & Silica & 0.07  & 1000  & Rasras  \cite{RasrasPTL2005}\\
    2005/07 & Delay, beamforming &Switched delay  & Polymer & 0.64   & 1750 & Howley  \cite{HowleyPTL2005,HowleyJLT2007}\\
    2005 & Delay lines &Photonic crystal  & SOI & 64   & - & Jiang  \cite{JiangSPIE2005}\\
    2007 & Beamforming &ORR, MZI & TriPleX & 0.55  & 700  & Zhuang  \cite{ZhuangPTL2007}\\
    2007/09 & MWP filter  &ORR, MZI & SOI & 0.25   & 7 & Rasras  \cite{RasrasJLT2007,RasrasJLT2009}\\
    2008 & Coherent receiver &PM, BPD & InP & -  & - & Ramaswamy  \cite{RamaswamyJLT2008}\\
    2008 & Delay &ORR & SiON & 0.35   & 570 & Melloni  \cite{MelloniOL2008}\\
    2008/09 & Differentiator, UWB  &ORR & SOI & -   & 40 & Liu  \cite{LiuOpex2008,LiuElectLett2009}\\
    2009/10 & UWB  &PPLN waveguide & LiNbO\textsubscript{3} & -   & - & Wang  \cite{WangOpex2009,WangPTL2010}\\
    2009 & RF phase shifter &ORR & SOI & -   & 20 & Chang  \cite{ChangPTL2009}\\
    2009 & RF spectrum analyzer &NL waveguide & Chalcogenide & 0.5  & 3000 & Pelusi \cite{PelusiNatPhotonics2009}\\
    2010 & Integrator &ORR & Silica & 0.06  & 47.5 & Ferrera \cite{FerreraNatCommunications2010}\\
    2010 & Delay &ORR & SOI & 4.5  & 7 & Cardenas \cite{CardenasOpex2010}\\
    2010 & Arb. waveform gen. & Add-drop ORR & SOI & 3.5   & 5 & Shen  \cite{ShenOpex2010}, Khan  \cite{KhanNatPhotonics2010}\\
    2010 & MWP filter   &MZI, delay & SOI & 0.9  & - & Toliver \cite{ToliverOFC2010}, Feng \cite{FengOpex2010}\\
    2010 & MWP filter   &ORR & SOI & 0.5  & 248 & Dong \cite{DongOpex2010_filter}\\
    2010 & Beamforming  &AWG, delay line  & Polymer & 0.06  & - & Yeniay \cite{YeniayPTL2010}\\
    2010 & Delay  &Photonic crystal  & GaAs & -  & - & Combrie \cite{CombrieCLEO2010}\\
    2010 & MWP filter   &ORR, SOA  & Hybrid silicon & - & - & Chen \cite{ChenMTT2010}\\
    2010/11 & MWP filter &ORR, SOA & InP & 1 & - &  Norberg \cite{NorbergPTL2010,NorbergJLT2011,GuzzonOpex2011}\\
    2010/11 & FM discrim., UWB & Add-drop ORR & TriPleX  & 1.2   & 150  & Marpaung  \cite{MarpaungOpex2010,MarpaungOpex2011}\\
    2010/11 & Beamforming, SCT &ORR & TriPleX & 0.6   & 700 & Zhuang \cite{ZhuangJLT2010}, Burla \cite{BurlaOpex2011}\\
    2011 & OSSB filter &RAMZI& TriPleX  & 0.1   & 70 & Zhuang  \cite{ZhuangOpex2011}\\
    2011 & FM discrim. &MZI, RAMZI & Silica & 0.045  & - & Wyrwas \cite{WyrwasMWP2011,WyrwasThesis2012}\\
    2011 & OPLL coh. receiver &PM, BPD & InP & 10  & - & Li \cite{LiPTL2011}, Bhardwaj \cite{BhardwajEL2011}\\
    2011 & OPLL coh. receiver &PM, BPD & InP & -  & - & Krishnamachari \cite{KrishnamachariMOTL2011}\\
    2011 & Delay &Bragg grating & SOI & 0.5  & - & Khan  \cite{KhanOpex2011}\\
    2011 & MWP filter &MZI, ORR & SOI & 0.5  & 300 & Ibrahim  \cite{IbrahimOpex2011,DjordjevicPTL2011}\\
    2011 & MWP filter &ORR & TriPleX & 0.029  & 2000 & Tien  \cite{TienOpex2011}\\
    2011 & MWP filter &MZI, ORR & SOI & 0.7  & 20 & Alipour  \cite{AlipourOpex2011}\\
    2011 & Arb. waveform gen. &MZI & Silica & 0.7  & 2000 & Samadi  \cite{SamadiOptComm2011}\\
    2011 & UWB &ORR & SOI & -  & 10 & Ding  \cite{YunhongIPC2011}\\
    2012 & Delay &ORR & SOI & - & 7 & Morton \cite{MortonPTL2012}\\
    2012 & Delay, MWP filter &SBS in waveguide & Chalcogenide & 0.3   & - & Pant \cite{PantOL2012}, Byrnes\cite{ByrnesCLEO2012}\\
    2012 & MWP filter &Microdisk & III-V/SOI & -  & 4.5 & Lloret  \cite{LloretOpex2012}\\
    2012 & UWB  &NL waveguide & SOI & -   & - & Yue  \cite{YueOL2012}\\
    2012 & Photonic ADC &Modulator, mux, PD & SOI & -  & - & Grein \cite{GreinCLEO2011}, Khilo  \cite{KhiloOpex2012}\\
    2012 & FM Discrim. &RAMZI & InP & -  & 150 & Fandi\~{n}o \cite{FandinoECIO2012}\\
    2012 & UWB &ORR & SOI & 5  & 20 & Mirshafiei \cite{MirshafieiPTL2012}\\
    \bottomrule
  \end{tabular}
\end{table*}

\section{High dynamic range microwave photonic link}
\label{sec:APL}

With the development of new photonic technologies and components, new types of MWP links have been reported. As mentioned in Subsection~\ref{subsec:FOM}, these investigations were driven by the need for higher performance links. Theoretical investigations have shown that an ideal Class-B photonic link would feature an SFDR in excess of 180~dB$\cdot$Hz for a relatively high photocurrent of 50~mA \cite{ZhangPTL2007}. In such an ideal Class-B photonic link, the input RF signal is half-wave rectified. Positive voltage is converted linearly to intensity and transmitted on one optical link. Negative voltage is transmitted as intensity over a second matched link. A balanced detector is used to subtract the two detected complementary half-wave rectified signals and restore the original RF signal with zero DC bias \cite{DarcieJLT2007}. In such a link a significant reduction of the shot noise and RIN can be expected due to the absence of the DC bias optical power. However, the realization of such a link has been found difficult, especially with IMDD scheme, either using MZMs \cite{DarciePTL2007} or directly modulated lasers \cite{MarpaungMWP2006}. 

\begin{figure}
  \includegraphics*[width=\linewidth]{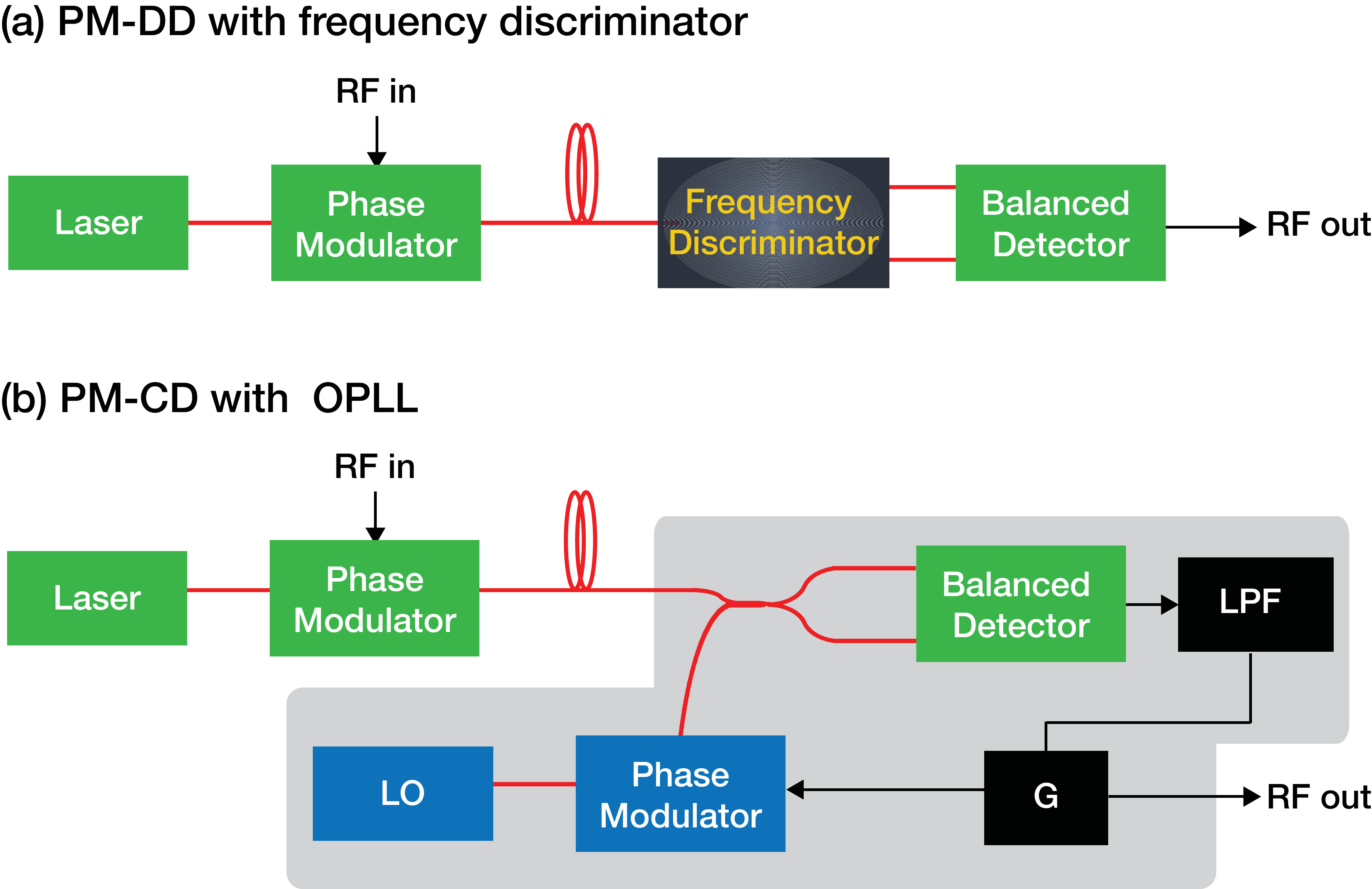}
  \caption{Two types of phase-modulated (PM) MWP links. (a) Direct detection (DD) with a frequency discriminator. (b) Coherent detection (CD) with an optical phase-locked loop (OPLL).}
  \label{MWPL}
\end{figure}

For this reason many have turned to phase or frequency modulation schemes to increase the MWP link performance and to eventually realize the ideal Class-B link. Phase modulation is highly attractive because it is intrinsically highly linear \textcolor{black}{(for example using conventional lithium niobate modulators)} and its operation does not require biasing. Frequency modulation is identical to phase modulation but with a modulation depth that is linearly dependent on the modulation frequency. Moreover, FM lasers have been demonstrated with high modulation efficiency, thereby promising a high link gain if implemented in MWP links \cite{WyrwasJLT2009}. The main challenge in phase or frequency modulation is to restore  the modulating microwave signals (i.e. demodulation) in a linear manner. Two options that have recently gained popularities are the direct detection scheme using a frequency discriminator (Figure~\ref{MWPL}~(a)) and coherent detection scheme using optical phase-locked loop (Figure~\ref{MWPL}~(b)). 

\subsection{Direct detection scheme with frequency discriminator}
\label{subsec:PMIM} 

In this approach, a phase-modulated signal is converted to intensity modulation (PM-IM conversion) using an optical discriminator, thereby allowing a simple direct detection scheme. This approach is attractive for the additional degree of freedom in tailoring the characteristic of the optical filter discriminator to enhance the MWP link performance. The photonic discriminator can be designed for increasing the link linearity and/or suppressing the noise in the MWP link. Different filter types have been proposed as the photonic discriminator, with the simplest one being a Mach-Zehnder interferometer (MZI) \cite{LaGassePTL1997,UrickMTT2007,McKinneyJLT2009}. The linearized version of the MZI filter approach has shown a very large SFDR (above ${125~\rm dB}\cdot{\rm Hz}^{2/3}$ at 5~GHz) but suffers a limited bandwidth. In another approach, Darcie et al. proposed the use of a phase modulator with a pair of fiber Bragg gratings as the frequency discriminators \cite{DarcieOFC2006,DarcieJLT2007,DriessenJLT2008}. The FBGs were custom designed to realize a linear slope for the PM-IM conversion for realizing the Class-B photonic link. However these FBGs and the required optical circulators are bulky and thus, preventing a compact discriminator. An SFDR of ${110~\rm dB}\cdot{\rm Hz}^{2/3}$ has been shown with this approach.

\begin{figure*} [ht!!]
  \sidecaption
  \includegraphics*[width=0.7\textwidth]{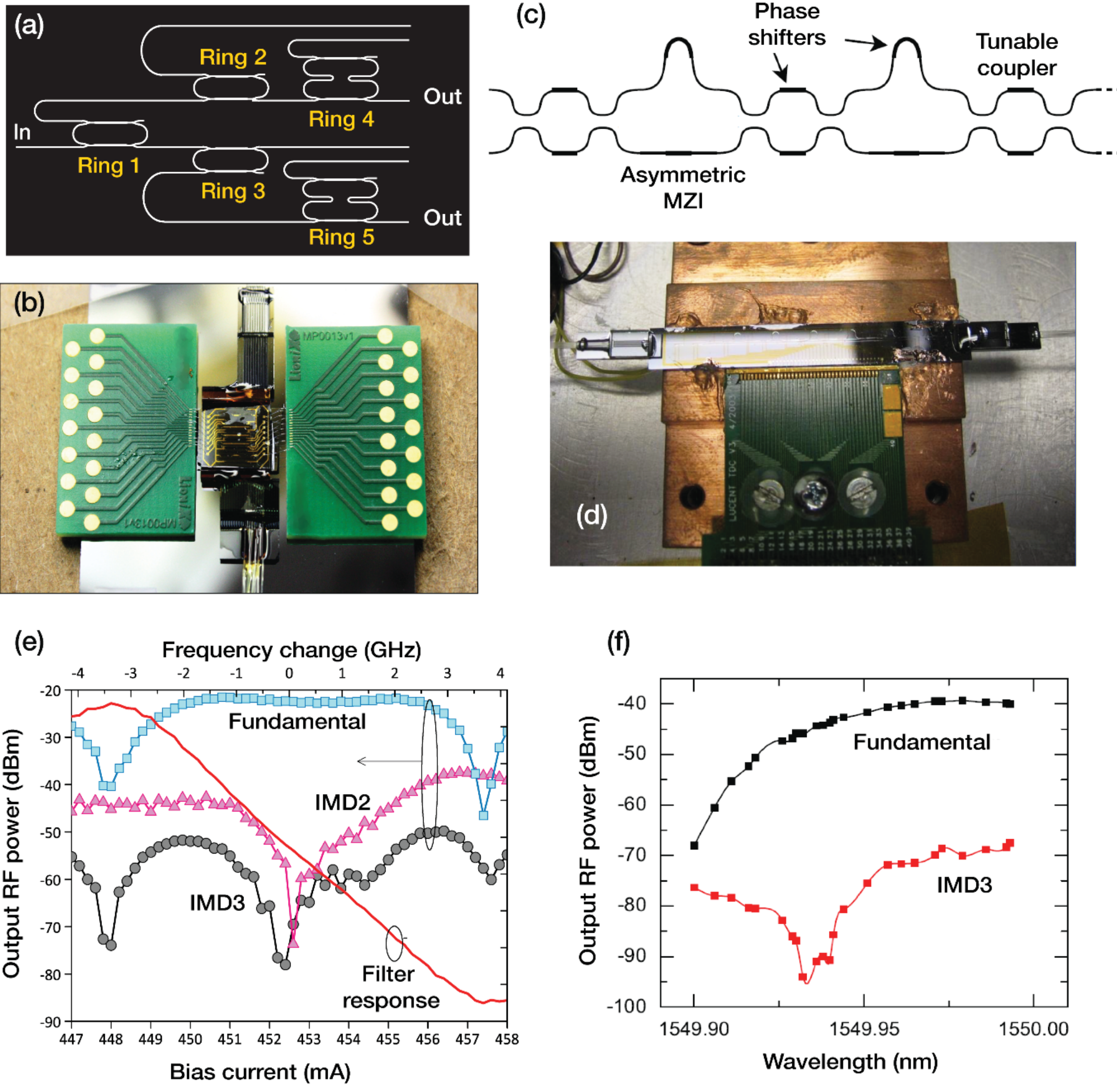}
  \caption{Two types of frequency discriminator chips recently reported, the ORR-based in TriPleX technology (a, b, e) and the cascaded MZIs-based in Silica (c, d, f). (a) and (c) schematic of the devices, (b) and (d) fabricated and packaged devices, (e) and (f) characterization results of the fundamental RF signals and the IMDs. (a, b, e) are taken from \cite{MarpaungOpex2010}, courtesy of the OSA, (c, d, f) are taken from \cite{WyrwasThesis2012}, courtesy of the University of California Berkeley.}
  \label{Discriminator}
\end{figure*}
 
To realize compact frequency discriminators many turn to photonic integrated circuits. The idea to use integrated photonic filter was initially proposed by Xie et al. \cite{XiePTL2002_grating,XiePTL2002_ring} in 2002. However the study did not lead to any device realization and experiment. In 2010 Marpaung et al. demonstrated the first PIC frequency discriminator for MWP link \cite{MarpaungMWP2010,MarpaungOpex2010}. The chip consists of five fully-reconfigurable optical ring resonators in add-drop configuration (Figure~\ref{Discriminator}~(a)). The chip was fabricated using the box-shape TriPleX\texttrademark\, waveguides (Figure~\ref{Discriminator}~(b)). Programmability of the chip is done using thermo-optical tuning mechanism.  The discriminator is used to show linear operation indicated by high IIP2 (46~dBm) and IIP3 (36~dBm) achieved at a single bias point (Figure~\ref{Discriminator}~(e)). For shot noise limited performance, the predicted SFDR is ${113~\rm dB}\cdot{\rm Hz}^{2/3}$ at 2~GHz. But high chip insertion loss prevents to achieve high SFDR due to the high amplified spontaneous emission (ASE) noise from EDFAs. Reducing the fiber-to-chip coupling and waveguide propagation losses will dramatically increase this SFDR. 

Subsequently, the discriminator chip consisting of cascaded Mach-Zehnder interferometers (MZIs) has recently been demonstrated \cite{WyrwasOFC2010,WyrwasMWP2011,WyrwasThesis2012}. The phase discriminator is a $6^{\rm{th}}$ order
finite impulse response (FIR) lattice filter Figure~\ref{Discriminator}~(c)) fabricated in a silica-on-silicon, planar lightwave circuit (PLC) process by Alcatel-Lucent Bell Labs (Figure~\ref{Discriminator}~(d)). It has 6 stages of symmetrical MZIs (switches) and asymmetrical MZIs (delay line interferometers) with an FSR of 120 GHz. The discriminator is tunable using chromium heaters deposited on the waveguides and is dynamically tuned to minimize the link distortion. At the optimal wavelength, the RF power input (two tones around 2~GHz) into the link is varied the IMD3 and fundamental powers are measured. The data is shown in Figure~\ref{Discriminator}~(f). For a photocurrent of 0.11 mA, the measured OIP3 is -19.5~dBm which is a 6.7 dB OIP3 performance improvement over an MZI with the same received photocurrent. For shot-noise limited noise performance, the link SFDR is ${112~\rm dB}\cdot{\rm Hz}^{2/3}$. If the photocurrent can be increased to 10~mA,  OIP3 increases to 19.7~dBm and the shot-noise limited SFDR is ${125~\rm dB}\cdot{\rm Hz}^{2/3}$.

Research activities in realizing the on-chip FM discriminators are expected to increase significantly. Recently a device which includes a tunable optical filter acting as a frequency discriminator and a high speed balanced photodetector integrated in the same chip has been proposed by researchers at the UPV Valencia, Spain \cite{FandinoECIO2012}. The filter has a cascade of two integrated ring-loaded Mach-Zehnder interferometers (RAMZIs) in each of the two branches. The filter is fabricated in InP technology. Deeply-etched (1.7~$\mu$m) rib waveguides with InGaAsP core are used to enable sharp bends (150~$\mu$m) and minimizing chip are to $6\times6$~mm$^2$. The performance of the MWP link using this chip has not yet been reported.

\subsection{Coherent detection with integrated optical phase locked loop}
\label{subsec:OPLL} 

The second approach to achieve linear phase demodulation is to use a coherent optical link and to detect the optical PM using an optical phase locked loop (OPLL) \cite{RamaswamyJLT2008,KrishnamachariMOTL2011,LiPTL2011,BhardwajEL2011}. This alternative is complex, but offers potentially high performance. The challenge of this approach is to realize a phase tracking receiver that can follow the linear phase modulation applied at the transmitter to ultimately realize a linear transmission. As shown in Figure~\ref{MWPL}~(b), the phase of the received signal is detected using a balanced detector by comparison with that of a local oscillator (LO) laser. The output signal is then reapplied to a receiver PM to modulate the LO phase and drive the loop such that the voltage driving the receiver PM is a replica of the transmitted signal. While attractive and simple in theory, this is difficult to achieve in practice, particularly at multi-GHz rates encountered in microwave photonics. Loop gain must be high to improve linearity over simpler approaches. This can be accomplished using either high optical power at the receiver or through electronic amplification. To ensure the loop stability a low pass filter (LPF) must be used. In practice, the loop delay must be a small fraction (e.g. 1/5) of the maximum RF frequency. This not only calls for monolithic integration of both the electronics and photonics, it also demands the elimination of any unwanted signal paths between the two.

\begin{figure}[ht]
  \includegraphics*[width=\linewidth]{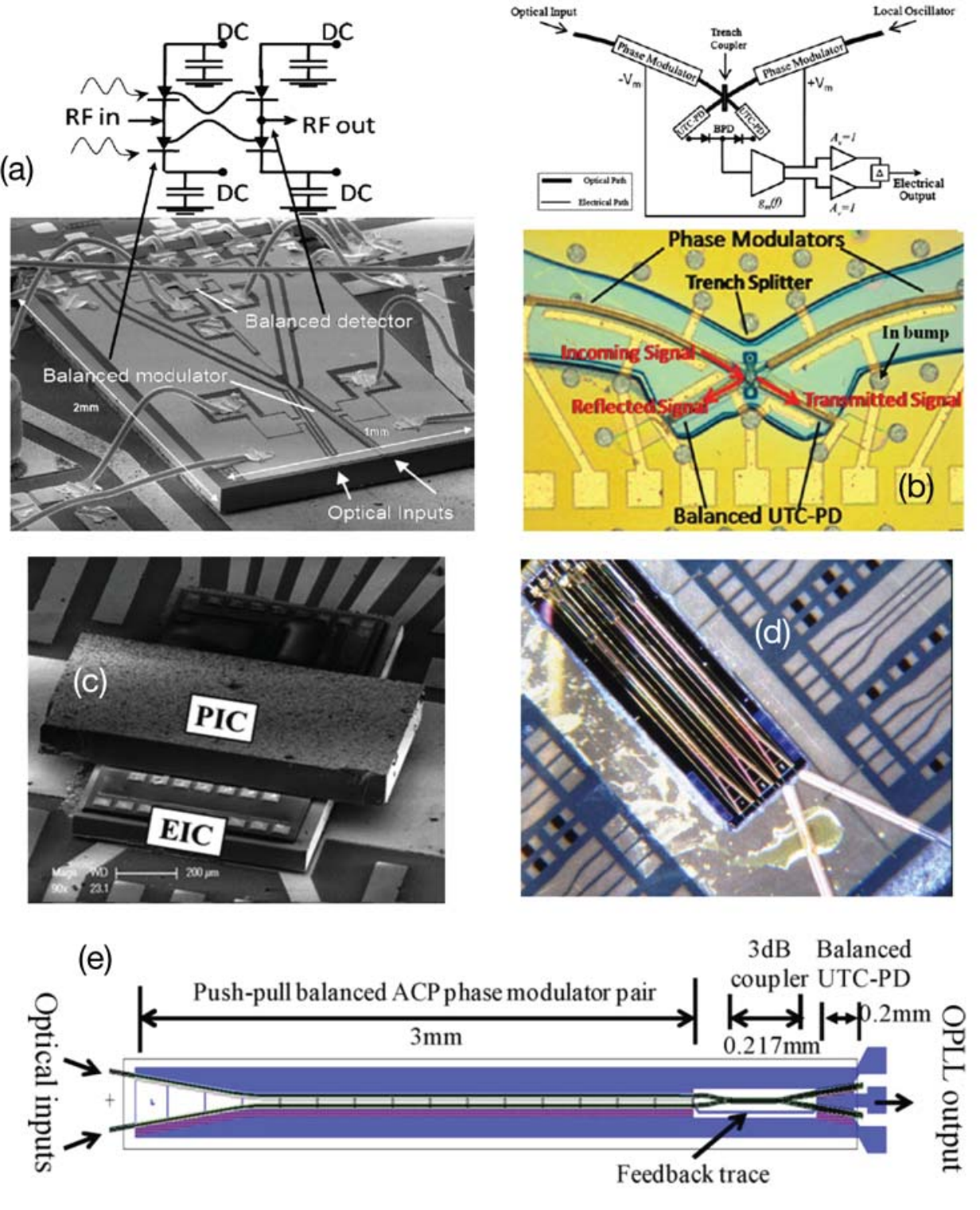}
  \caption{OPLL realizations in InP for coherent detection MWP link. (a) SEM and block diagram of integrated optoelectronic receiver reported in \cite{RamaswamyJLT2008}, courtesy of the IEEE. Layout of the PIC and (c) flip-chip bonded
PIC and EIC reported in \cite{KrishnamachariMOTL2011}, courtesy of Wiley. Photograph (d) and layout (e) of the ACP-OPLL reported in \cite{LiPTL2011}, (d) courtesy of the SPIE, (e) courtesy of the IEEE. }
  \label{OPLL}
\end{figure}

In \cite{RamaswamyJLT2008}, an InP photonic integration platform was fabricated  consisting of a balanced uni-traveling carrier (UTC) photodetector pair [13], a $2\times2$ waveguide multimode interference (MMI) coupler and
tracking phase modulators in a balanced configuration. The schematic and the SEM image of the realized device is shown in Figure~\ref{OPLL}~(a). The tracking optical phase modulators are driven differentially so as to add opposite-sign phase shifts to the incoming signal and LO resulting in a cancellation of even-order nonlinearities and common-mode noise. Additionally, driving the modulators in a differential fashion doubles the drive voltage presented to the modulator thereby doubling the available phase swing. The photonic integrated circuit (PIC) was wirebonded to the electronic integrated circuit (EIC) used to provide feedback gain and filtering. Using this device a 3-dB loop bandwidth of 1.45~GHz was demonstrated, and SFDR of ${125~\rm dB}\cdot{\rm Hz}^{2/3}$ at 300~MHz and ${113~\rm dB}\cdot{\rm Hz}^{2/3}$ at 1~GHz were achieved. The reduced SFDR at 1 GHz is due to the large loop delay ($\approx$ 35~ps) of this receiver with a substantial portion coming from the wirebonds. 

To reduce the loop delay, the same group proposed a novel ultra compact coherent receiver PIC containing two push-pull phase modulators, a balanced UTC photodetector pair and an ultrashort frustrated total internal reflection (FTIR)
trench coupler. Moreover, smaller delay can be obtained by compact integration of the PIC and EIC via flip-chip bonding \cite{KrishnamachariMOTL2011}.The fully fabricated PIC is shown in Figure~\ref{OPLL}~(b). The flip-chip bonded PIC and EIC are shown in Figure~\ref{OPLL}~(c). Using this device an SFDR of ${122~\rm dB}\cdot{\rm Hz}^{2/3}$ at 300~MHz has been achieved. 

A different approach to control such a short loop propagation delay has been proposed by Li et al. \cite{LiJLT2009}.  The configuration consists of the so-called attenuation-counterpropagating (ACP) in-loop phase modulator where the optical and microwave fields propagate in the opposite direction and the microwave field is strongly attenuated. This unique configuration eliminates the propagation delay of the in-loop phase modulator at the expense of a tolerable decrease in bandwidth. The proof of concept of this ACP-OPLL was experimentally demonstrated using an MZ-like structure fabricated in LiNbO\textsubscript{3} butt-coupled to a bulk UTC BPD. A standard two-tone test was performed to probe the linearity of the device and the SFDR was measured to be ${134~\rm dB}\cdot{\rm Hz}^{2/3}$ at the frequency of 100~MHz. In an attempt to extend the high SFDR to a higher frequency, a monolithically integrated ACP-OPLL consisting of the phase modulators and the UTC BPD was fabricated InP-based material platform \cite{LiPTL2011,BhardwajEL2011}. The photograph and the layout of the realized device are shown in Figure~\ref{OPLL}~(d-e). The OPLL was designed to reach the OIP3 $>$ 40~dBm and the SFDR of ${140~\rm dB}\cdot{\rm Hz}^{2/3}$ at a bandwidth beyond 2.7~GHz \cite{LiPTL2011}. however, due to a faulty BPD, the loop showed a small bandwidth ($<$~200~MHz) and relatively low OIP3 (13~dBm). Thus the SFDR was limited to ${124.5~\rm dB}\cdot{\rm Hz}^{2/3}$ at 150~MHz. The predicted shot-noise limited SFDR is ${130.1~\rm dB}\cdot{\rm Hz}^{2/3}$ at 150~MHz.

A quick comparison of the two detection schemes lead to a conclusion that the frequency discriminator approach have shown better performance at higher frequencies. The operating frequency actually is not a limiting factor for these discriminators, since it is bounded by the filters FSR which can be relatively large. The OPLL approach with PIC is very promising for very large SFDR, up to ${140~\rm dB}\cdot{\rm Hz}^{2/3}$, but in the current implementation is currently limited to low frequencies.   

\textcolor{black}{Following the growth of optical in-phase/quadrature (I/Q) receivers for digital applications, there is a novel type of phase modulated link architecture that uses coherent I/Q detection followed by signal digitization and digital signal processing (DSP) \cite{ClarkPTL2007,ClarkMTT2010}. An SFDR as high as ${126.8~\rm dB}\cdot{\rm Hz}^{2/3}$ at the frequency of 1~GHz has been demonstrated using this technique \cite{ClarkMTT2010}. Although currently the receiver architecture has been demonstrated using discrete components, the link is expected to benefit hugely from the merging of electronic and photonic integration to create a high performance receiver circuit.} 

\section{Microwave photonic filters}
\label{sec:MWPfilter}

\begin{figure} [hb!]
  \includegraphics*[width=\linewidth]{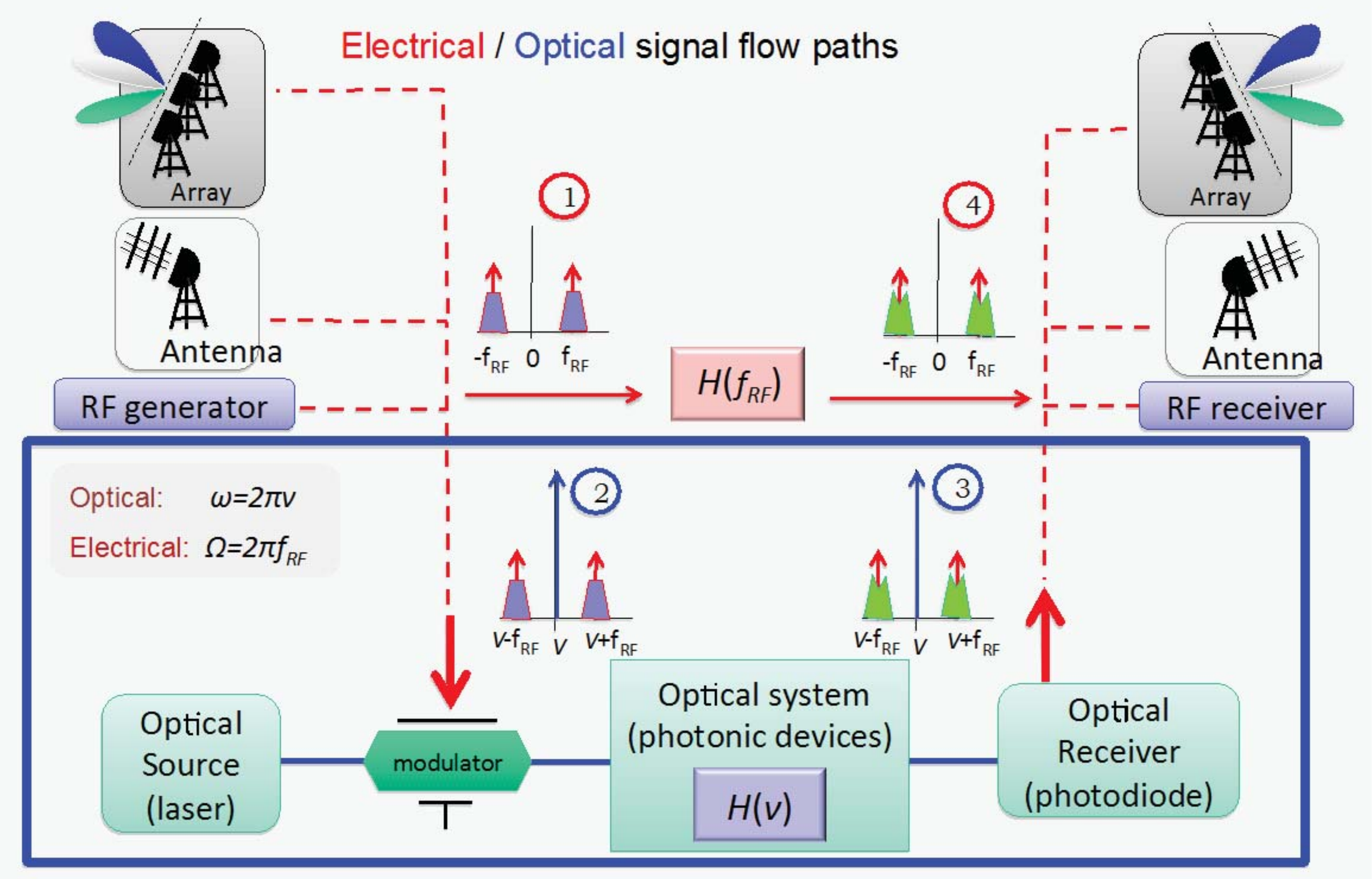}
  \caption{Generic reference model of a Microwave Photonic filter.}
  \label{MPF1}
\end{figure}

A microwave photonic filter \cite{MinasianMTT2006,CapmanyJLT2005,CapmanyJLT2006}, is a photonic subsystem designed with the aim of carrying equivalent tasks to those of an ordinary microwave filter within a radio frequency (RF) system or link, bringing supplementary advantages inherent to photonics such as low loss, high bandwidth, immunity to electromagnetic interference (EMI), and also providing features which are very difficult or even impossible to achieve with traditional technologies, such as fast tunability, and reconfigurability. The term microwave is freely used throughout the literature to designate either RF, microwave, or millimeter-wave signals. Figure~\ref{MPF1} shows a generic reference layout of an MWP filter.

\begin{figure}
  \includegraphics*[width=\linewidth]{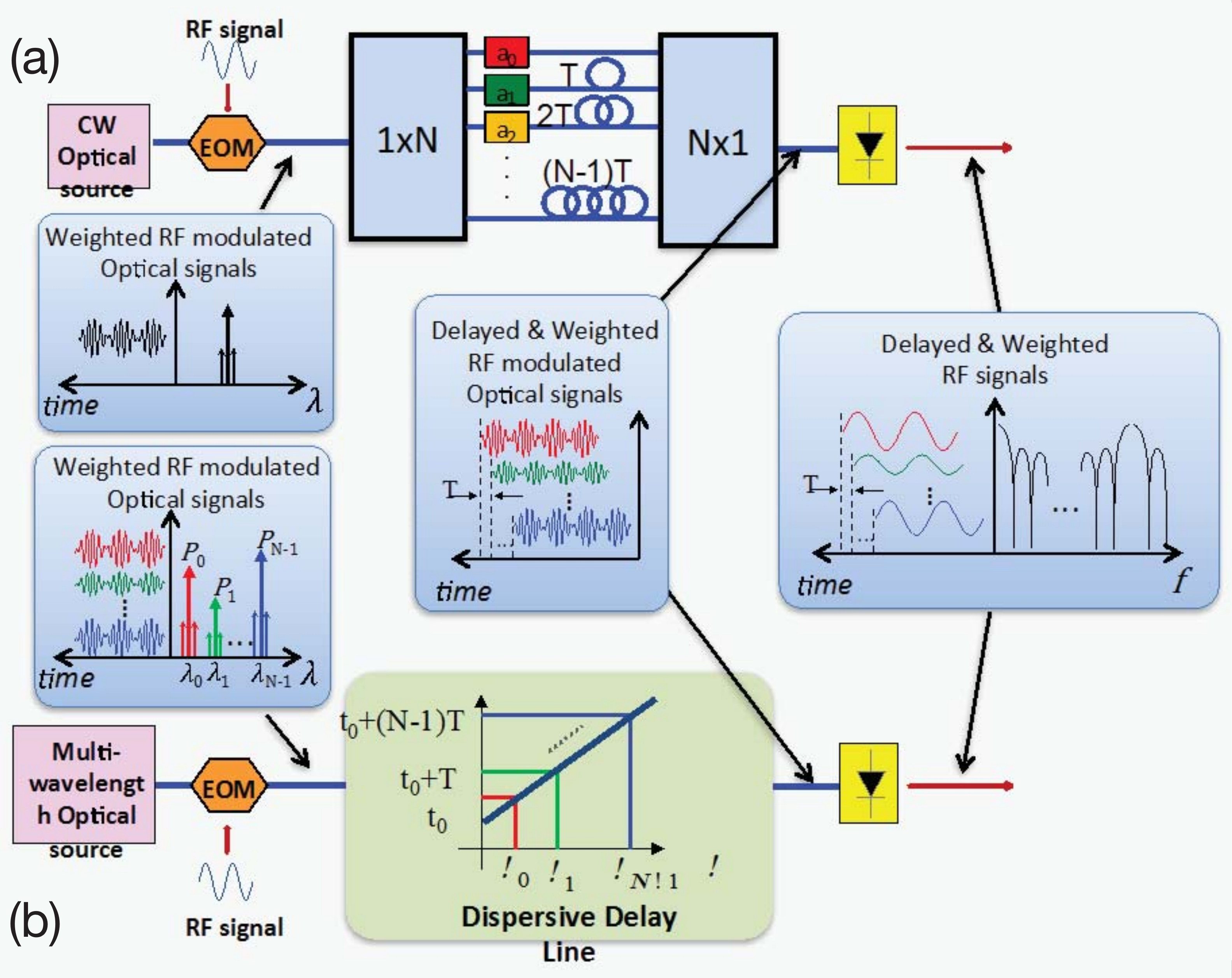}
  \caption{General schematics of a discrete-time FIR MWP.  (a) Traditional approach based on a single optical source in combination with multiple delay lines. (b) A more compact approach based on a multi-wavelength optical source combined with a single dispersive element.}
  \label{MPF2}
\end{figure}

An input RF signal (with spectrum sideband centered at frequency $\pm f_{\mathrm{RF}}$, shown in point 1) coming from a generator or detected by means of a single or an array of antennas is used to modulate the output of an optical source which upconverts its spectrum to the optical region of the spectrum (point 2), such that the sidebands are now centered at $\nu\pm f_{\mathrm{RF}}$ , where $\nu$ represents the central frequency of the optical source. The combined optical signal is then processed by an optical system composed of several photonic devices and characterized by an optical field transfer function $H\left(\nu\right)$. The mission of the optical system is to modify the spectral characteristics of the sidebands so at its outputs they are modified according to a specified requirement as shown in point 3. Finally, an optical detector is employed to downconvert the processed sidebands again to the RF part of the spectrum by suitable beating with the optical carrier so the recovered RF signal, now processed (as shown in point 4) is ready to be sent to a RF receiver or to be re-radiated. The overall performance of the filter is characterized by an end-to-end electrical transfer function $H\left(f_{\mathrm{RF}}\right)$  which is shown in Figure~\ref{MPF1} and links the input and output RF signals.
The most powerful and versatile approach for the implementation of MWP filters is that based on discrete-time signal processing \cite{CapmanyJLT2005} where a number of weighted and delayed samples of the RF signal are produced in the optical domain and combined upon detection. In particular, finite impulse response (FIR) \cite{CapmanyJLT2006} filters combine at their output a finite set of delayed and weighted replicas or taps of the input optical signal while infinite impulse response (IRR) are based on recirculating cavities to provide an infinite number of weighted and delayed replicas of the input optical signal \cite{CapmanyJLT2006}. For instance and taking as an example a FIR configuration, the electronic transfer function is given by:

\begin{equation}
  \label{eq:MPFeq1}
  H\left(f_{\mathrm{RF}}\right) = 
  \sum_{k=0}^{N-1} a_{k}\,e^{-j 2 \pi k f_{\mathrm{RF}} T} \,,
\end{equation}
\nolinebreak
where $a_{k}=\left|a_{k}\right|\,e^{-j k \phi}$  represents the weight of the $k$-th sample, and  $T$ the time delay between consecutive samples.  Note that Eq.\,\eqref{eq:MPFeq1} implies that the filter is periodic in the frequency domain. The period, known as free spectral range (FSR) is given by $f_{\mathrm{FSR}}=1/T$. The usual implementation of this concept in the context of microwave photonics can follow two approaches as shown in Figure~\ref{MPF2}. In the first one (Figure~\ref{MPF2}a), the delays between consecutive samples are obtained, for instance, by means of a set of   optical fibers or waveguides where the length of the fiber/waveguide in the $k$-th tap is $c\left(k-1\right)T/n$, being $c$ and $n$ the light velocity in the vacuum and the refractive index respectively.  This simple scheme does not allow tuning, as this would require changing the value of $T$ . An alternative approach (Figure~\ref{MPF2}b) is based on the combination of a dispersive delay line and different optical carriers where the value of the basic delay $T$ is changed by tuning the wavelength separation among the carriers, thereby allowing tunability \cite{CapmanyJLT2005}. While in the first case the weight of the $k$-th tap, represented by $a_{k}$ , can be changed by inserting loss/gain devices in the fiber coils, with the second approach $a_{k}$ is readily adjusted by changing the optical power emitted by the optical sources \cite{CapmanyJLT2005}. 

Finally, MWP filters can operate under \textsl{incoherent regime}, where sample coefficients in Eq.\,\eqref{eq:MPFeq1} correspond to optical intensities and are thus positive or under \textsl{coherent regime} where the taps in Eq.\,\eqref{eq:MPFeq1} can be complex-valued in general. In the first case the basic delay  $T$ is much greater than the coherence time~\footnote{The coherence time is defined as the measure of temporal coherence of a light source,  expressed as the time over which the field correlation decays} of the optical source that feeds the filter while in the second is much smaller.   

\subsection{Requirements of microwave photonic filters}
\label{subsec:filterrequirements}

\begin{figure}
  \includegraphics*[width=\linewidth]{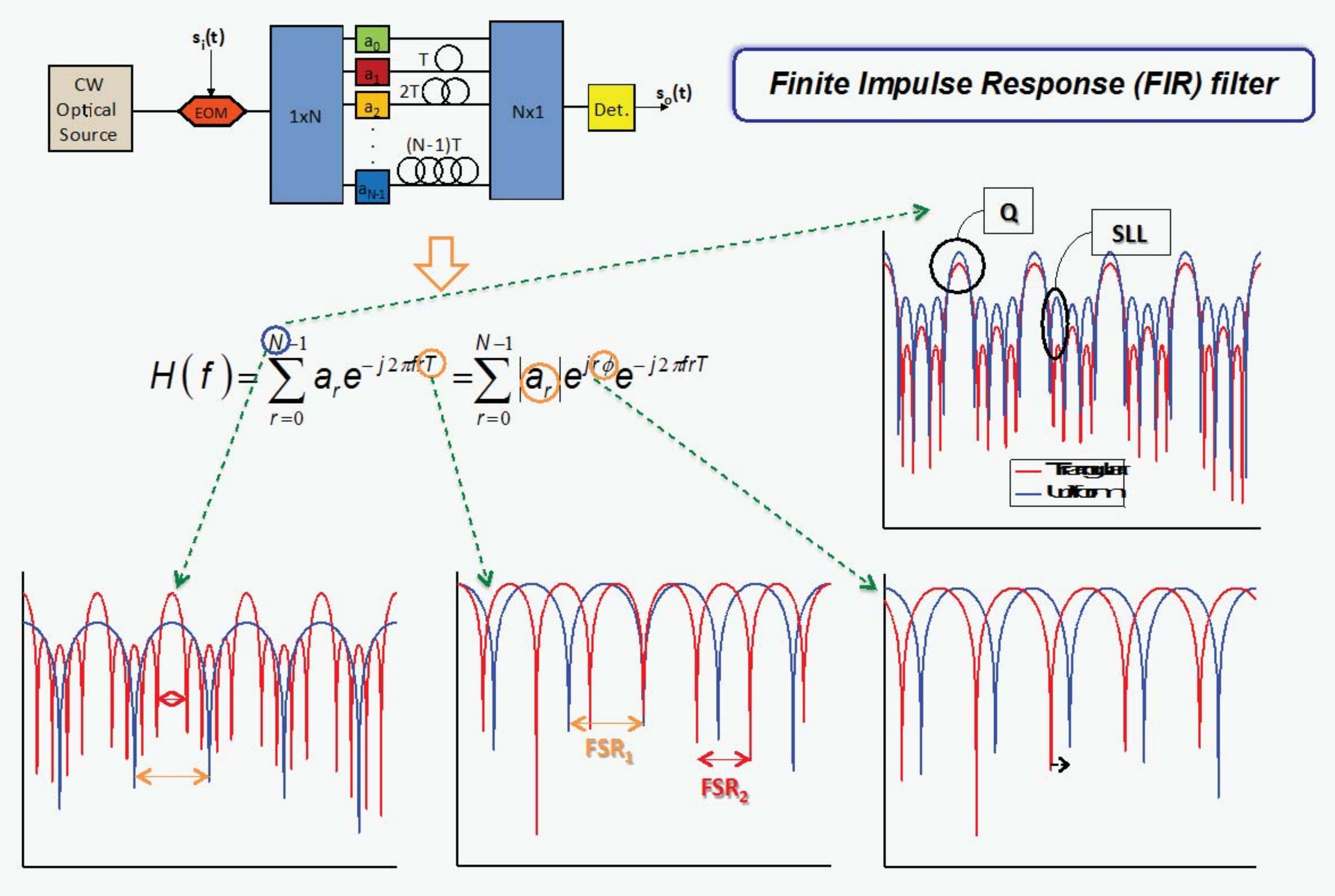}
  \caption{Illustration of the requirements on sample parameters to achieve MWP filter tunability, reconfigurability, selectivity.}
  \label{MPF3}
\end{figure}

MWP filter flexibility in terms of tunability, reconfigurability and selectivity is achieved by acting over the different parameters characterizing the samples in Eq.\,\eqref{eq:MPFeq1} with a variety of techniques having been reported in the literature \cite{MoraOL2003,CapmanyOL2003,CapmanyOpex2005,SupradeepaNatPhotonics2012}. The effect of the relevant parameters in Eq.\,\eqref{eq:MPFeq1} on the filter response is illustrated in Figure~\ref{MPF3}.

The number of samples $N$ dictates whether the filter is either a notch $\left(N=2\right)$, or a bandpass $\left(N>2\right)$ type. As mentioned above, $T$ fixes the spectral period, thus  changing $T$ results in compressing or stretching the spectral response. This is a technique usually employed in the literature for tuning the notch or bandpass-positions of a MWP filter. Fast tuning can be achieved by changing the wavelength separation between adjacent carriers in the scheme of Figure~\ref{MPF2}b with a current record value in the range of 40~ns \cite{SupradeepaNatPhotonics2012}. The phase of the tap coefficients allow the tuning of the spectral response without actually stretching or compressing it. The implementation of phase values depends on the approach followed for its implementation. For incoherent MWP filters, a photonic RF-phase shifter is required which can be implemented in a variety of technologies, including, stimulated Brillouin scattering, coherent population oscillations in SOA devices and passive configurations based on ring cavities and resonators. All of the above can provide the required  phase-shift dynamic range, but only the last two are prone for integration and can provide switching speed below one microsecond. For coherent filters, the phase shifts are optically provided by photonic components. The law followed by the tap coefficient moduli dictates the filter shape (reconfiguration). Filters featuring different windowing functions, both static and dynamically reconfigurable structures have been reported in the literature where tap amplitude setting has been achieved using different techniques, including spatial light modulators, SOAs, and also by fixing the output power of laser modes. Finally, the filter selectivity is dictated by the number of samples which determine the quality factor   and the main to secondary sidelobe (SSL) rejection ratio. FIR schemes using multiwavelength sources can provide from 40 to over 60 samples with the current record featuring a SSL value of 61~dB.

\subsection{Coherent filters}
\label{subsec:coherent}

\begin{figure}
  \includegraphics*[width=\linewidth]{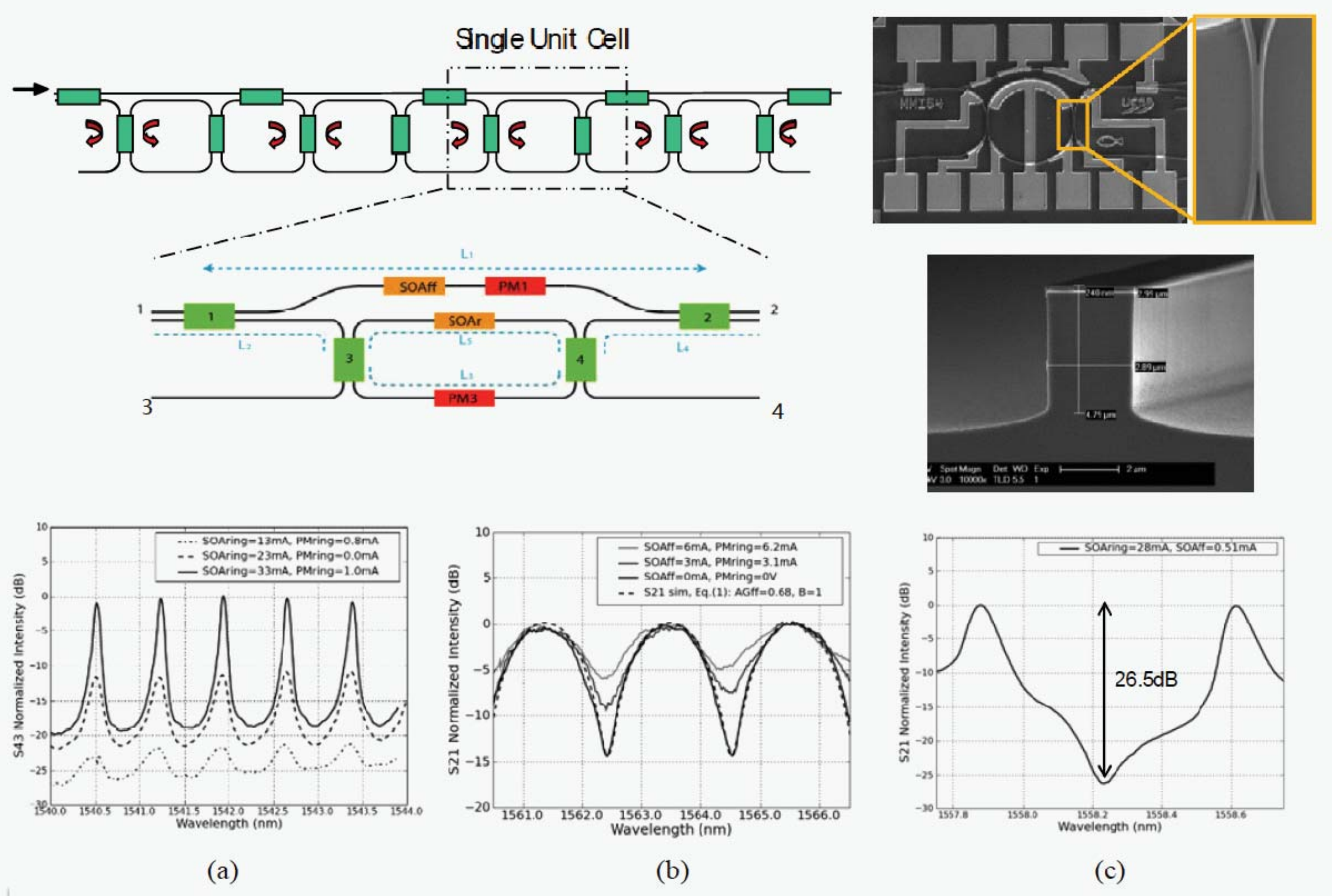}
  \caption{Integrated InP-InGaAsP first-order MWP coherent filter providing one pole and one zero reported in \cite{NorbergJLT2011}. (Upper) 
  schematic of a unit cell configuration and SEM images of the contacts and waveguides. (Lower) measured transfer functions for one pole (left), one zero (center) and one pole-one zero (right) configurations (courtesy of the IEEE).}
  \label{MPF4}
\end{figure}

As far as integrated MWP coherent filtering is concerned, several groups have reported results \cite{RasrasJLT2007,RasrasJLT2009,TuJLT2010,NorbergPTL2010,ChenMTT2010,GuzzonOpex2011,NorbergJLT2011,GuzzonJQE2012,FengOpex2010,DongOpex2010_filter,DjordjevicPTL2011,IbrahimOpex2011,AlipourOpex2011}. Many of the preliminary approaches have been based mainly on single cavity ring resonators. A few however have also focused on more elaborated designs involving more than one cavity and programmable features. These filters can be useful particularly when the RF information has already been modulated onto the lightwave carrier and it might be advisable to perform some prefiltering in the optical domain prior to the receiver. Representative results from one cavity filters can be found in \cite{NorbergPTL2010,ChenMTT2010,NorbergJLT2011}. For instance, \cite{NorbergJLT2011} reports the results for a unit cell, shown in the upper of Figure~\ref{MPF4} that could be an element of  more complex lattice filters.   

\begin{figure}
  \includegraphics*[width=\linewidth]{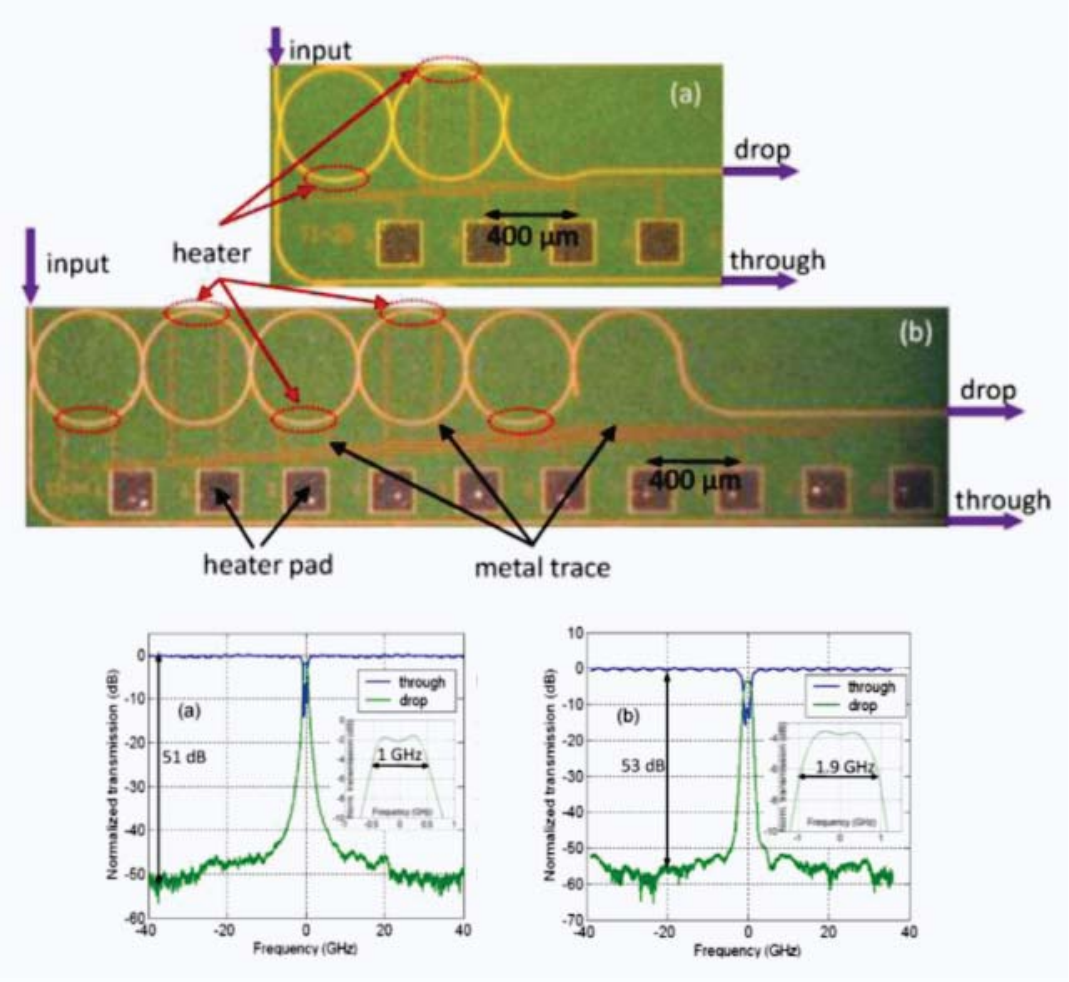}
  \caption{Integrated two (upper) and five (middle) cavity Silicon MWP coherent filters. Details of measured passbands (lower) for the two (left) and five (right) cavity structures as reported in \cite{FengOpex2010} (courtesy of the OSA).}
  \label{MPF5}
\end{figure}

This unit cell, integrated in InP-InGaAsP is composed of two forward paths and contains one ring. By selectively biasing one semiconductor optical amplifier (SOA) and phase modulators placed in the arms of the unit cell, filters with a single pole, a single zero  or a combination of both can be programmed as shown in the lower part of Figure~\ref{MPF4}. In particular and for the design reported in \cite{NorbergJLT2011} the frequency tuning range spans around 100 GHz. A hybrid version incorporating silicon waveguides has also been reported \cite{ChenMTT2010} that combines III-V quantum well layers bonded with low loss passive silicon waveguides.  Low loss waveguides allow for long loop delays while III-V quantum devices provide active tuning capability. The same group involved in \cite{GuzzonJQE2012} is now reporting results of more complex designs involving second and third order filters as well as other different unit cell configurations \cite{GuzzonOpex2011}.
 	A more complex design, this time in Silicon, has been also recently presented \cite{FengOpex2010,DongOpex2010_filter} where 1-2 GHz-bandwidth filters with very high extinction ratios (~50 dB) have been demonstrated. The silicon waveguides employed to construct these filters have propagation losses of $~0.5$ dB/cm. Each ring of a filter is thermally controlled by metal heaters situated on the top of the ring. With a power dissipation of $~72$ mW, the ring resonance can be tuned by one free spectral range, resulting in wavelength-tunable optical filters. Both second-order and fifth-order ring resonators have been demonstrated, which can find ready application in microwave/radio frequency signal processing. The upper part of Figure~\ref{MPF5} shows the filter layouts while their spectral response can be found in the lower part of the same figure.

\subsection{Incoherent filters}
\label{subsec:incoherent}

\begin{figure}
  \includegraphics*[width=\linewidth]{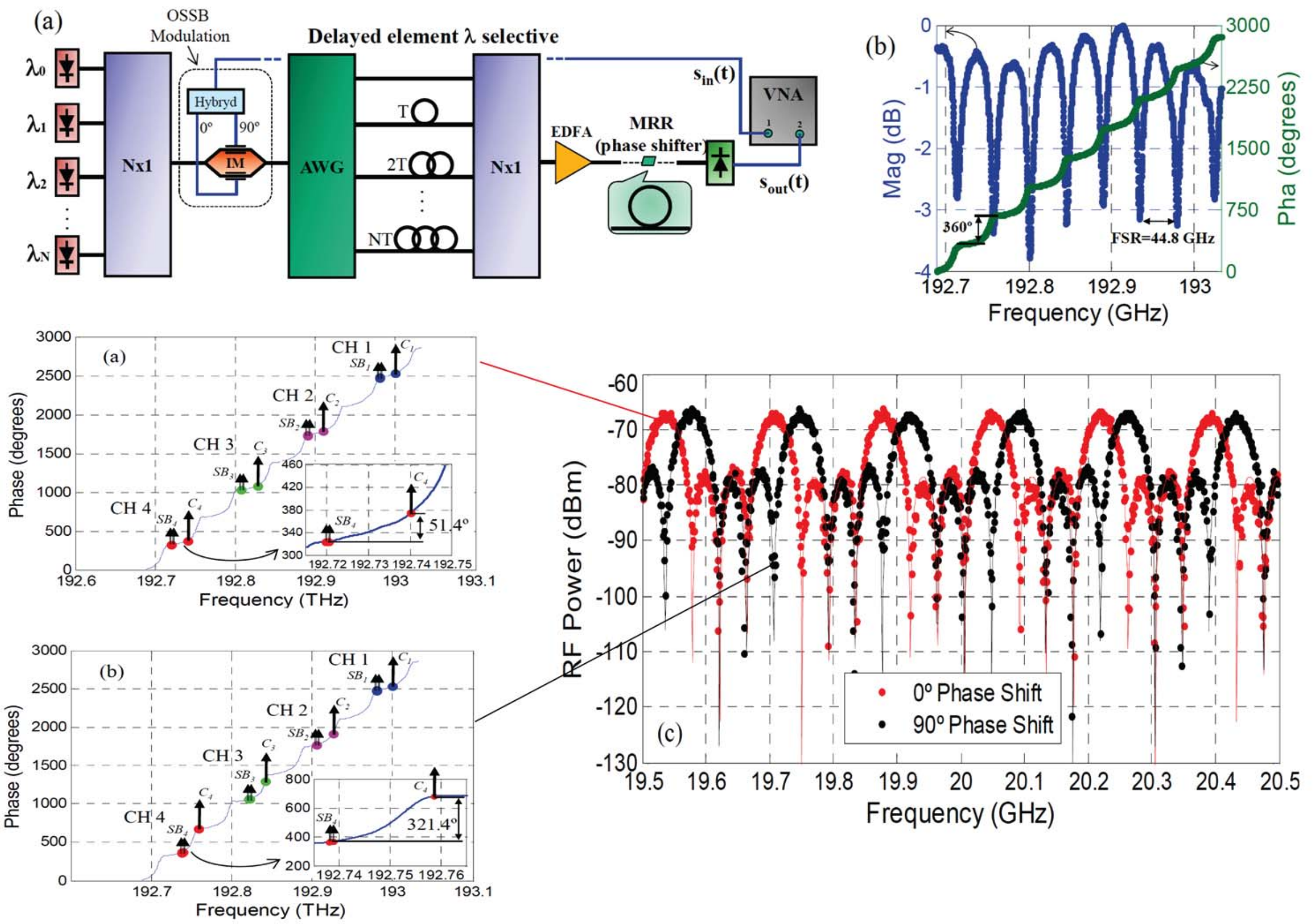}
  \caption{Tunable incoherent MWP filter based on multiple phase shifters implemented by periodic resonances of an integrated SOI ring resonator. (Upper left) filter configuration. (Upper right) amplitude and phase response of the SOI ring resonator. (Lower left) spectral locations of the four optical carriers and subcarriers to achieve respectively 0 and $90\textDegree$ phase shifts. (Lower right) Filter transfer functions corresponding to the two cases (0 and $90\textDegree$ phase shifts) showing tunability.}
  \label{MPF6}
\end{figure}

Work in integrated MWP incoherent filters as been reported as well by various groups \cite{MunozJLT2002,PastorOL2003,PoloPTL2003,XuePTL2009,LloretOpex2011}. Initially research efforts focused on the use of integrated array waveguide grating (AWG) devices \cite{MunozJLT2002} to perform a variety of functions, including spectral slicing to provide low-cost multiple input optical carriers \cite{PastorOL2003} or selective true time delays \cite{PoloPTL2003}. More recent efforts have focused towards the implementation of complex-valued sample filters by means of exploiting several techniques to integrate MWP phase shifters. For instance in \cite{XuePTL2009} a two-tap tunable notch filter configuration is proposed where phase shifting by means of coherent population oscillations in SOA devices followed by optical filtering. In another approach \cite{LloretOpex2011}, the periodic spectrum of a integrated SOI ring resonator is employed as a multicarrier tunable and independent phase shifter. The configuration for a 4-tap filter is shown in Figure~\ref{MPF6}.

Here the basic differential delay between samples is implemented in the first stage while an independent phase coefficient for each tap is selected in the ring resonator by fine tuning of the wavelength of each carrier. A similar configuration based on a hybrid InP-SOI tunable phase shifter has also been recently reported \cite{LloretOpex2012}. In this case tuning is achieved not by changing the source wavelength but by carrier injection into the III-V microdisk. A more versatile configuration which can provide both phase and optical delay line tuning has been recently reported \cite{BurlaOpex2011}. It consists of  a reconfigurable optical delay line (ODL) with a separate carrier tuning (SCT) \cite{MortonPTL2009} unit and an optical sideband filter on a single CMOS compatible photonic chip. The processing functionalities are carried out with optical ring resonators as building blocks demonstrating reconfigurable microwave photonic filter operation in a bandwidth over 1 GHz.
Most of the incoherent MWP filters reported so far require a dispersive delay line which is usually implemented by either a dispersive fiber link or a linearly chirped fiber Bragg grating which being bulky devices, prevent a complete integration of the filter on a chip. The ultimate and most challenging limitation towards the full implementation of integrated microwave photonic signal processors is therefore the availability of a dispersive delay line with a footprint compatible with the chip size providing at the same time the group delay variation required by high-frequency RF applications. A challenging and attractive approach is that based on a photonic crystal (PhC) waveguide which, if suitably designed, can fulfill the above requirements introducing moderate losses. Researchers have recently demonstrated for the first time both notch and bandpass microwave filters based on such component. Tuning over 0 - 50 GHz spectral range is demonstrated by adjusting the optical delay. The underlying technological achievement is a low-loss 1.5~mm long photonic crystal waveguide capable of generating a controllable delay up to 170~ps still with limited signal attenuation and degradation. Owing to its very small footprint, more complex and elaborate filter functions are potentially feasible with this technology.

\section{Optical delay line and beamforming}
\label{sec:delay} 

Delaying and phase shifting RF signals are the basic functionalities for more complicated signal processing functionalities. In this section we review the integrated MWP techniques that have been proposed for tunable optical delay, phase shifting and beamforming. 
    
\subsection{Time delay and phase shifter}
\label{subsec:delaytechnique} 
Reconfigurable optical delay lines (ODL) and wideband tunable phase shifters have primary importance in a number of MWP signal processing applications like optical beamforming and MWP filter. The simplest way for generating delay in the optical domain is through physical length of optical fibers. However, this can become rather bulky. For this reason, integrated photonic solutions are used. A number of approach have been reported over the years. For example, optical switches can be used to provide discretely tunable delay by means of selecting waveguides with different propagation length. This approach has been demonstrated using devices in silica \cite{RasrasPTL2005} and in polymer \cite{HowleyPTL2005}. Others proposed tunable delay based on optical filters \cite{LenzJQE2006}. For example cascaded ORRs have been demonstrated for tunable delays in silica \cite{RasrasPTL2005}, TriPleX \cite{ZhuangPTL2007}, silicon oxynitride (SiON) \cite{MelloniOL2008}, and SOI \cite{CardenasOpex2010,MortonPTL2012}. Others used integrated Bragg-gratings in SOI that can be either electrically \cite{KhanOpex2011} or thermally tuned \cite{GiuntoniOpex2012}. 

Besides delay, phase shifting is also attractive for a number of signal processing applications. For narrowband phase shift, SOI ring resonators have been used as widely tunable RF phase shifters \cite{ChangPTL2009,LloretOpex2011}. Others used semiconductor waveguides in SOAs \cite{XueOL2009,XueOpex2010}, or a microdisk in hybrid III-V/SOI platform \cite{LloretOpex2012}. 

\begin{figure} [ht!]
  \includegraphics*[width=\linewidth]{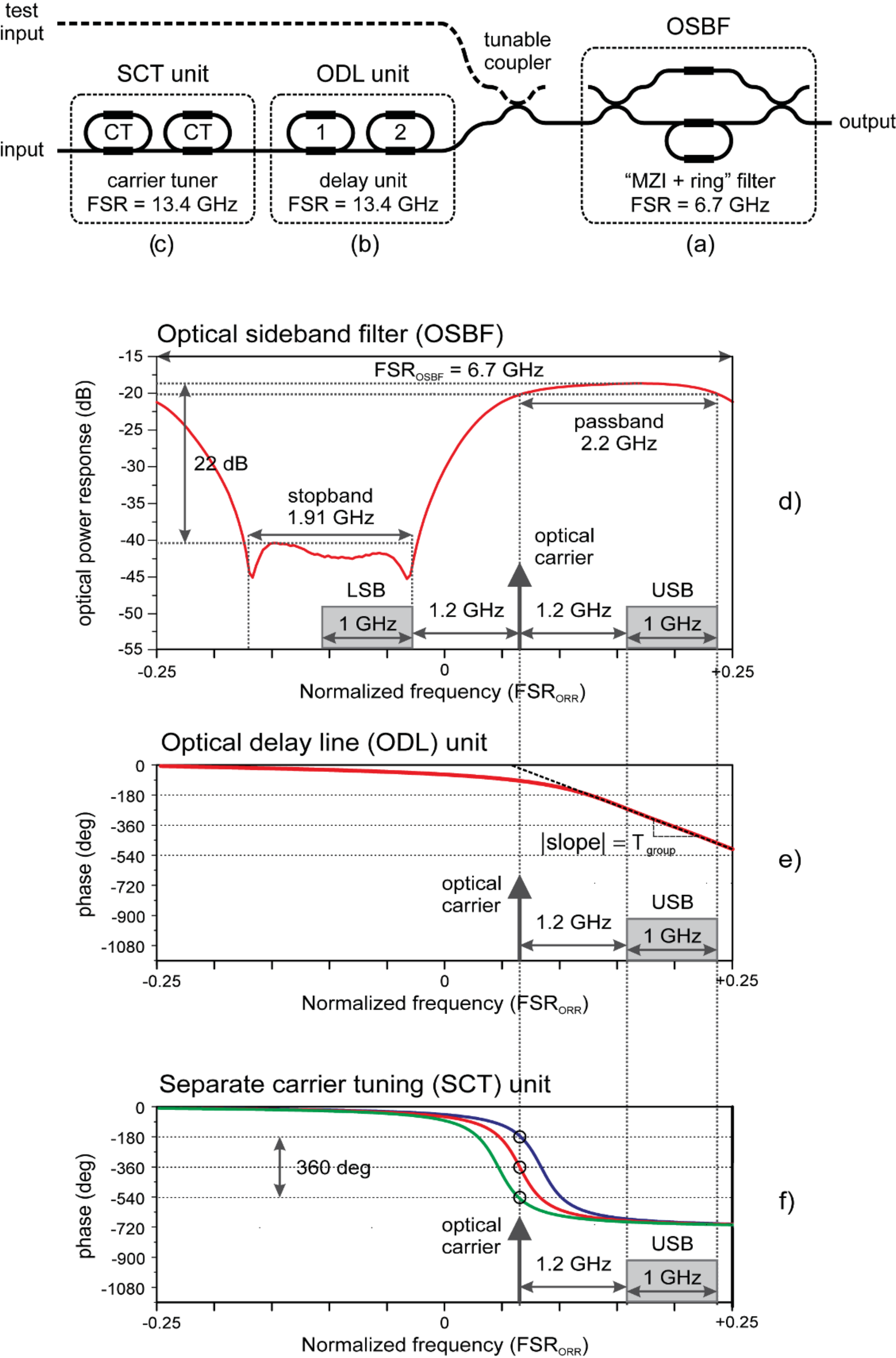}
  \caption{A signal processor based on separate carrier tuning scheme consisting of (a) an optical sideband filter (OSBF), (b) optical delay lines (ODL), and (c) carrier phase tuner (SCT). (d-f) denote the measured responses of the OSBF, ODL and SCT, respectively (from \cite{BurlaOpex2011}, courtesy of the OSA).}
  \label{SCT}
\end{figure}

To have a complete signal processing capabilities, it is attractive to obtain both delay and phase shift at different signal frequency components. This is widely known as the separate carrier tuning (SCT) scheme, proposed by Morton and Khurgin \cite{MortonPTL2009}. Chin et al. demonstrated this functionality in optical fiber using the stimulated Brillouin scattering (SBS) effect \cite{ChinOpex2010}. Burla et al. \cite{BurlaOpex2011} demonstrated the SCT scheme together with  optical single sideband filtering monolithically integrated in a single chip.  The processor consists of a reconfigurable optical delay line, a separate carrier tuning unit and an optical sideband filter. The optical sideband filter, a Mach-Zehnder interferometer loaded with an optical ring resonator in one of its arms, removes one of the radio frequency sidebands of a double-sideband intensity-modulated optical carrier. The ODL and separate carrier tuning unit are individually implemented using a pair of cascaded optical ring resonators. Varying the group delay of the signal sideband by tuning the resonance frequencies and the coupling factor of the optical ring resonators in ODL, while also applying a full $0-2\pi$ carrier phase shift in separate carrier tuning, allowed the demonstration of a two-tap microwave photonic filter whose notch position can be shifted by 360\textdegree over a bandwidth of 1~GHz. The principle of this SCT scheme is depicted in Figure~\ref{SCT}.

\subsection{Optical beamforming}
\label{subsec:OBFN}

In a phased-array antenna, the beam is formed by adjusting the phase relationship between a number of radiating elements \cite{SeedsMWP2002}. For wideband signals, a particular problem arise for  phased arrays: if  a  constant phase shift is produced from element to element, the beam pointing  is different for different frequency  components-a  phenomenon called beam squinting \cite{NgJLT1991,FrigyesMTT1995}. It turns out that this squint  can  be compensated for by  using  (variable) delay lines rather than  phase shifters. The use of optical techniques to generate true time delay for phased array applications has been the subject of extensive research over the past two decades. A compilation of these techniques can be found in \cite{CapmanyNatPhoton2007,SeedsMWP2002,SeedsMWP2006,YaoMWP2009}. One of the most well-known concept is the fiber optic prism where the dispersion property of the fiber-optic link was used to create variable delays for variable source wavelengths \cite{Frankel1995}. As the laser wavelength is tuned, the differential delay between fiber paths changes, thus steering the antenna beam. The concept was later on revisited and demonstrated by Zmuda et al. in 1997 using fiber Bragg-gratings \cite{ZmudaPTL1997}.

\begin{figure*} [ht!!]
  \sidecaption
  \includegraphics*[width=1\textwidth]{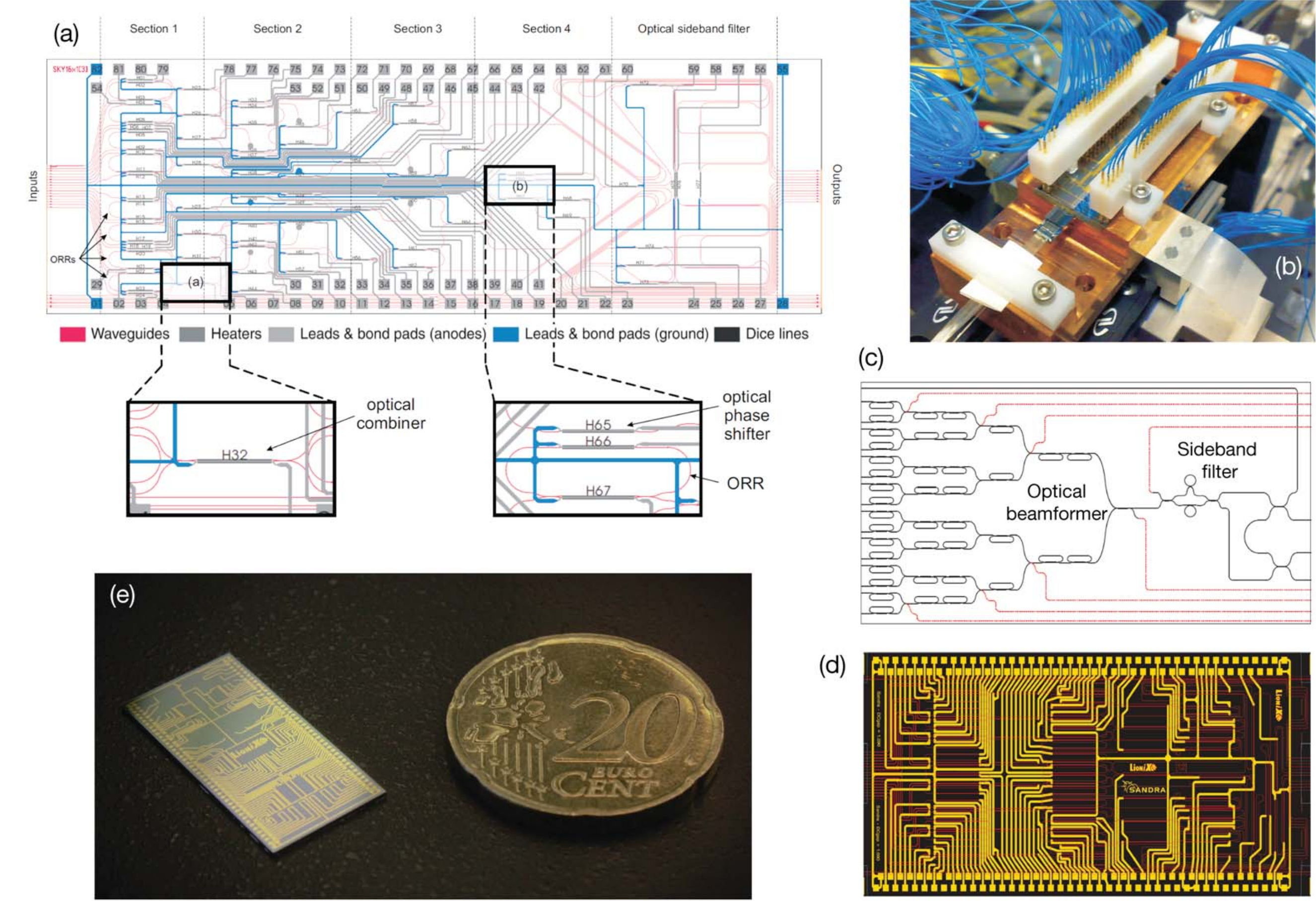}
  \caption{(a) Overview of the 16$\times$1 OBFN chip layout (13~mm$\times$70~mm). Waveguides, heaters for electro-optical tuning, bond pads, and lines for electrical control are visible. The insets provide a closer view of the tunable combiner of the phase shifter, and of the ORR. (b) Photograph of the packaged chip. \cite{BurlaAO2012}, courtesy of the OSA.  (c) Functional design of the $16\times1$ photonic beamformer showing the ORR delay elements and the sideband filter. (d) Chip layout of the BFN  showing the optical waveguides, the heaters layout and the electrical wiring. The chip dimension is 22~mm$\times$7~mm. (e) The 16x1 photonic BFN chip pictured with a 20 Euro-Cent coin for size comparison.}
  \label{OBFN}
\end{figure*} 

Beamformers based on fibers can be rather bulky. To reduce the footprint of the beamformer, as well as to obtain precise time delay, many turn to integrated photonic solutions \cite{AckermanIMS1992,NgSPIE1994,NgPTL1994,HorikawaIMS1995,HorikawaOFC1996,StulemeijerPTL1999,GrosskopfFIO2003,HowleyJLT2007,ZhuangPTL2007,ZhuangJLT2010,CardenasOpex2010,MortonPTL2012,BurlaAO2012}. In general three categories of integrated photonic beamformer have been demonstrated over the years: wideband beamformer based on discretely tunable TTD \cite{AckermanIMS1992,NgPTL1994,NgSPIE1994,HorikawaIMS1995,HorikawaOFC1996,HowleyJLT2007}, wideband beamformer based on continuously tunable TTD \cite{ZhuangPTL2007,ZhuangJLT2010,CardenasOpex2010,MortonPTL2012,BurlaAO2012} and narrowband beamformer based on optical phase shifters \cite{HorikawaMTT1995,StulemeijerPTL1999,GrosskopfFIO2003,GrosskopfAntenna2003}.      

As early as 1992, Ackerman et al. \cite{AckermanIMS1992} proposed an integrated optical switches in LiNbO\textsubscript{3} to form a 6-bit TTD unit for 2-6~GHz radar application. However, in this approach the delays were induced by optical fibers. A monolithic integrated beamformer was proposed by Ng et al. \cite{NgPTL1994,NgSPIE1994} using a PIC in GaAs. This approach used curved GaAlAs/GaAs rib-waveguides with propagation losses of 1~dB/cm integrated monolithically with GaAs based detector switches to form 4 (2-bit) switched delay lines. The switched delay lines approach was also demonstrated in silica PLC by Horikawa et al. \cite{HorikawaIMS1995,HorikawaOFC1996}. Two architectures were proposed, a cascaded and a parallel switches configurations. The switching is done using a $2\times2$ thermo-optic switches and the loss in the delay lines was 0.1~dB/cm. Finally, polymer materials have been considered for the switched-delay beamformer. In \cite{HowleyJLT2007} the realization of a  packaged 4-bit TTD device composed of monolithically integrated polymer waveguide delay lines and five $2\times2$ polymer total internal reflection (TIR) thermo optic switches was reported. The polymer waveguides exhibited a single-mode behavior with a measured propagation loss of 0.45~dB/cm at the wavelength of 1.55~$\mu$m. The delay lines used waveguides with bend radius of 1.75~mm. The dimension of the TTD device is 21.7-mm long by 13.7-mm wide. Beamforming at the X-band frequency range (8-12~GHz) was demonstrated. 

Narrowband beamforming can be achieved using optical phase shifters instead of TTDs. Using a coherent detection scheme phase and amplitude of an optical signal can be directly transferred to a microwave signal by mixing this signal with an optical local oscillator signal. In this way, modulation of phase of a microwave signal can be performed using optical phase shifters. Initial works in this category include the self heterodyning beamformer based on LiNbO\textsubscript{3} proposed by Horikawa et al. in 1995 \cite{HorikawaMTT1995}. InP-based PIC has also been considered \cite{StulemeijerPTL1999}. This beamformer controls the amplitude and phase of the RF signals using phase modulators and variable attenuators. Although the beamformer is narrowband, it featured an ultra compact footprint of $8.5\times8$~mm$^2$ for a $16\times1$ beamformer. A similar concept of narrowband beamforming was also implemented in silica PLC where thermo-optic effect is used to achieve the desired phase shifting \cite{GrosskopfFIO2003}. Amplitude control was performed using PLC type MZIs with two 3~dB multimode interferometer (MMI) couplers (50:50 ratio) and independent thermo-optic phase shifters on both waveguide arms. The chip size of the realized eight-channel OBFN is about 4~mm$\times$65~mm. Beamforming at 60~GHz was demonstrated \cite{GrosskopfAntenna2003}.      

For practical applications like satellite communications, a wideband continuously tunable beamformer is required. Such beamformer based on cascaded tunable ORRs was proposed by the researchers in the University of Twente \cite{ZhuangPTL2007,MeijerinkJLT2010,ZhuangJLT2010,MarpaungEuCAP2011,MarpaungMWP2011,BurlaAO2012,MarpaungEuCAP2011,MarpaungMWP2011}.  In \cite{ZhuangPTL2007}, a state-of-the-art ring resonator-based $1\times8$ beamformer chip has been proposed. A binary tree topology is used for the network such that a different number of ORRs is cascaded for delay generation at each output. The beamformer was fabricated in TriPleX\texttrademark\,waveguide technology. With this beamformer, a linearly increasing delay up to 1.2~ns in a 2.5~GHz bandwidth was demonstrated. In \cite{MeijerinkJLT2010,ZhuangJLT2010}, the suitability of the beamformer for satellite communications was investigated. The intended application was communications at the Ku-band frequency range (10-12.5~GHz). A coherent beamforming architecture using optical single sideband-suppressed carrier (OSSB-SC) modulation and balanced coherent detection was proposed and the performance was analyzed \cite{MeijerinkJLT2010}. The OSSB-SC signal was generated using a sideband filter integrated on the same chip. A $1\times8$ beamformer with 8~ORRs was fabricated using the box-shape TriPleX\texttrademark\,waveguides with a propagation loss of 0.6~dB/cm \cite{ZhuangJLT2010}. The ORRs have a minimum bending radius of 700~$\mu$m. The total chip footprint measured 66.0~mm$\times$12.8~mm.

In \cite{BurlaAO2012}, a $16\times1$ beamformer with a high degree of complexity was reported for phased-array antenna in radio astronomy application. The beamformer consists of 20~ORRs, more than 25~MZI tunable couplers and an OSBF. The chip was also fabricated in the box-shape TriPleX\texttrademark\,waveguides that exhibit propagation loss of 0.2~dB/cm. The chip footprint is 70.0~mm$\times$13~mm. The schematic of the chip showing the optical waveguides layout the heaters for thermo-optical tuning and the realized chip (packaged) are shown in Figures~\ref{OBFN}~(a) and (b). The work was focused on the system integration and the experimental demonstration of the beamformer with the phased array antenna. The measurements show a wideband, continuous beamsteering operation over a steering angle of 23.5 degrees and an instantaneous bandwidth of 500~MHz limited only by the measurement setup.         

A novel $16\times1$ beamformer design was reported in \cite{MarpaungMWP2011}. The beamformer was designed to meet the requirements of a Ku-band PAA system with instantaneous bandwidth of more than 4~GHz and a maximum time delay of 290~ps. For the total PAA system, 32 of these beamformers will be used to beamform a large scale PAA of 2048 antenna elements. The complete system level analysis of this PAA system using the optical beamformers was reported in \cite{MarpaungEuCAP2011}. The novel beamformer consists of 40~ORRs. From the system level simulations, it was concluded that to have a good noise figure and SNR at the receiver output, the waveguide propagation loss should not exceed 0.2~dB/cm. For this reason, and for the sake of foot-print reduction, the beamformer was fabricated using the double-stripe TriPleX\texttrademark\,waveguide technology. As reported in \cite{ZhuangOpex2011}, these waveguides feature a propagation loss as low as 0.1~dB/cm while maintaining a tight bending radius down to 75~$\mu$m. For the beamformer reported in \cite{MarpaungMWP2011,MarpaungEuCAP2011}, the bend radius of the ORRs was chosen to be 125~$\mu$m. This enables a highly complex beamformer with a small total footprint of 22~mm$\times$7~mm, which is a size reduction of nearly 10 times of the beamformer reported in \cite{BurlaAO2012}. The schematic, layout and the photograph of this beamformer are shown in Figures~\ref{OBFN}~(c)-(e). For an implementation in an actual PAA system, an important aspect of system stability and reliability must be addressed. For this reason, work towards hybrid RF and photonic integration of the passive optical beamformer in TriPleX\texttrademark\, technology, an array of surface normal electroabsorption modulators in InP platform and RF front-end is ongoing \cite{MarpaungMWP2011,MarpaungEuCAP2011}.

\section{Microwave signal generation}
\label{sec:generation}

Microwave signal generation techniques have enjoyed enormous progress over the past five years. The aim in such research activities is generating ultra broad bandwidth RF waves with arbitrary and reconfigurable phase and amplitude or to generate extremely stable and pure microwave carriers. In this section we review the integrated MWP approaches in arbitrary waveform generation, ultrawideband (UWB) pulse shaping and stable carrier generation using optoelectronic oscillators (OEO).   

\begin{figure*} [ht!!]
 \sidecaption
  \includegraphics*[width=1\textwidth]{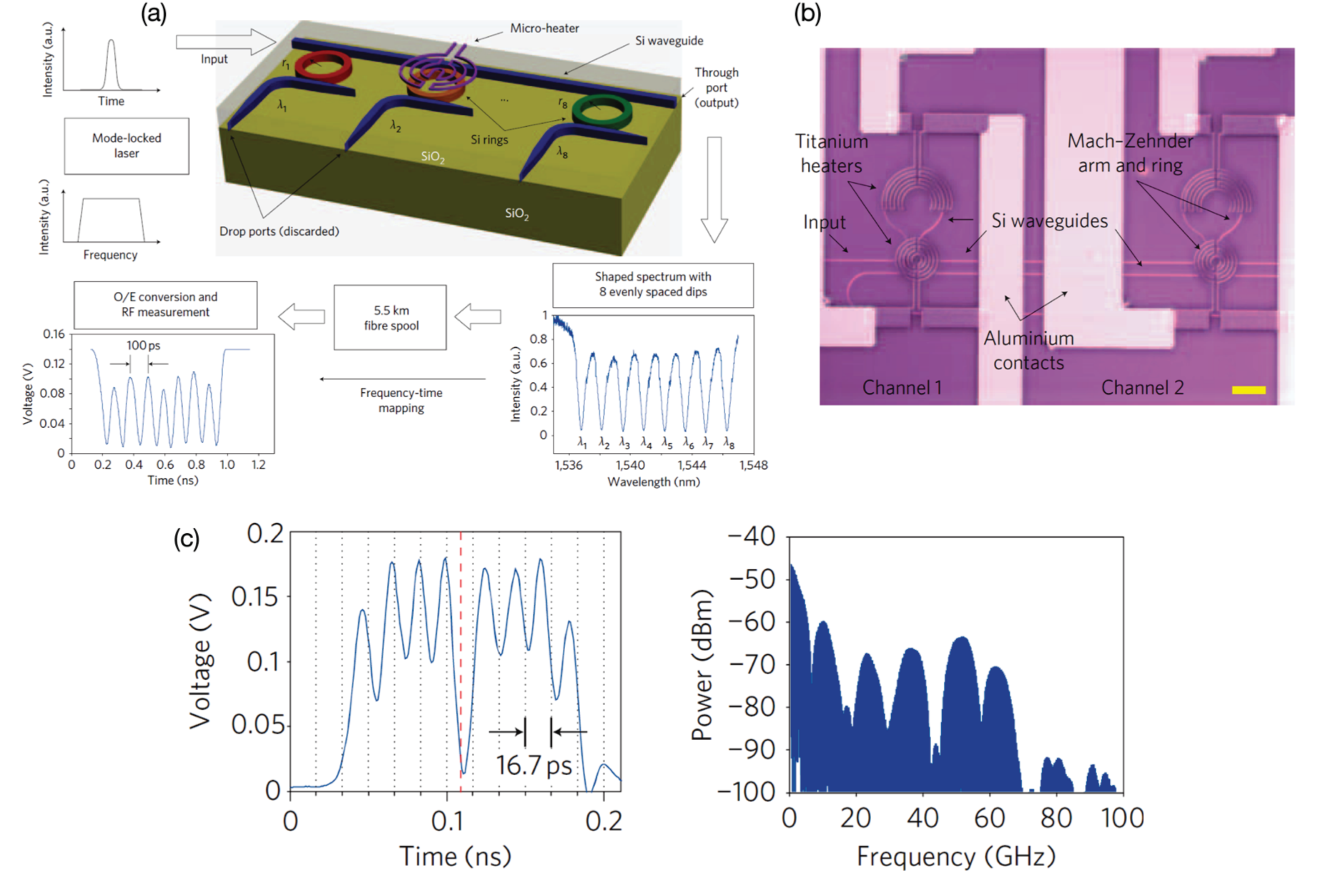}
  \caption{On-chip arbitrary waveform generation using cascaded of 8 microring resonators in SOI platform. (a) Waveform generator system. (b) Optical image of the pulse shaper. (c) Example of waveform generation at 60~GHz (from \cite{KhanNatPhotonics2010}, courtesy of the Macmillan Publishers Ltd).}
  \label{AWG}
\end{figure*}  

\subsection{Arbitrary waveform generation}
\label{subsec:arbitrary}

Microwave arbitrary waveform very  useful for pulsed radar, modern instrumentation systems and UWB communications. However, current electronic arbitrary  waveform generation (AWG) is limited in frequency and bandwidth. The current state-of-the-art electronics AWG operates at a maximum bandwidth of 5.6~GHz and a frequency up to 9.6~GHz~\footnote{Tektronix AWG7122C. See: http://www.tek.com/signal-generator/awg7000-arbitrary-waveform-generator}. Photonics, on the other hand,  has become a promising  solution for generating high-frequency  microwave waveforms \cite{CapmanyNatPhoton2007, YaoOptComm2011}. A  variety of MWP techniques have been proposed during the past few years that can generate microwave waveforms in the  gigahertz and multiple gigahertz region. These include direct space-to-time mapping \cite{McKinneyOL2002}, spectral shaping and wavelength-to-time mapping \cite{ChouPTL2003,LinUWB2005} and temporal pulse shaping. With these techniques, frequency-chirped, and phase coded microwave waveforms have been demonstrated. 

But the above mentioned techniques have been  implemented using complex bulk optic devices which can be expensive, complicated and bulky. Another option is to use all-fiber pulse-shaper. For example, a linearly chirped  microwave waveform can be generated  by spectral shaping using chirped FBGs followed by frequency-to-time mapping in a dispersive device \cite{WangPTL2008}. But fiber-based devices lack of programmability which is necessary in arbitrary waveform generation.  

An on-chip integrated pulse shaper is thus a desirable solution to overcome the limitations commonly associated with conventional bulk optics pulse shapers. Recently, researchers at Purdue University demonstrated an integrated ultrabroadband arbitrary microwave waveform generator that incorporates a fully programmable spectral shaper fabricated on a silicon photonic chip \cite{ShenOpex2010,KhanNatPhotonics2010}. The spectral shaper is a reconfigurable filter consisting of eight add-drop microring resonators on a silicon photonics platform. Ultra-compact cross-section (500~nm$\times$250~nm) silicon nanowires have been used to fabricate the rings \cite{ShenOpex2010}. The typical bending radius used was 5~$\mu$m and the waveguide propagation loss was around 3.5~dB/cm \cite{XiaoOpex2007}. This spectral shaper programmability is achieved by thermally tuning both the resonant frequencies and the coupling strengths of the microring resonators. A cartoon of the spectral shaper and an optical image showing two channels of ring resonators with micro-heaters are shown in Figure~\ref{AWG}~(a) and (b), respectively.

The principle of the photonic arbitrary microwave waveform generation system implemented here is shown in Figure~\ref{AWG}~(a).The spectral shaper is used to modify the spectrum emitted from a mode-locked laser. The shaped spectrum then undergoes wavelength-to-time mapping in a dispersive device, which in this case is a length (5.5~km) of optical fiber,  before being converted to the electrical domain a microwave waveform using a high-speed photodetector. By incorporating the spectral shaper into a, a variety of different waveforms are generated, including those with an apodized amplitude profile, multiple $\pi$ phase shifts (Figure~\ref{AWG}~(c)), two-tone waveforms and frequency-chirped waveforms at the central frequency of 60~GHz.

In another demonstration of arbitrary waveform generation using PIC,  a planar lightwave circuits (PLCs) fabricated on silica-on-silicon is used to generate pulse trains at 40 GHz and 80 GHz with flat-top, Gaussian, and apodized profiles \cite{SamadiOptComm2011}. The pulse shaper is a 12 tap finite impulse response (FIR) filter that performs both phase and amplitude filtering. It is implemented as 12 stages of cascaded Mach-Zehnder interferometers each with an FSR of 80 GHz, which corresponds to a temporal tap separation of 12.5~ps.  The waveguide cross-section used for the PLC fabrication is designed to be 3.5~$\mu$m$\times$3.5~$\mu$m. The curved portions of the circuit have a radius of at least 2~mm in order to minimize bending losses. The propagation loss of the waveguides is 0.7~dB/cm.

\begin{figure*} [ht!!]
 \sidecaption
  \includegraphics*[width=0.7\textwidth]{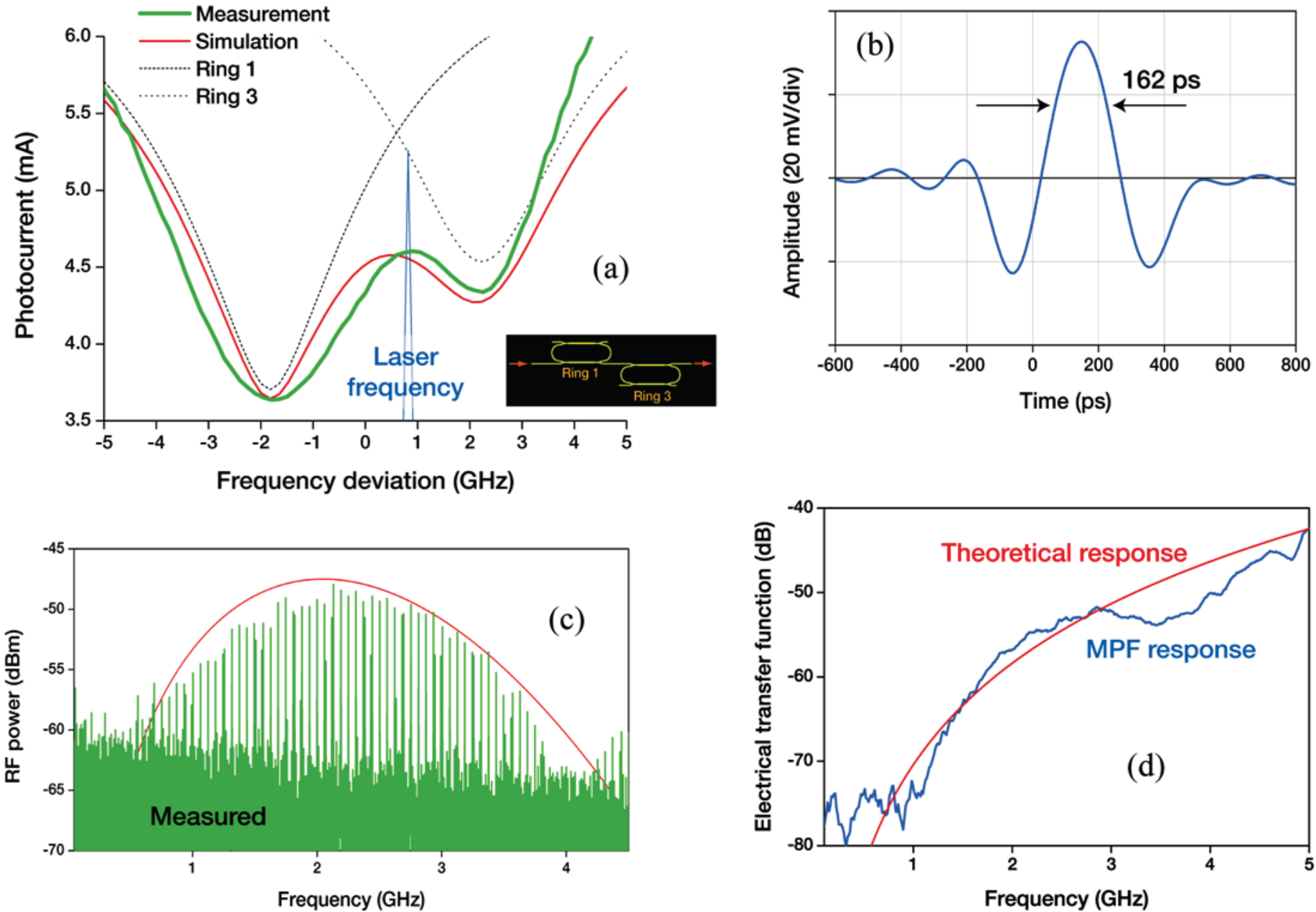}
  \caption{Measurement results on the doublet generation with a cascade of two ORRs. (a) Measured two ORRs through responses depicted together with simulation results. (b) Waveforms of the generated doublet. (c) Power spectral density of the generated doublet compared with the theoretical response. (d) Comparison of the theoretical and the measured transfer function of the microwave photonic filter synthesized for the doublet generation (from \cite{MarpaungOpex2011}, courtesy of the OSA).}
  \label{UWB}
\end{figure*}  

\subsection{Impulse radio UWB pulse shaping}
\label{subsec:UWB} 

In the last five years numerous techniques have been proposed for the so-called photonic generation of impulse-radio UWB (IR-UWB) pulses. In this approach UWB signals are generated and later on distributed in the optical domain to increase the reach of the UWB transmission, similar to the more general concept of radio over fiber. The generated pulses are usually the variants of Gaussian monocycles, doublets or in some occasion higher-order derivatives of the basic Gaussian pulses. These techniques usually aim at generating the pulses which power spectral densities (PSD) satisfy the regulation (i.e. spectral mask) specified by the U.S. Federal Communications Commission (FCC) for indoor UWB systems. Different techniques that have been proposed for IR-UWB pulses generation, such as spectral shaping combined with frequency-to-time mapping, nonlinear biasing of an MZM, spectral filtering using MWP delay-line filters \cite{BoleaOpex2009}, or using PM-IM conversion to achieve temporal differentiation of the input electrical signals. A comprehensive review on these techniques is given in \cite{YaoJLT2007}. 

Integrated photonics technologies have also been exploited in the approach of IR-UWB signal generation \cite{LiuElectLett2009,YunhongIPC2011,MarpaungOpex2011,MirshafieiPTL2012,YueOL2012,WangOpex2009,WangPTL2010}. Liu et al. reported the use of an SOI ORR as a temporal differentiator to shape input Gaussian electrical pulses into their first-order derivatives (i.e., monocycles) \cite{LiuElectLett2009}. Using the a similar technique, a silicon add-drop ORR has been used to generate monocycles from a 12.5 Gb/s NRZ signals \cite{YunhongIPC2011}. However, Gaussian monocycles cannot fill the FCC spectral mask with high efficiency. For this reason, higher order derivatives of the Gaussian pulses are often desired. Marpaung et al. recently reported the use a cascade of two add-drop ORRs in TriPleX technology to generate the second-order derivatives (doublets and modified doublets) of the input Gaussian pulses used to modulate the phase of the optical carrier \cite{MarpaungOpex2011}. The PM-IM transfer using the cascaded ORRs forms an MWP bandpass filter that shapes the input Gaussian spectrum accordingly. This is the integrated photonics implementation of the MWP delay-line filtering technique. An example of the spectral filtering to generate the Gaussian doublet is shown in Figure~\ref{UWB}.  
  
An important aspect in IR-UWB pulse shaping is to fill the spectral mask with a high power efficiency. Very recently, Mirshafiei et al. have demonstrated an output pulse obtained from a linear combination of a Gaussian pulse and its copy, filtered using a silicon ORR. By careful adjustment of the amplitude and relative time delay between the pulses, an output pulse with a power efficiency of 52\% has been obtained \cite{MirshafieiPTL2012}. 

Recently, on-chip nonlinear optics have been used for IR-UWB generation. These include a monocycle generation based on two-photon absorption in a silicon waveguide \cite{YueOL2012} and monocycles generations exploiting the parametric attenuation effect of sum-frequency generation (SFG) \cite{WangOpex2009} or using the quadratic nonlinear interaction seeded by dark pulses \cite{WangPTL2010} in a periodically poled lithium niobate (PPLN) waveguide. We expect to see more techniques based on on-chip nonlinear techniques for AWG and IR-UWB. 

\subsection{Optoelectronic oscillator and optical comb generation}
\label{subsec:OEO} 

\begin{figure}
  \includegraphics*[width=\linewidth]{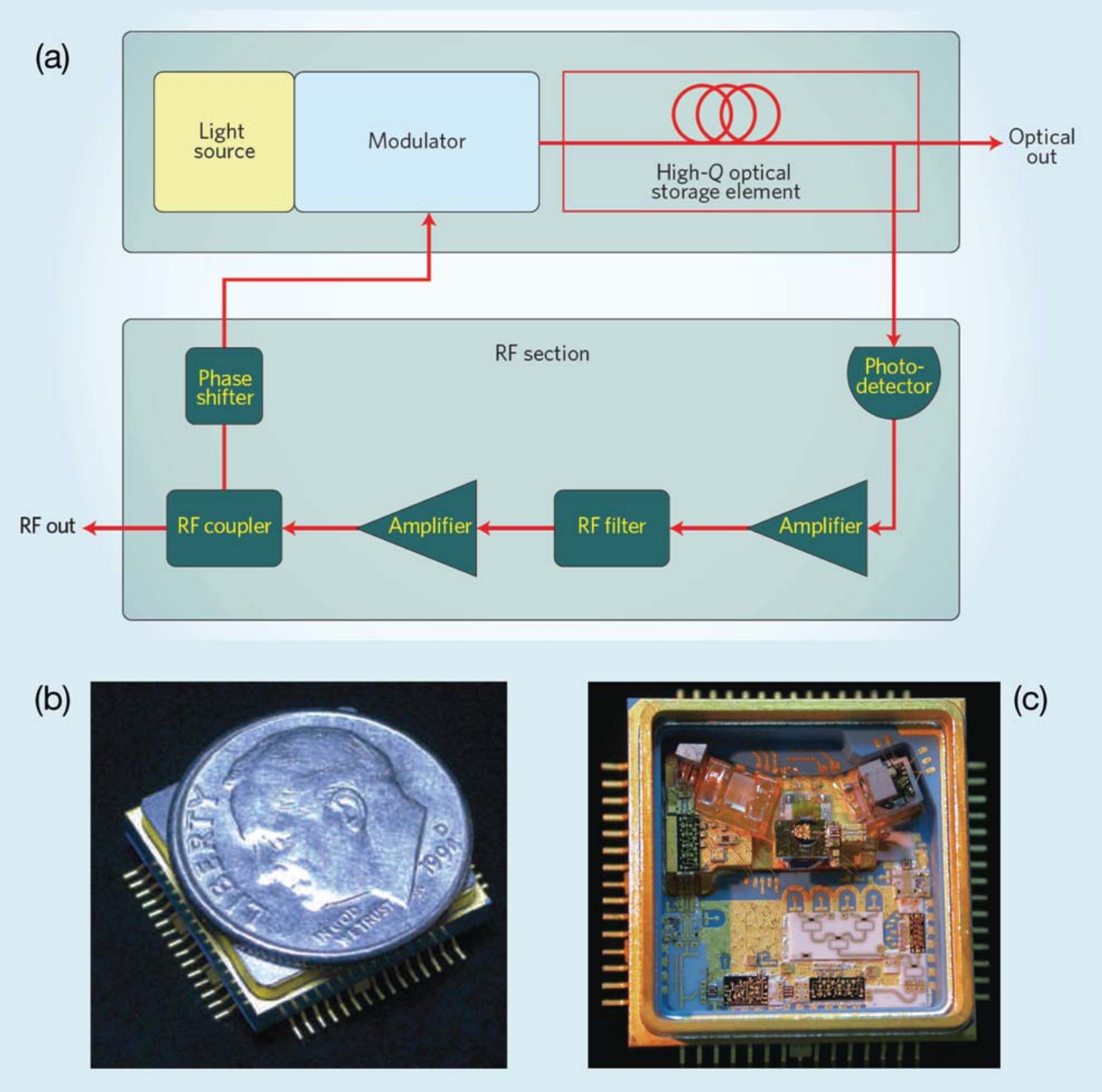}
  \caption{(a) Block diagram of a generic OEO. (b) and (c) Miniature OEO based on a lithium niobate WGMR (from \cite{MalekiNatPhotonics2011}, courtesy of the Macmillan Publishers Ltd).}
  \label{OEO}
\end{figure}

In 1986 Steve Yao and Lute Maleki, two researchers then at the Jet Propulsion Laboratory proposed a new type of high-performance oscillator known as the optoelectronic oscillator (OEO) \cite{YaoJOSAB1996}. The typical configuration of an OEO, which is shown Figure~\ref{OEO}~(a), is based on the use of optical waveguides and resonators, which exhibit significantly lower loss than their electronic counterparts. Typically, light from a laser is modulated and passed through a long length of optical fiber before reaching a photodetector. The output of the photodetector is amplified, filtered, adjusted for phase and then feedback to the modulator providing self-sustained oscillation if the overall round trip gain is larger than the loss and the circulating waves can be combined in phase. 

Optoelectronic oscillators (OEOs) are thus ultra-pure microwave generators based on optical energy storage instead of high finesse radio-frequency (RF) resonators. These oscillators have many specific advantages, such has exceptionally low phase noise, and versatility of the output frequency (only limited by the RF bandwidth of the optoelectronic components). Such ultra-pure microwaves are indeed needed in a wide range of applications, including time-frequency metrology, frequency synthesis, and aerospace engineering.

The spectral purity of the signal in the OEO is directly related to the Q-factor of the loop and so far,  most OEOs utilize a long length of fiber to achieve high spectral purity. A disadvantage of using a fiber loop is the production of \textsl{super modes} that appear in the phase noise spectrum caused by the propagation of waves multiple times around the OEO loop. In addition, a fiber delay line is bulky, so that the oscillators can not be considered as an optimal solution for the implementation of transportable microwave source. Along the same line, this bulky delay line element has to be temperature-stabilized, a feedback control process which is energy consuming. One solution to circumvent all these disadvantages has been proposed which also leads to the possibility of full integration of OEOs. It consists in replacing the optical fiber loop by a high Q-factor cavity implemented by means of a whispering gallery mode resonator (WGMR) \cite{MalekiNatPhotonics2011,LiangOL2010,DevganOEOPTL2010,VolyanskiyOpex2010}. WGMRs ranging in size from a few hundred micrometers to a few millimeters can be fabricated from a wide variety of optically transparent materials reaching Q-factors in the range of $3\times10^{11}$. OEOs based on high-Q WGMRs made from electro-optic materials can provide high performance in a miniaturized package smaller than a coin, as shown in Figure~\ref{OEO}~(b) and (c) \cite{MalekiNatPhotonics2011}, operating in frequency ranges from 10 to 40 GHz and featuring instantaneous linewidths below 200~Hz. Furthermore, in this configuration, the resonator, serves both as the high-Q element and as the modulator in the OEO loop. 

Optical Comb generation on a chip \cite{KippenbergScience2011} is also of  great interest in microwave photonics as it enables several applications such as the precise measurement of optical frequencies through direct referencing to microwave atomic clocks \cite{FosterOpex2011} and the production of multiple taps for high sidelobe rejection signal processors \cite{FerdousNatPhotonics2011}. Combs have been traditionally generated using mode-locked ultrafast laser sources but recently the generation of optical frequency combs through the nonlinear process of continuous-wave optical-parametric oscillation using micro-scale resonators has attracted significant interest since these devices have the potential to yield highly compact and frequency agile comb sources. In particular \cite{FosterOpex2011} reports the generation of optical frequency combs from a highly-robust CMOS-compatible integrated microresonator optical parametric oscillator where both the microresonator and the coupling waveguide are fabricated monolithically in a single silicon nitride layer using electron-beam lithography and subsequently clad with silica. This approach brings the advantage of providing a fully-monolithic and sealed device with coupling and operation that is insensitive to the surrounding environment.

\section{Other emerging applications}
\label{sec:other}

A number of emerging and exciting applications have surfaced in the past years. They are beyond the scope of this paper, but are worthy to mention here due to their potential. These approach have taken the full advantage of the availability of PIC technologies. The first application is fundamental computing functions like differentiation and integration. Recently, photonic differentiators using SOI integrated Bragg-gratings \cite{RutkowskaOpex2011} and all-optical temporal integration using an SOI four-port microring resonator \cite{FerreraNatCommunications2010,FerreraOpex2011} have been reported. These devices have applications in real time analysis of differential equations.  

On chip microwave frequency conversion is another technique that recently received increasing interest. In this approach, RF frequency mixing for signal upconversion and downconversion is performed in a photonic integrated circuit. In \cite{GutierrezOL2012}, the RF mixer is realized in silicon electro-optical MZ modulator enhanced via slow-light propagation. An upconversion from 1~GHz to 10.25~GHz was demonstrated.  In \cite{JinPTL2012}, Jin et al. proposed the so-called RF photonic link on chip PIC which can operate in  a linear mode (as an MWP link) or a mixer mode for microwave frequency conversion. The device used for the mixing is the ACP-OPLL receiver reported earlier \cite{LiPTL2011}. Frequency downconversion from 1.05~GHz to 50~MHz was demonstrated.        

A very exciting emerging application is the photonic analog-to-digital conversion (ADC). Photonic ADCs have been actively investigated over the last decades; an overview and classification of photonic ADCs can be found in an excellent review by Valley \cite{ValleyOpex2007}. Recently, a chip incorporating the core optical components of the photonic ADC (a modulator, wavelength demultiplexers, and photodetectors) has been fabricated in silicon photonics and shown to produce 3.5~ENOB for a 10~GHz input \cite{GreinCLEO2011,KhiloOpex2012}. 

The final application is frequency measurement and spectrum analysis. The recent year have shown increase in techniques for instantaneous microwave frequency measurement (IFM). Wide bandwidth is the main attraction to do IFM system with photonics, compared to purely electrical solution. An added value will be fast reconfigurability, lightweight, transparency to microwave frequencies and small form factor. Such systems can readily be applied in the military and security applications \cite{JacobsOFC2007}. As for the spectrum analysis, on-chip RF spectrum analyzer with THz bandwidth based on nonlinear optics have been demonstrated both in chalcogenide \cite{PelusiNatPhotonics2009} and SOI \cite{CorcoranOpex2010} waveguides. These devices will find application in high speed optical communications.

\section{Prospective: what's next for integrated MWP?}
\label{subsec:prospect}

We believe that integrated MWP has just started to bloom and is set to have a bright future. Continuing the current trend, we believe that MWP filters will continue as the leading signal processing applications. We expect to see more demonstration of PIC based MWP filters, involving resonators and/or photonic crystals.  We also expect to see more use of nonlinear optics for MWP signal processing. For example four-wave mixing (FWM) for enhancing the gain of MWP link \cite{WallCLEO2012} and for MWP filtering \cite{SupradeepaNatPhotonics2012,VidalPJ2012} have been reported. Moreover, SBS on chip based on chalcogenide has just been reported \cite{PantOpex2011}. This has been demonstrated for delay line \cite{PantOL2012} and MWP filter \cite{ByrnesCLEO2012}. It is expected that this phenomena will also be used for beamforming and MWP link. Recently the SBS on silicon chip has been predicted \cite{GaetaNatPhoton2012}. In the near future this will also be useful for integrated MWP. From the modulation technique point of view,  the use of phase or frequency modulation is predicted to show significant rise in interest. Recently, a beamformer based on PM has been proposed \cite{XueOptComm2011}, as well as radio over fiber link \cite{GasullaOpex2012}. More systems will also take advantage of simultaneous phase and intensity modulation \cite{UrickAVFOP2011}. These systems will also use FM discriminator to synchronize the two modulation scheme and to enhanced the link gain and SFDR. In the technology level, it is expected to see efforts in reducing the loss of optical waveguides even further as well as reducing the power consumption for tuning and reconfigurability, either for thermal tuning \cite{DongOpex2010_power}, or tuning using other cladding materials like liquid crystals \cite{DeCortOL2011} or chalcogenides \cite{MelloniSPIE2012}. \textcolor{black}{To realize high performance circuits, many MWP applications will take the advantage of a mixture of optical and electronic devices.  It is anticipated that future work will merge these two technologies and be dealing with the issues of creating integrated circuits involving both technologies.} Finally, the field will see an increase in implementation of PICs for the generation and processing of THz signals and waveforms, as reported in \cite{SteedJSTQE2011,BenYooTerahertz2012}.   

\begin{acknowledgement}
DM and CR acknowledge the support of the European Commission via the 7th Framework Program SANDRA project (Large Scale Integrating Project for the FP7 Topic AAT.2008.4.4.2). SS and JC acknowledge the GVA-2008-092 PROMETEO project Microwave Photonics. 

DM and CR would like to thank Leimeng Zhuang, Maurizio Burla and Reza Khan for their contributions to the article. S.S and J.C would like to thank Ivana Gasulla, Juan Lloret and Juan Sancho for their constant support and collaboration.
\end{acknowledgement}

\begin{biographies}
  \authorbox{David}{David~Marpaung}{received his Ph.\,D. degree in electrical engineering from the University of Twente, the Netherlands in 2009. The topic of his PhD thesis was the performance enhancement of analog photonic links. From August~2009 until July~2012 he was a postdoctoral researcher in the University of Twente within the framework of the European Commission FP7 funded project SANDRA. He was leading the research activities on microwave photonic system integration for a large scale photonic beamformer in a phased array antenna system for satellite communications. Since August 2012 he is a Research Fellow in the Centre for Ultrahigh Bandwidth Devices for Optical Systems (CUDOS), School of Physics University of Sydney, Australia, working on photonic chip-based nonlinear signal processing for high speed coherent optical communication systems. His main research interest is the use of integrated photonic circuits for microwave photonic signal processing and high capacity optical communications.}
  \authorbox{Chris}{Chris~Roeloffzen}{received the M.Sc. degree in applied physics and Ph.D. degree in electrical engineering from the University of Twente, Enschede, The Netherlands, in 1998 and 2002, respectively. From 1998 to 2002 he was engaged with research on integrated optical add-drop demultiplexers in Silicon Oxinitride waveguide technology, in the Integrated Optical Microsystems Group at the University of Twente. In 2002 he became an Assistant Professor in the Telecommunication Engineering Group at the University of Twente. He is now involved with research and education on microwave photonic systems. He is founder of the company SATRAX BV, a spin off of the University of Twente.}
  \authorbox{Rene}{Ren\'{e}~Heideman}{obtained his M.Sc and his PhD degree in Applied Physics at the University of Twente. After his post-doc positions he applied his extensive know-how in the industry. Since 2001, he is co-
founder and CTO of LioniX BV. He is an expert in the field of MST, based more than 20 years of experience. He specializes in Integrated Optics (IO), covering both (bio-) chemical sensing and telecom applications. He is (co)author of more than 150 papers and holds more than 20 patents in the IO-field, on 10 different subjects. He participates in several Dutch steering committees (IOP Photonic Devices, STW Generic Technologies for Integrated Photonics and MinacNed (Microsystems and Nanotechnology)) as well as several EU projects.}
  \authorbox{Arne}{Arne~Leinse}{received a M.Sc. degree from the University of Twente in the integrated Optical Microsystems group in 2001. In this same group he started his PhD work on the topic of active microring resonators for various optical applications. His PhD work was carried out in the framework of a European project (IST 2000-28018 "`Next generation Active Integrated optic Sub-systems"') and his thesis was titled: "`Polymeric microring resonator based electro-optic modulator". In 2005 he joined LioniX BV where he is now as a project-/account manager involved in several integrated optical projects (e.g. the Senter Novem Smartmix Memphis project (Merging Electronics and Micro\&nano-PHotonics in Integrated Systems) in which more than 20 partners participate.}
  \authorbox{Salvador}{Salvador~Sales}{received the degree of Ingeniero de Telecomunicaci\'{o}n and the Ph.D. in telecomunicaci\'{o}n from the Universidad Polit\'{e}cnica de Valencia, Spain. He is Professor at the Departamento de Comunicaciones, Universidad Polit\'{e}cnica de Valencia Spain. He is also working in the ITEAM Research Institute. He is currently the coordinator of the Ph.D. Telecomunicaci\'{o}n students of the Universidad Polit\'{e}cnica de Valencia. He has been Faculty Vicedean of the UPVLC in 1998 and Deputy Director of the Departamento de Comunicaciones in 2004�2008. He is co-author of more than 80 journal papers and 150 international conferences. He has been collaborating and leading some national and European research projects since 1997. His main research interests include optoelectronic signal processing for optronic and microwave systems, optical delay lines, fiber Bragg gratings, WDM and SCM ligthwave systems and semiconductor optical amplifiers. Prof. Sales received the Annual Award of the Spanish Telecommunication Engineering Association to the best Ph.D. on optical communications.}
  \authorbox{Jose}{Jos\'{e}~Capmany}{was born in Madrid, Spain, on December 15 1962. He received the Ingeniero de Telecomunicacion degree from the Universidad Polit\'{e}cnica de Madrid (UPM) in 1987 and the Licenciado en Ciencias F\'{i}sicas in 2009 from UNED. He holds a PhD in Electrical Engineering from UPM and a PhD in Quantum Physics from the Universidad de Vigo.
Since 1991 he is with the Departamento de Comunicaciones, Universidad Polit\'{e}cnica de Valencia (UPV), where he started the activities on optical communications and photonics, founding the Optical Communications Group (www.gco.upv.es). He has been an Associate Professor from 1992 to 1996, and Full Professor in optical communications, systems, and networks since 1996. In parallel, he has been Telecommunications Engineering Faculty Vice-Dean from 1991 to 1996, and Deputy Head of the Communications Department since 1996. Since 2002, he is the Director of the ITEAM Research Institute, Universidad Polit\'{e}cnica de Valencia. His research activities and interests cover a wide range of subjects related to optical communications including optical signal processing, ring resonators, fiber gratings, RF filters, SCM, WDM, and CDMA transmission, wavelength conversion, optical bistability and more recently quantum cryptography and quantum information processing using photonics. He has published over 420 papers in international refereed journals and conferences and has been a member of the Technical Programme Committees of the European Conference on Optical Communications (ECOC), the Optical Fiber Conference (OFC), the Integrated Optics and Optical Communications Conference (IOOC), CLEO Europe, and the Optoelectronics and Communications Conference (OECC). Professor Capmany has also carried out activities related to professional bodies and is the Founder and current Chairman of the LEOS Spanish Chapter, and a Fellow of the Institution of Electrical and Electronic Engineers (IEEE), the Optical Society of America (OSA) and the Institution of Electrical Engineers (IEE). He has acted as a reviewer for over 25 SCI journals in the field of photonics and telecommunications.
Professor Capmany is the recipient of the Extraordinary Engineering Doctorate Prize of the Universidad Polit\'{e}cnica de Madrid and the Extaordinary Physics Laurea Prize from UNED. He is an associate Editor of IEEE Photonics Technology Letters. He is a founder and Chief innovation Officer at VLC Photonics.}
\end{biographies}
%

\bibliographystyle{lpr}
\bibliography{IEEEabrv,LPR-IMWP}

\providecommand{\WileyBibTextsc}{}
\let\textsc\WileyBibTextsc
\providecommand{\othercit}{}
\providecommand{\jr}[1]{#1}
\providecommand{\etal}{~et~al.}


\begin{thebibliography}{[100]}

\bibitem{CapmanyNatPhoton2007}
 \textsc{J.~Capmany} and  \textsc{D.~Novak},
 \jr{{Nat. Photonics}} \textbf{{1}}({6}), {319--330} ({2007}).


\bibitem{SeedsMWP2002}
 \textsc{A.~Seeds},
 \jr{{IEEE} Trans. Microw. Theory Tech.} \textbf{50}(3), 877--887 (2002).


\bibitem{SeedsMWP2006}
 \textsc{A.\,J. Seeds} and  \textsc{K.\,J. Williams},
 \jr{J. Lightw. Technol.} \textbf{24}(12), 4628--4641 (2006).


\bibitem{YaoMWP2009}
 \textsc{J.~Yao},
 \jr{J. Lightw. Technol.} \textbf{27}(3), 314--335 (2009).


\othercit
\bibitem{CoxBook2004}
 \textsc{C.\,H. Cox},
\emph{Analog Optical Links : Theory and Practice} (Cambridge University Press,
  Cambridge, 2004).


\bibitem{CoxMTT1997}
 \textsc{C.~Cox},  \textsc{E.~Ackerman},  \textsc{R.~Helkey},  and
  \textsc{G.~Betts},
 \jr{{IEEE} Trans. Microw. Theory Tech.} \textbf{45}(8), 1375--1383 (1997).


\bibitem{CoxMTT2006}
 \textsc{C.~Cox},  \textsc{E.~Ackerman},  \textsc{G.~Betts},  and
  \textsc{J.~Prince},
 \jr{{IEEE} Trans. Microw. Theory Tech.} \textbf{54}(2), 906--920 (2006).


\bibitem{UrickElectLett2006}
 \textsc{V.~Urick},  \textsc{M.~Rogge},  \textsc{F.~Bucholtz},  and
  \textsc{K.~Williams},
 \jr{Electron. Lett.} \textbf{42}(9), 552--553 (2006).


\othercit
\bibitem{AckermanIMS2007}
 \textsc{E.~Ackerman},  \textsc{G.~Betts},  \textsc{W.~Burns},
  \textsc{J.~Campbell},  \textsc{C.~Cox},  \textsc{N.~Duan},
  \textsc{J.~Prince},  \textsc{M.~Regan},  and  \textsc{H.~Roussell},
in: {Proceedings} of the {IEEE MTT-S International Microwave Symposium (IMS
  2007)}, {Honolulu, HI}, {} (2007),  pp.\,51--54.


\bibitem{KarimPTL2007}
 \textsc{A.~Karim} and  \textsc{J.~Devenport},
 \jr{{IEEE} Photon. Technol. Lett.} \textbf{19}(5), 312--314 (2007).


\bibitem{McKinneyPTL2007}
 \textsc{J.~McKinney},  \textsc{M.~Godinez},  \textsc{V.~Urick},
  \textsc{S.~Thaniyavarn},  \textsc{W.~Charczenko},  and
  \textsc{K.~Williams},
 \jr{{IEEE} Photon. Technol. Lett.} \textbf{19}(7), 465--467 (2007).


\othercit
\bibitem{UrickIMS2011}
 \textsc{V.~Urick},  \textsc{J.~McKinney},  \textsc{J.~Diehl},  and
  \textsc{K.~Williams},
in: {Proceedings} of the {IEEE MTT-S International Microwave Symposium (IMS
  2011)}, {Baltimore, MD}, {} (2011),  pp.\,1--4.


\bibitem{KolnerAO1987}
 \textsc{B.~Kolner} and  \textsc{D.~Dolfi},
 \jr{{Appl. Optics}} \textbf{{26}}({17}), {3676--3680} ({1987}).


\bibitem{Cummings1998}
 \textsc{U.~Cummings} and  \textsc{W.~Bridges},
 \jr{J. Lightw. Technol.} \textbf{16}(8), 1482--1490 (1998).


\bibitem{DarcieJLT2007}
 \textsc{T.\,E. Darcie},  \textsc{J.~Zhang},  \textsc{P.\,F. Driessen},  and
  \textsc{J.\,J. Eun},
 \jr{J. Lightwave Technol.} \textbf{25}(1), 157--164 (2007).


\othercit
\bibitem{UrickSPIE2012}
 \textsc{V.\,J. Urick},  \textsc{J.\,F. Diehl},  \textsc{M.\,N. Draa},
  \textsc{J.\,D. McKinney},  and  \textsc{K.\,J. Williams},
in: {SPIE Proceedings} of the {RF and Millimeter-Wave Photonics II} (SPIE, ),
  p.\,825904.


\bibitem{Roman1998}
 \textsc{J.~Roman},  \textsc{L.~Nichols},  \textsc{K.~Wiliams},
  \textsc{R.~Esman},  \textsc{G.~Tavik},  \textsc{M.~Livingston},  and
  \textsc{M.~Parent},
 \jr{{IEEE} Trans. Microw. Theory Tech.} \textbf{46}(12), 2317--2323 (1998).


\bibitem{Montebugnoli2005}
 \textsc{S.~Montebugnoli},  \textsc{M.~Boschi},  \textsc{F.~Perini},
  \textsc{P.~Faccin},  \textsc{G.~Brunori},  and  \textsc{E.~Pirazzini},
 \jr{{Microw. Opt. Technol. Lett.}} \textbf{{46}}({1}), {48--54} ({2005}).


\bibitem{LimJLT2010}
 \textsc{C.~Lim},  \textsc{A.~Nirmalathas},  \textsc{M.~Bakaul},
  \textsc{P.~Gamage},  \textsc{K.\,L. Lee},  \textsc{Y.~Yang},
  \textsc{D.~Novak},  and  \textsc{R.~Waterhouse},
 \jr{J. Lightwave Technol.} \textbf{28}(4), 390--405 (2010).


\bibitem{YaoNatPhoton2010}
 \textsc{J.~Yao},
 \jr{{Nat. Photonics}} \textbf{{4}}({2}), {79--80} ({2010}).


\bibitem{YaoOptComm2011}
,
 \jr{Opt. Commun.} \textbf{284}(15), 3723 -- 3736 (2011).


\othercit
\bibitem{JacobsOFC2007}
 \textsc{E.~Jacobs},  \textsc{R.~Olsen},  \textsc{J.~Rodgers},
  \textsc{D.~Evans},  \textsc{T.~Weiner},  and  \textsc{C.~Lin},
in: {Proceedings} of the {Optical Fiber Communication and the National Fiber
  Optic Engineers Conference (OFC/NFOEC)}, {Anaheim,\,CA}, {USA}, {} (2007).


\bibitem{GasullaPhotonicsJournal2011}
 \textsc{I.~Gasulla},  \textsc{J.~Lloret},  \textsc{J.~Sancho},
  \textsc{S.~Sales},  and  \textsc{J.~Capmany},
 \jr{{IEEE Photonics J.}} \textbf{{3}}({2}), {311--315} ({2011}).


\bibitem{CapmanyNatPhoton2011}
 \textsc{J.~Capmany},  \textsc{I.~Gasulla},  and  \textsc{S.~Sales},
 \jr{{Nat. Photonics}} \textbf{{5}}({12}), {731--733} ({2011}).


\othercit
\bibitem{WoodwardMWP2011}
 \textsc{T.~Woodward},  \textsc{A.~Agarwal},  \textsc{T.~Banwell},
  \textsc{P.~Toliver},  \textsc{B.~Luff},  \textsc{D.~Feng},  \textsc{P.~Dong},
   \textsc{D.~Lee},  \textsc{N.\,N. Feng},  and  \textsc{M.~Asghari},
in: {Proceedings} of the {IEEE Topical Meeting on Microwave Photonics (MWP
  2011)}, {Singapore}, {} (2011),  pp.\,377 --380.


\othercit
\bibitem{ColdrenMWP2010}
 \textsc{L.~Coldren},
in: {Proceedings} of the {IEEE Topical Meeting on Microwave Photonics (MWP
  2010)}, {Montreal}, {Canada}, {} (2010).


\bibitem{JalaliMicrowaveMag2006}
 \textsc{B.~Jalali},  \textsc{M.~Paniccia},  and  \textsc{G.~Reed},
 \jr{IEEE Microwave Magazine} \textbf{7}(3), 58 --68 (2006).


\bibitem{SorefJSTQE2006}
 \textsc{R.~Soref},
 \jr{{IEEE} J. Sel. Topics Quantum Electron.} \textbf{12}(6), 1678 --1687
  (2006).


\bibitem{JalaliJLT2006}
 \textsc{B.~Jalali} and  \textsc{S.~Fathpour},
 \jr{J. Lightw. Technol.} \textbf{24}(12), 4600 --4615 (2006).


\bibitem{LiangElectLett2009}
 \textsc{D.~Liang} and  \textsc{J.~Bowers},
 \jr{Electron. Lett.} \textbf{45}(12), 578--581 (2009).


\bibitem{ColdrenJLT2011}
 \textsc{L.~Coldren},  \textsc{S.~Nicholes},  \textsc{L.~Johansson},
  \textsc{S.~Ristic},  \textsc{R.~Guzzon},  \textsc{E.~Norberg},  and
  \textsc{U.~Krishnamachari},
 \jr{J. Lightw. Technol.} \textbf{29}(4), 554 --570 (2011).


\bibitem{SmitLPR2012}
 \textsc{M.~Smit},  \textsc{J.~van\,der Tol},  and  \textsc{M.~Hill},
 \jr{Laser \& Photonics Reviews} \textbf{6}(1), 1--13 (2012).


\bibitem{KishJSTQE2011}
 \textsc{F.~Kish},  \textsc{D.~Welch},  \textsc{R.~Nagarajan},
  \textsc{J.~Pleumeekers},  \textsc{V.~Lal},  \textsc{M.~Ziari},
  \textsc{A.~Nilsson},  \textsc{M.~Kato},  \textsc{S.~Murthy},
  \textsc{P.~Evans},  \textsc{S.~Corzine},  \textsc{M.~Mitchell},
  \textsc{P.~Samra},  \textsc{M.~Missey},  \textsc{S.~DeMars},
  \textsc{R.~Schneider},  \textsc{M.~Reffle},  \textsc{T.~Butrie},
  \textsc{J.~Rahn},  \textsc{M.~Van~Leeuwen},  \textsc{J.~Stewart},
  \textsc{D.~Lambert},  \textsc{R.~Muthiah},  \textsc{H.~Tsai},
  \textsc{J.~Bostak},  \textsc{A.~Dentai},  \textsc{K.~Wu},  \textsc{H.~Sun},
  \textsc{D.~Pavinski},  \textsc{J.~Zhang},  \textsc{J.~Tang},
  \textsc{J.~McNicol},  \textsc{M.~Kuntz},  \textsc{V.~Dominic},
  \textsc{B.~Taylor},  \textsc{R.~Salvatore},  \textsc{M.~Fisher},
  \textsc{A.~Spannagel},  \textsc{E.~Strzelecka},  \textsc{P.~Studenkov},
  \textsc{M.~Raburn},  \textsc{W.~Williams},  \textsc{D.~Christini},
  \textsc{K.~Thomson},  \textsc{S.~Agashe},  \textsc{R.~Malendevich},
  \textsc{G.~Goldfarb},  \textsc{S.~Melle},  \textsc{C.~Joyner},
  \textsc{M.~Kaufman},  and  \textsc{S.~Grubb},
 \jr{{IEEE} J. Sel. Topics Quantum Electron.} \textbf{17}(6), 1470 --1489
  (2011).


\bibitem{OrcuttOpex2011}
 \textsc{J.\,S. Orcutt},  \textsc{A.~Khilo},  \textsc{C.\,W. Holzwarth},
  \textsc{M.\,A. Popovi\'{c}},  \textsc{H.~Li},  \textsc{J.~Sun},
  \textsc{T.~Bonifield},  \textsc{R.~Hollingsworth},  \textsc{F.\,X.
  K\"{a}rtner},  \textsc{H.\,I. Smith},  \textsc{V.~Stojanovi\'{c}},  and
  \textsc{R.\,J. Ram},
 \jr{Opt. Express} \textbf{19}(3), 2335--2346 (2011).


\bibitem{OrcuttOpex2012}
 \textsc{J.\,S. Orcutt},  \textsc{B.~Moss},  \textsc{C.~Sun},  \textsc{J.~Leu},
   \textsc{M.~Georgas},  \textsc{J.~Shainline},  \textsc{E.~Zgraggen},
  \textsc{H.~Li},  \textsc{J.~Sun},  \textsc{M.~Weaver},
  \textsc{S.~Uro\v{s}evi\'{c}},  \textsc{M.~Popovi\'{c}},  \textsc{R.\,J. Ram},
   and  \textsc{V.~Stojanovi\'{c}},
 \jr{Opt. Express} \textbf{20}(11), 12222--12232 (2012).


\bibitem{KoehlOPN2011}
 \textsc{S.~Koehl},  \textsc{A.~Liu},  and  \textsc{M.~Paniccia},
 \jr{Opt. Photon. News} \textbf{22}(3), 24--29 (2011).


\bibitem{DumonEpixfab}
 \textsc{P.~Dumon},  \textsc{W.~Bogaerts},  \textsc{R.~Baets},  \textsc{J.\,M.
  Fedeli},  and  \textsc{L.~Fulbert},
 \jr{Electron. Lett.} \textbf{45}(12), 581 --582 (2009).


\bibitem{Jeppix}
 \textsc{X.~Leijtens},
 \jr{IET Optoelectronics} \textbf{5}(5), 202 --206 (2011).


\bibitem{HochbergNatPhotonics2010}
 \textsc{M.~Hochberg} and  \textsc{T.~Baehr-Jones},
 \jr{{Nat. Photonics}} \textbf{{4}}({8}), {492--494} ({2010}).


\othercit
\bibitem{AurrionIPC2011}
 \textsc{G.~Fish},  \textsc{S.~Nicholes},  \textsc{V.~Kaman},  and
  \textsc{A.~Fang},
in: {Proceedings} of the {2011 IEEE Photonics Conference}, {Arlington, \,VA},
  {USA}, {} (2011),  pp.\,87 --88.


\bibitem{StulemeijerPTL1999}
 \textsc{J.~Stulemeijer},  \textsc{F.~van Vliet},  \textsc{K.~Benoist},
  \textsc{D.~Maat},  and  \textsc{M.~Smit},
 \jr{{IEEE} Photon. Technol. Lett.} \textbf{11}(1), 122 --124 (1999).


\bibitem{NorbergPTL2010}
 \textsc{E.~Norberg},  \textsc{R.~Guzzon},  \textsc{S.~Nicholes},
  \textsc{J.~Parker},  and  \textsc{L.~Coldren},
 \jr{{IEEE} Photon. Technol. Lett.} \textbf{22}(2), 109 --111 (2010).


\bibitem{NorbergJLT2011}
 \textsc{E.~Norberg},  \textsc{R.~Guzzon},  \textsc{J.~Parker},
  \textsc{L.~Johansson},  and  \textsc{L.~Coldren},
 \jr{J. Lightw. Technol.} \textbf{29}(11), 1611 --1619 (2011).


\bibitem{GuzzonOpex2011}
 \textsc{R.\,S. Guzzon},  \textsc{E.\,J. Norberg},  \textsc{J.\,S. Parker},
  \textsc{L.\,A. Johansson},  and  \textsc{L.\,A. Coldren},
 \jr{{Opt. Express}} \textbf{{19}}({8}), {7816--7826} ({2011}).


\bibitem{GuzzonJQE2012}
 \textsc{R.~Guzzon},  \textsc{E.~Norberg},  and  \textsc{L.~Coldren},
 \jr{{IEEE} J. Quantum Electron.} \textbf{48}(2), 269 --278 (2012).


\bibitem{LiPTL2011}
 \textsc{Y.~Li},  \textsc{A.~Bhardwaj},  \textsc{R.~Wang},  \textsc{S.~Jin},
  \textsc{L.~Coldren},  \textsc{J.~Bowers},  and  \textsc{P.~Herczfeld},
 \jr{{IEEE} Photon. Technol. Lett.} \textbf{23}(20), 1475 --1477 (2011).


\bibitem{BhardwajEL2011}
 \textsc{A.~Bhardwaj},  \textsc{Y.~Li},  \textsc{R.~Wang},  \textsc{S.~Jin},
  \textsc{P.~Herczfeld},  \textsc{J.~Bowers},  and  \textsc{L.~Coldren},
 \jr{Electron. Lett.} \textbf{47}(19), 1090 --1092 (2011).


\bibitem{KrishnamachariMOTL2011}
 \textsc{U.~Krishnamachari},  \textsc{S.~Ristic},  \textsc{A.~Ramaswamy},
  \textsc{L.\,A. Johansson},  \textsc{C.\,H. Chen},  \textsc{J.~Klamkin},
  \textsc{A.~Bhardwaj},  \textsc{M.\,J. Rodwell},  \textsc{J.\,E. Bowers},  and
   \textsc{L.\,A. Coldren},
 \jr{{Microw. Opt. Technol. Lett.}} \textbf{{53}}({10}), {2343--2345} ({2011}).


\bibitem{HimenoJSTQE1998}
 \textsc{A.~Himeno},  \textsc{K.~Kato},  and  \textsc{T.~Miya},
 \jr{{IEEE} J. Sel. Topics Quantum Electron.} \textbf{4}(6), 913 --924 (1998).


\bibitem{AdarJLT1994}
 \textsc{R.~Adar},  \textsc{M.~Serbin},  and  \textsc{V.~Mizrahi},
 \jr{J. Lightw. Technol.} \textbf{12}(8), 1369 --1372 (1994).


\othercit
\bibitem{HorikawaIMS1995}
 \textsc{K.~Horikawa},  \textsc{I.~Ogawa},  \textsc{H.~Ogawa},  and
  \textsc{T.~Kitoh},
in: {Proceedings} of the {IEEE MTT-S International Microwave Symposium (IMS
  1995)}, {Vancouver, BC, Canada}, {} (1995),  pp.\,65 --68 vol.1.


\othercit
\bibitem{HorikawaOFC1996}
 \textsc{K.~Horikawa},  \textsc{I.~Ogawa},  \textsc{T.~Kitoh},  and
  \textsc{H.~Ogawa},
in: {Proceedings} of the {Optical Fiber Communications (OFC)}, {San Jose,\,CA},
  {USA}, {} (1996).


\bibitem{GrosskopfFIO2003}
 \textsc{G.~Grosskopf},
 \jr{Fiber and Integrated Optics} \textbf{22}(1), 35--46 (2003).


\bibitem{RasrasPTL2005}
 \textsc{M.~Rasras},  \textsc{C.~Madsen},  \textsc{M.~Cappuzzo},
  \textsc{E.~Chen},  \textsc{L.~Gomez},  \textsc{E.~Laskowski},
  \textsc{A.~Griffin},  \textsc{A.~Wong-Foy},  \textsc{A.~Gasparyan},
  \textsc{A.~Kasper},  \textsc{J.\,L. Grange},  and  \textsc{S.~Patel},
 \jr{{IEEE} Photon. Technol. Lett.} \textbf{17}(4), 834 --836 (2005).


\bibitem{LaGassePTL1997}
 \textsc{M.~LaGasse} and  \textsc{S.~Thaniyavaru},
 \jr{{IEEE} Photon. Technol. Lett.} \textbf{9}(5), 681 --683 (1997).


\othercit
\bibitem{WyrwasMWP2011}
 \textsc{J.~Wyrwas},  \textsc{M.~Rasras},  \textsc{Y.~Chen},
  \textsc{M.~Cappuzzo},  \textsc{E.~Chen},  \textsc{L.~Gomez},
  \textsc{F.~Klemens},  \textsc{R.~Keller},  \textsc{M.~Earnshaw},
  \textsc{F.~Padro},  \textsc{C.~Bolle},  \textsc{R.~Peach},
  \textsc{C.~Middleton},  \textsc{R.~DeSalvo},  and  \textsc{M.~Wu},
in: {Proceedings} of the {IEEE Topical Meeting on Microwave Photonics (MWP
  2011)}, {Singapore}, {} (2011),  pp.\,41 --44.


\othercit
\bibitem{WyrwasThesis2012}
 \textsc{J.~Wyrwas},
Linear, Low Noise Microwave Photonic Systems using Phase and Frequency
  Modulation,
PhD thesis, EECS Department, University of California, Berkeley, May 2012.


\bibitem{SamadiOptComm2011}
 \textsc{P.~Samadi},  \textsc{L.\,R. Chen},  \textsc{C.~Callender},
  \textsc{P.~Dumais},  \textsc{S.~Jacob},  and  \textsc{D.~Celo},
 \jr{{Opt. Commun.}} \textbf{{284}}({15, SI}), {3737--3741} ({2011}).


\othercit
\bibitem{CallenderSPIE2012}
 \textsc{C.\,L. Callender},  \textsc{P.~Dumais},  \textsc{C.~Blanchetiere},
  \textsc{S.~Jacob},  \textsc{C.~Ledderhof},  \textsc{C.\,W. Smelser},
  \textsc{K.~Yadav},  and  \textsc{J.~Albert},
in: {SPIE Proceedings} of the {Optical Components and Materials IX} (SPIE, ),
  p.\,82570P.


\bibitem{RongNatPhoton2007}
 \textsc{H.~Rong},  \textsc{S.~Xu},  \textsc{Y.\,H. Kuo},  \textsc{V.~Sih},
  \textsc{O.~Cohen},  \textsc{O.~Raday},  and  \textsc{M.~Paniccia},
 \jr{{Nat. Photonics}} \textbf{{1}}({4}), {232--237} ({2007}).


\bibitem{HochbergNatPhoton2012}
 \textsc{T.~Baehr-Jones},  \textsc{T.~Pinguet},  \textsc{P.~Lo~Guo-Qiang},
  \textsc{S.~Danziger},  \textsc{D.~Prather},  and
  \textsc{M.~Hochberg},
 \jr{{Nat. Photonics}} \textbf{{6}}({4}), {206--208} ({2012}).


\bibitem{FischerPTL1996}
 \textsc{U.~Fischer},  \textsc{T.~Zinke},  \textsc{J.\,R. Kropp},
  \textsc{F.~Arndt},  and  \textsc{K.~Petermann},
 \jr{{IEEE} Photon. Technol. Lett.} \textbf{8}(5), 647 --648 (1996).


\bibitem{DongOpex2010_loss}
 \textsc{P.~Dong},  \textsc{W.~Qian},  \textsc{S.~Liao},  \textsc{H.~Liang},
  \textsc{C.\,C. Kung},  \textsc{N.\,N. Feng},  \textsc{R.~Shafiiha},
  \textsc{J.~Fong},  \textsc{D.~Feng},  \textsc{A.\,V. Krishnamoorthy},  and
  \textsc{M.~Asghari},
 \jr{{Opt. Express}} \textbf{{18}}({14}), {14474--14479} ({2010}).


\bibitem{RasrasJLT2009}
 \textsc{M.\,S. Rasras},  \textsc{K.\,Y. Tu},  \textsc{D.\,M. Gill},
  \textsc{Y.\,K. Chen},  \textsc{A.\,E. White},  \textsc{S.\,S. Patel},
  \textsc{A.~Pomerene},  \textsc{D.~Carothers},  \textsc{J.~Beattie},
  \textsc{M.~Beals},  \textsc{J.~Michel},  and  \textsc{L.\,C.
  Kimerling},
 \jr{{J. Lightwave Technol.}} \textbf{{27}}({12}), {2105--2110} ({2009}).


\bibitem{IbrahimOpex2011}
 \textsc{S.~Ibrahim},  \textsc{N.\,K. Fontaine},  \textsc{S.\,S. Djordjevic},
  \textsc{B.~Guan},  \textsc{T.~Su},  \textsc{S.~Cheung},  \textsc{R.\,P.
  Scott},  \textsc{A.\,T. Pomerene},  \textsc{L.\,L. Seaford},  \textsc{C.\,M.
  Hill},  \textsc{S.~Danziger},  \textsc{Z.~Ding},  \textsc{K.~Okamoto},  and
  \textsc{S.\,J.\,B. Yoo},
 \jr{Opt. Express} \textbf{19}(14), 13245--13256 (2011).


\bibitem{KhanOpex2011}
 \textsc{S.~Khan},  \textsc{M.\,A. Baghban},  and  \textsc{S.~Fathpour},
 \jr{Opt. Express} \textbf{19}(12), 11780--11785 (2011).


\bibitem{GiuntoniOpex2012}
 \textsc{I.~Giuntoni},  \textsc{D.~Stolarek},  \textsc{D.\,I. Kroushkov},
  \textsc{J.~Bruns},  \textsc{L.~Zimmermann},  \textsc{B.~Tillack},  and
  \textsc{K.~Petermann},
 \jr{Opt. Express} \textbf{20}(10), 11241--11246 (2012).


\bibitem{XiaNatPhoton2007}
 \textsc{F.~Xia},  \textsc{L.~Sekaric},  and  \textsc{Y.~Vlasov},
 \jr{{Nat. Photonics}} \textbf{{1}}({1}), {65--71} ({2007}).


\bibitem{XiaoOpex2007}
 \textsc{S.~Xiao},  \textsc{M.\,H. Khan},  \textsc{H.~Shen},  and
  \textsc{M.~Qi},
 \jr{Opt. Express} \textbf{15}(22), 14467--14475 (2007).


\bibitem{GnanEL2008}
 \textsc{M.~Gnan},  \textsc{S.~Thorns},  \textsc{D.~Macintyre},
  \textsc{R.~De~La~Rue},  and  \textsc{M.~Sorel},
 \jr{Electron. Lett.} \textbf{44}(2), 115 --116 (2008).


\bibitem{BogaertsJLT2009}
 \textsc{W.~Bogaerts},  \textsc{R.~Baets},  \textsc{P.~Dumon},
  \textsc{V.~Wiaux},  \textsc{S.~Beckx},  \textsc{D.~Taillaert},
  \textsc{B.~Luyssaert},  \textsc{J.~Van~Campenhout},  \textsc{P.~Bienstman},
  and  \textsc{D.~Van~Thourhout},
 \jr{J. Lightw. Technol.} \textbf{23}(1), 401 -- 412 (2005).


\bibitem{CardenasOpex2009}
 \textsc{J.~Cardenas},  \textsc{C.\,B. Poitras},  \textsc{J.\,T. Robinson},
  \textsc{K.~Preston},  \textsc{L.~Chen},  and  \textsc{M.~Lipson},
 \jr{Opt. Express} \textbf{17}(6), 4752--4757 (2009).


\bibitem{YegnanarayananPTL1997}
 \textsc{S.~Yegnanarayanan},  \textsc{P.~Trinh},  \textsc{F.~Coppinger},  and
  \textsc{B.~Jalali},
 \jr{{IEEE} Photon. Technol. Lett.} \textbf{9}(5), 634 --635 (1997).


\bibitem{RasrasJLT2007}
 \textsc{M.~Rasras},  \textsc{D.~Gill},  \textsc{S.~Patel},  \textsc{K.\,Y.
  Tu},  \textsc{Y.\,K. Chen},  \textsc{A.~White},  \textsc{A.~Pomerene},
  \textsc{D.~Carothers},  \textsc{M.~Grove},  \textsc{D.~Sparacin},
  \textsc{J.~Michel},  \textsc{M.~Beals},  and  \textsc{L.~Kimerling},
 \jr{J. Lightw. Technol.} \textbf{25}(1), 87 --92 (2007).


\bibitem{DongOpex2010_filter}
 \textsc{P.~Dong},  \textsc{N.\,N. Feng},  \textsc{D.~Feng},  \textsc{W.~Qian},
   \textsc{H.~Liang},  \textsc{D.\,C. Lee},  \textsc{B.\,J. Luff},
  \textsc{T.~Banwell},  \textsc{A.~Agarwal},  \textsc{P.~Toliver},
  \textsc{R.~Menendez},  \textsc{T.\,K. Woodward},  and
  \textsc{M.~Asghari},
 \jr{Opt. Express} \textbf{18}(23), 23784--23789 (2010).


\othercit
\bibitem{ToliverOFC2010}
 \textsc{P.~Toliver},  \textsc{R.~Menendez},  \textsc{T.~Banwell},
  \textsc{A.~Agarwal},  \textsc{T.\,K. Woodward},  \textsc{N.\,N. Feng},
  \textsc{P.~Dong},  \textsc{D.~Feng},  \textsc{W.~Qian},  \textsc{H.~Liang},
  \textsc{D.\,C. Lee},  \textsc{B.\,J. Luff},  and  \textsc{M.~Asghari},
in: {Proceedings} of the {Optical Fiber Communication and the National Fiber
  Optic Engineers Conference (OFC/NFOEC)}, {San Diego,\,CA}, {USA}, {} (2010).


\bibitem{FengOpex2010}
 \textsc{N.\,N. Feng},  \textsc{P.~Dong},  \textsc{D.~Feng},  \textsc{W.~Qian},
   \textsc{H.~Liang},  \textsc{D.\,C. Lee},  \textsc{J.\,B. Luff},
  \textsc{A.~Agarwal},  \textsc{T.~Banwell},  \textsc{R.~Menendez},
  \textsc{P.~Toliver},  \textsc{T.\,K. Woodward},  and
  \textsc{M.~Asghari},
 \jr{Opt. Express} \textbf{18}(24), 24648--24653 (2010).


\bibitem{CardenasOpex2010}
 \textsc{J.~Cardenas},  \textsc{M.\,A. Foster},  \textsc{N.~Sherwood-Droz},
  \textsc{C.\,B. Poitras},  \textsc{H.\,L.\,R. Lira},  \textsc{B.~Zhang},
  \textsc{A.\,L. Gaeta},  \textsc{J.\,B. Khurgin},  \textsc{P.~Morton},  and
  \textsc{M.~Lipson},
 \jr{Opt. Express} \textbf{18}(25), 26525--26534 (2010).


\bibitem{MortonPTL2012}
 \textsc{P.\,A. Morton},  \textsc{J.~Cardenas},  \textsc{J.\,B. Khurgin},  and
  \textsc{M.~Lipson},
 \jr{{IEEE} Photon. Technol. Lett.} \textbf{24}(6), 512 --514 (2012).


\bibitem{ShenOpex2010}
 \textsc{H.~Shen},  \textsc{M.\,H. Khan},  \textsc{L.~Fan},  \textsc{L.~Zhao},
  \textsc{Y.~Xuan},  \textsc{J.~Ouyang},  \textsc{L.\,T. Varghese},  and
  \textsc{M.~Qi},
 \jr{Opt. Express} \textbf{18}(17), 18067--18076 (2010).


\bibitem{KhanNatPhotonics2010}
 \textsc{M.\,H. Khan},  \textsc{H.~Shen},  \textsc{Y.~Xuan},  \textsc{L.~Zhao},
   \textsc{S.~Xiao},  \textsc{D.\,E. Leaird},  \textsc{A.\,M. Weiner},  and
  \textsc{M.~Qi},
 \jr{{Nat. Photonics}} \textbf{{4}}({2}), {117--U30} ({2010}).


\othercit
\bibitem{YunhongIPC2011}
 \textsc{Y.~Ding},  \textsc{C.~Peucheret},  \textsc{J.~Xu},  \textsc{H.~Hou},
  \textsc{X.~Zhang},  and  \textsc{D.~Huang},
in: {Proceedings} of the {2011 IEEE Photonics Conference}, {Arlington, \,VA},
  {USA}, {} (2011),  pp.\,258 --259.


\bibitem{YueOL2012}
 \textsc{Y.~Yue},  \textsc{H.~Huang},  \textsc{L.~Zhang},  \textsc{J.~Wang},
  \textsc{J.\,Y. Yang},  \textsc{O.\,F. Yilmaz},  \textsc{J.\,S. Levy},
  \textsc{M.~Lipson},  and  \textsc{A.\,E. Willner},
 \jr{Opt. Lett.} \textbf{37}(4), 551--553 (2012).


\bibitem{MirshafieiPTL2012}
 \textsc{M.~Mirshafiei},  \textsc{S.~LaRochelle},  and
  \textsc{L.~Rusch},
 \jr{{IEEE} Photon. Technol. Lett.} \textbf{PP}(99), 1 (2012).


\bibitem{ZhuangPTL2007}
 \textsc{L.~Zhuang},  \textsc{C.~Roeloffzen},  \textsc{R.~Heideman},
  \textsc{A.~Borreman},  \textsc{A.~Meijerink},  and  \textsc{W.~van
  Etten},
 \jr{{IEEE} Photon. Technol. Lett.} \textbf{19}(15), 1130 --1132 (2007).


\bibitem{MeijerinkJLT2010}
 \textsc{A.~Meijerink},  \textsc{C.\,G.\,H. Roeloffzen},
  \textsc{R.~Meijerink},  \textsc{L.~Zhuang},  \textsc{D.\,A.\,I. Marpaung},
  \textsc{M.\,J. Bentum},  \textsc{M.~Burla},  \textsc{J.~Verpoorte},
  \textsc{P.~Jorna},  \textsc{A.~Hulzinga},  and  \textsc{W.~van Etten},
 \jr{J. Lightwave Technol.} \textbf{28}(1), 3--18 (2010).


\bibitem{ZhuangJLT2010}
 \textsc{L.~Zhuang},  \textsc{C.\,G.\,H. Roeloffzen},  \textsc{A.~Meijerink},
  \textsc{M.~Burla},  \textsc{D.\,A.\,I. Marpaung},  \textsc{A.~Leinse},
  \textsc{M.~Hoekman},  \textsc{R.\,G. Heideman},  and  \textsc{W.~van
  Etten},
 \jr{J. Lightwave Technol.} \textbf{28}(1), 19--31 (2010).


\othercit
\bibitem{MarpaungEuCAP2011}
 \textsc{D.~Marpaung},  \textsc{L.~Zhuang},  \textsc{M.~Burla},
  \textsc{C.~Roeloffzen},  \textsc{J.~Verpoorte},  \textsc{H.~Schippers},
  \textsc{A.~Hulzinga},  \textsc{P.~Jorna},  \textsc{W.~Beeker},
  \textsc{A.~Leinse},  \textsc{R.~Heideman},  \textsc{B.~Noharet},
  \textsc{Q.~Wang},  \textsc{B.~Sanadgol},  and  \textsc{R.~Baggen},
in: {Proceedings} of the {5th European Conference on Antennas and Propagation
  (EuCAP 2011)}, {Rome}, {Italy}, {} (2011),  pp.\,2623 --2627.


\othercit
\bibitem{MarpaungMWP2011}
 \textsc{D.~Marpaung},  \textsc{L.~Zhuang},  \textsc{M.~Burla},
  \textsc{C.~Roeloffzen},  \textsc{B.~Noharet},  \textsc{Q.~Wang},
  \textsc{W.~Beeker},  \textsc{A.~Leinse},  and  \textsc{R.~Heideman},
in: {Proceedings} of the {IEEE Topical Meeting on Microwave Photonics (MWP
  2011)}, {Singapore}, {} (2011),  pp.\,458 --461.


\bibitem{BurlaAO2012}
 \textsc{M.~Burla},  \textsc{C.\,G.\,H. Roeloffzen},  \textsc{L.~Zhuang},
  \textsc{D.~Marpaung},  \textsc{M.\,R. Khan},  \textsc{P.~Maat},
  \textsc{K.~Dijkstra},  \textsc{A.~Leinse},  \textsc{M.~Hoekman},  and
  \textsc{R.~Heideman},
 \jr{Appl. Opt.} \textbf{51}(7), 789--802 (2012).


\othercit
\bibitem{MarpaungMWP2010}
 \textsc{D.~Marpaung},  \textsc{C.~Roeloffzen},  \textsc{R.~Timens},
  \textsc{A.~Leinse},  and  \textsc{M.~Hoekman},
in: {Proceedings} of the {IEEE Topical Meeting on Microwave Photonics (MWP
  2010)}, {Montreal}, {Canada}, {} (2010),  pp.\,131 --134.


\bibitem{MarpaungOpex2011}
 \textsc{D.~Marpaung},  \textsc{L.~Chevalier},  \textsc{M.~Burla},  and
  \textsc{C.~Roeloffzen},
 \jr{Opt. Express} \textbf{19}(25), 24838--24848 (2011).


\bibitem{ZhuangOpex2011}
 \textsc{L.~Zhuang},  \textsc{D.~Marpaung},  \textsc{M.~Burla},
  \textsc{W.~Beeker},  \textsc{A.~Leinse},  and  \textsc{C.~Roeloffzen},
 \jr{Opt. Express} \textbf{19}(23), 23162--23170 (2011).


\bibitem{BurlaOpex2011}
 \textsc{M.~Burla},  \textsc{D.~Marpaung},  \textsc{L.~Zhuang},
  \textsc{C.~Roeloffzen},  \textsc{M.\,R. Khan},  \textsc{A.~Leinse},
  \textsc{M.~Hoekman},  and  \textsc{R.~Heideman},
 \jr{Opt. Express} \textbf{19}(22), 21475--21484 (2011).


\othercit
\bibitem{HeidemanSPIE2009}
 \textsc{R.~Heideman},  \textsc{A.~Leinse},  \textsc{W.~Hoving},
  \textsc{R.~Dekker},  \textsc{D.~Geuzebroek},  \textsc{E.~Klein},
  \textsc{R.~Stoffer},  \textsc{C.~Roeloffzen},  \textsc{L.~Zhuang},  and
  \textsc{A.~Meijerink},
in: {SPIE Proceedings} of the {Photonics Packaging, Integration, and
  Interconnects IX} (SPIE, ),  p.\,72210R.


\bibitem{HeidemanJSTQE2012}
 \textsc{R.~Heideman},  \textsc{M.~Hoekman},  and
  \textsc{F.~Schreuder},
 \jr{{IEEE} J. Sel. Topics Quantum Electron.} \textbf{PP}(99), 1 (2012).


\bibitem{TienOpex2010}
 \textsc{M.\,C. Tien},  \textsc{J.\,F. Bauters},  \textsc{M.\,J.\,R. Heck},
  \textsc{D.\,J. Blumenthal},  and  \textsc{J.\,E. Bowers},
 \jr{Opt. Express} \textbf{18}(23), 23562--23568 (2010).


\bibitem{BautersOpex2011_1}
 \textsc{J.\,F. Bauters},  \textsc{M.\,J.\,R. Heck},  \textsc{D.~John},
  \textsc{D.~Dai},  \textsc{M.\,C. Tien},  \textsc{J.\,S. Barton},
  \textsc{A.~Leinse},  \textsc{R.\,G. Heideman},  \textsc{D.\,J. Blumenthal},
  and  \textsc{J.\,E. Bowers},
 \jr{Opt. Express} \textbf{19}(4), 3163--3174 (2011).


\bibitem{TienOpex2011}
 \textsc{M.\,C. Tien},  \textsc{J.\,F. Bauters},  \textsc{M.\,J.\,R. Heck},
  \textsc{D.\,T. Spencer},  \textsc{D.\,J. Blumenthal},  and  \textsc{J.\,E.
  Bowers},
 \jr{Opt. Express} \textbf{19}(14), 13551--13556 (2011).


\bibitem{DaiOpex2011}
 \textsc{D.~Dai},  \textsc{Z.~Wang},  \textsc{J.\,F. Bauters},  \textsc{M.\,C.
  Tien},  \textsc{M.\,J.\,R. Heck},  \textsc{D.\,J. Blumenthal},  and
  \textsc{J.\,E. Bowers},
 \jr{Opt. Express} \textbf{19}(15), 14130--14136 (2011).


\bibitem{BautersOpex2011_2}
 \textsc{J.\,F. Bauters},  \textsc{M.\,J.\,R. Heck},  \textsc{D.\,D. John},
  \textsc{J.\,S. Barton},  \textsc{C.\,M. Bruinink},  \textsc{A.~Leinse},
  \textsc{R.\,G. Heideman},  \textsc{D.\,J. Blumenthal},  and  \textsc{J.\,E.
  Bowers},
 \jr{Opt. Express} \textbf{19}(24), 24090--24101 (2011).


\bibitem{DaiLSA2012}
 \textsc{D.~Dai},  \textsc{J.~Bauters},  and  \textsc{J.\,E. Bowers},
 \jr{{Light Sci. Appl.}} \textbf{{1}}({e1}), {1--12} ({2012}).


\bibitem{MorichettiJLT2007}
 \textsc{F.~Morichetti},  \textsc{A.~Melloni},  \textsc{M.~Martinelli},
  \textsc{R.~Heideman},  \textsc{A.~Leinse},  \textsc{D.~Geuzebroek},  and
  \textsc{A.~Borreman},
 \jr{J. Lightw. Technol.} \textbf{25}(9), 2579 --2589 (2007).


\bibitem{Oldenbeuving2012}
 \textsc{R.\,M. Oldenbeuving},  \textsc{E.\,J. Klein},  \textsc{H.\,L.
  Offerhaus},  \textsc{C.\,J. Lee},  \textsc{H.~Song},  and  \textsc{K.\,J.
  Boller},
 \jr{arXiv 1204.0353v1, [physics.optics]}.


\othercit
\bibitem{NgSPIE1994}
 \textsc{W.\,W. Ng},  \textsc{D.~Yap},  \textsc{A.\,A. Narayanan},
  \textsc{T.\,P. Liu},  and  \textsc{R.\,R. Hayes},
in: {SPIE Proceedings} of the {Optoelectronic Signal Processing for
  Phased-Array Antennas IV} (SPIE, ),  pp.\,114--123.


\othercit
\bibitem{CombrieCLEO2010}
 \textsc{S.~Combrie},  \textsc{J.~Bourderionnet},  \textsc{P.~Colman},
  \textsc{D.~Dolfi},  and  \textsc{A.~De~Rossi},
in: {Proceedings} of the {2010 Conference on Lasers and Electro-Optics (CLEO)
  and Quantum Electronics and Laser Science Conference (QELS)}, {San Jose,
  \,CA}, {USA}, {} (2010).


\bibitem{HorikawaMTT1995}
 \textsc{K.~Horikawa},  \textsc{Y.~Nakasuga},  and  \textsc{H.~Ogawa},
 \jr{{IEEE} Trans. Microw. Theory Tech.} \textbf{43}(9), 2395 --2401 (1995).


\othercit
\bibitem{MitchellLEOS2007}
 \textsc{A.~Mitchell},
in: {Proceedings} of the {20th Annual Meeting of the IEEE Lasers and
  Electro-Optics Society (LEOS 2007)}, {Lake Buena Vista}, {USA}, {} (2007).


\bibitem{IlchenkoPTL2008}
 \textsc{V.\,S. Ilchenko},  \textsc{A.\,B. Matsko},  \textsc{I.~Solomatine},
  \textsc{A.\,A. Savchenkov},  \textsc{D.~Seidel},  and
  \textsc{L.~Maleki},
 \jr{{IEEE} Photon. Technol. Lett.} \textbf{20}(19), 1600 --1612 (2008).


\bibitem{WangOpex2009}
 \textsc{J.~Wang},  \textsc{Q.~Sun},  \textsc{J.~Sun},  and
  \textsc{W.~Zhang},
 \jr{Opt. Express} \textbf{17}(5), 3521--3530 (2009).


\bibitem{WangPTL2010}
 \textsc{J.~Wang} and  \textsc{J.~Sun},
 \jr{{IEEE} Photon. Technol. Lett.} \textbf{22}(3), 140 --142 (2010).


\bibitem{WijayantoElectLett2012}
 \textsc{Y.~Wijayanto},  \textsc{H.~Murata},  and  \textsc{Y.~Okamura},
 \jr{Electron. Lett.} \textbf{48}(1), 36--38 (2012).


\bibitem{YeniayJLT2004}
 \textsc{A.~Yeniay},  \textsc{R.~Gao},  \textsc{K.~Takayama},  \textsc{R.~Gao},
   and  \textsc{A.\,F. Garito},
 \jr{J. Lightwave Technol.} \textbf{22}(1), 154 (2004).


\bibitem{HowleyPTL2005}
 \textsc{B.~Howley},  \textsc{Y.~Chen},  \textsc{X.~Wang},  \textsc{Q.~Zhou},
  \textsc{Z.~Shi},  \textsc{Y.~Jiang},  and  \textsc{R.~Chen},
 \jr{{IEEE} Photon. Technol. Lett.} \textbf{17}(9), 1944 --1946 (2005).


\bibitem{HowleyJLT2007}
 \textsc{B.~Howley},  \textsc{X.~Wang},  \textsc{M.~Chen},  and
  \textsc{R.~Chen},
 \jr{J. Lightw. Technol.} \textbf{25}(3), 883 --890 (2007).


\bibitem{YeniayPTL2010}
 \textsc{A.~Yeniay} and  \textsc{R.~Gao},
 \jr{{IEEE} Photon. Technol. Lett.} \textbf{22}(21), 1565 --1567 (2010).


\bibitem{MaddenOpex2007}
 \textsc{S.\,J. Madden},  \textsc{D.\,Y. Choi},  \textsc{D.\,A. Bulla},
  \textsc{A.\,V. Rode},  \textsc{B.~Luther-Davies},  \textsc{V.\,G. Ta'eed},
  \textsc{M.\,D. Pelusi},  and  \textsc{B.\,J. Eggleton},
 \jr{Opt. Express} \textbf{15}(22), 14414--14421 (2007).


\bibitem{EggletonNatPhotonics2011}
 \textsc{B.\,J. Eggleton},  \textsc{B.~Luther-Davies},  and
  \textsc{K.~Richardson},
 \jr{{Nat. Photonics}} \textbf{{5}}({3}), {141--148} ({2011}).


\bibitem{EggletonLPR2012}
 \textsc{B.~Eggleton},  \textsc{T.~Vo},  \textsc{R.~Pant},  \textsc{J.~Schr},
  \textsc{M.~Pelusi},  \textsc{D.~Yong~Choi},  \textsc{S.~Madden},  and
  \textsc{B.~Luther-Davies},
 \jr{Laser \& Photonics Reviews} \textbf{6}(1), 97--114 (2012).


\bibitem{PantOpex2011}
 \textsc{R.~Pant},  \textsc{C.\,G. Poulton},  \textsc{D.\,Y. Choi},
  \textsc{H.~Mcfarlane},  \textsc{S.~Hile},  \textsc{E.~Li},
  \textsc{L.~Thevenaz},  \textsc{B.~Luther-Davies},  \textsc{S.\,J. Madden},
  and  \textsc{B.\,J. Eggleton},
 \jr{Opt. Express} \textbf{19}(9), 8285--8290 (2011).


\bibitem{PantOL2012}
 \textsc{R.~Pant},  \textsc{A.~Byrnes},  \textsc{C.\,G. Poulton},
  \textsc{E.~Li},  \textsc{D.\,Y. Choi},  \textsc{S.~Madden},
  \textsc{B.~Luther-Davies},  and  \textsc{B.\,J. Eggleton},
 \jr{Opt. Lett.} \textbf{37}(5), 969--971 (2012).


\bibitem{PelusiNatPhotonics2009}
 \textsc{M.~Pelusi},  \textsc{F.~Luan},  \textsc{T.\,D. Vo},
  \textsc{M.\,R.\,E. Lamont},  \textsc{S.\,J. Madden},  \textsc{D.\,A. Bulla},
  \textsc{D.\,Y. Choi},  \textsc{B.~Luther-Davies},  and  \textsc{B.\,J.
  Eggleton},
 \jr{{Nat. Photonics}} \textbf{{3}}({3}), {139--143} ({2009}).


\othercit
\bibitem{ByrnesCLEO2012}
 \textsc{A.~Byrnes},  \textsc{R.~Pant},  \textsc{C.~Poulton},  \textsc{E.~Li},
  \textsc{D.~Choi},  \textsc{S.~Madden},  \textsc{B.~Luther-Davies},  and
  \textsc{B.~Eggleton},
in: {Proceedings} of the {2012 Conference on Lasers and Electro-Optics (CLEO)
  and Quantum Electronics and Laser Science Conference (QELS)}, {San Jose,
  \,CA}, {USA}, {} (2012),  p.\,CTu2A.6.


\bibitem{NgPTL1994}
 \textsc{W.~Ng},  \textsc{D.~Yap},  \textsc{A.~Narayanan},  and
  \textsc{A.~Walston},
 \jr{{IEEE} Photon. Technol. Lett.} \textbf{6}(2), 231 --234 (1994).


\bibitem{TangOptEng2000}
 \textsc{S.~Tang},  \textsc{B.~Li},  \textsc{N.~Jiang},  \textsc{D.~An},
  \textsc{Z.~Fu},  \textsc{L.~Wu},  and  \textsc{R.\,T. Chen},
 \jr{Optical Engineering} \textbf{39}(3), 643--651 (2000).


\othercit
\bibitem{JiangSPIE2005}
 \textsc{Y.~Jiang},  \textsc{W.~Jiang},  \textsc{X.~Chen},  \textsc{L.~Gu},
  \textsc{B.~Howley},  and  \textsc{R.\,T. Chen},
in: {SPIE Proceedings} of the {Photonic Crystal Materials and Devices III}
  (SPIE, ),  p.\,825904.


\bibitem{RamaswamyJLT2008}
 \textsc{A.~Ramaswamy},  \textsc{L.~Johansson},  \textsc{J.~Klamkin},
  \textsc{H.\,F. Chou},  \textsc{C.~Sheldon},  \textsc{M.~Rodwell},
  \textsc{L.~Coldren},  and  \textsc{J.~Bowers},
 \jr{J. Lightw. Technol.} \textbf{26}(1), 209 --216 (2008).


\bibitem{MelloniOL2008}
 \textsc{A.~Melloni},  \textsc{F.~Morichetti},  \textsc{C.~Ferrari},  and
  \textsc{M.~Martinelli},
 \jr{Opt. Lett.} \textbf{33}(20), 2389--2391 (2008).


\bibitem{LiuOpex2008}
 \textsc{F.~Liu},  \textsc{T.~Wang},  \textsc{L.~Qiang},  \textsc{T.~Ye},
  \textsc{Z.~Zhang},  \textsc{M.~Qiu},  and  \textsc{Y.~Su},
 \jr{Opt. Express} \textbf{16}(20), 15880--15886 (2008).


\bibitem{LiuElectLett2009}
 \textsc{F.~Liu},  \textsc{T.~Wang},  \textsc{Z.~Zhang},  \textsc{M.~Qiu},  and
   \textsc{Y.~Su},
 \jr{{Electron. Lett.}} \textbf{{45}}({24}), {1247--1248} ({2009}).


\bibitem{ChangPTL2009}
 \textsc{Q.~Chang},  \textsc{Q.~Li},  \textsc{Z.~Zhang},  \textsc{M.~Qiu},
  \textsc{T.~Ye},  and  \textsc{Y.~Su},
 \jr{{IEEE} Photon. Technol. Lett.} \textbf{{21}}({1-4}), {60--62} ({2009}).


\bibitem{FerreraNatCommunications2010}
 \textsc{M.~Ferrera},  \textsc{Y.~Park},  \textsc{L.~Razzari},  \textsc{B.\,E.
  Little},  \textsc{S.\,T. Chu},  \textsc{R.~Morandotti},  \textsc{D.\,J.
  Moss},  and  \textsc{J.~Aza\~{n}a},
 \jr{{Nat. Commun.}} \textbf{{1}}({June}) ({2010}).


\bibitem{ChenMTT2010}
 \textsc{H.\,W. Chen},  \textsc{A.\,W. Fang},  \textsc{J.\,D. Peters},
  \textsc{Z.~Wang},  \textsc{J.~Bovington},  \textsc{D.~Liang},  and
  \textsc{J.\,E. Bowers},
 \jr{{IEEE Trans. Microw. Theory Tech.}} \textbf{{58}}({11, Part 2, SI}),
  {3213--3219} ({2010}).


\bibitem{MarpaungOpex2010}
 \textsc{D.~Marpaung},  \textsc{C.~Roeloffzen},  \textsc{A.~Leinse},  and
  \textsc{M.~Hoekman},
 \jr{Opt. Express} \textbf{18}(26), 27359--27370 (2010).


\bibitem{DjordjevicPTL2011}
 \textsc{S.~Djordjevic},  \textsc{L.~Luo},  \textsc{S.~Ibrahim},
  \textsc{N.~Fontaine},  \textsc{C.~Poitras},  \textsc{B.~Guan},
  \textsc{L.~Zhou},  \textsc{K.~Okamoto},  \textsc{Z.~Ding},
  \textsc{M.~Lipson},  and  \textsc{S.~Yoo},
 \jr{{IEEE} Photon. Technol. Lett.} \textbf{23}(1), 42 --44 (2011).


\bibitem{AlipourOpex2011}
 \textsc{P.~Alipour},  \textsc{A.\,A. Eftekhar},  \textsc{A.\,H. Atabaki},
  \textsc{Q.~Li},  \textsc{S.~Yegnanarayanan},  \textsc{C.\,K. Madsen},  and
  \textsc{A.~Adibi},
 \jr{Opt. Express} \textbf{19}(17), 15899--15907 (2011).


\bibitem{LloretOpex2012}
 \textsc{J.~Lloret},  \textsc{G.~Morthier},  \textsc{F.~Ramos},
  \textsc{S.~Sales},  \textsc{D.\,V. Thourhout},  \textsc{T.~Spuesens},
  \textsc{N.~Olivier},  \textsc{J.\,M. F\'{e}d\'{e}li},  and
  \textsc{J.~Capmany},
 \jr{Opt. Express} \textbf{20}(10), 10796--10806 (2012).


\othercit
\bibitem{GreinCLEO2011}
 \textsc{M.\,E. Grein},  \textsc{S.~Spector},  \textsc{A.~Khilo},
  \textsc{A.\,H. Nejadmalayeri},  \textsc{M.\,Y. Sander},  \textsc{M.~Peng},
  \textsc{J.~Wang},  \textsc{C.\,M. Sorace},  \textsc{M.\,W. Geis},
  \textsc{M.\,M. Willis},  \textsc{D.\,M. Lennon},  \textsc{T.~Lyszczarz},
  \textsc{E.\,P. Ippen},  and  \textsc{F.\,X. Kaertner},
in: {Proceedings} of the {2011 Conference on Lasers and Electro-Optics (CLEO)
  and Quantum Electronics and Laser Science Conference (QELS)}, {Baltimore,
  \,MD}, {USA}, {} (2011),  p.\,CThI1.


\bibitem{KhiloOpex2012}
 \textsc{A.~Khilo},  \textsc{S.\,J. Spector},  \textsc{M.\,E. Grein},
  \textsc{A.\,H. Nejadmalayeri},  \textsc{C.\,W. Holzwarth},  \textsc{M.\,Y.
  Sander},  \textsc{M.\,S. Dahlem},  \textsc{M.\,Y. Peng},  \textsc{M.\,W.
  Geis},  \textsc{N.\,A. DiLello},  \textsc{J.\,U. Yoon},
  \textsc{A.~Motamedi},  \textsc{J.\,S. Orcutt},  \textsc{J.\,P. Wang},
  \textsc{C.\,M. Sorace-Agaskar},  \textsc{M.\,A. Popovi\'{c}},
  \textsc{J.~Sun},  \textsc{G.\,R. Zhou},  \textsc{H.~Byun},  \textsc{J.~Chen},
   \textsc{J.\,L. Hoyt},  \textsc{H.\,I. Smith},  \textsc{R.\,J. Ram},
  \textsc{M.~Perrott},  \textsc{T.\,M. Lyszczarz},  \textsc{E.\,P. Ippen},  and
   \textsc{F.\,X. K\"{a}rtner},
 \jr{Opt. Express} \textbf{20}(4), 4454--4469 (2012).


\othercit
\bibitem{FandinoECIO2012}
 \textsc{J.~Fandi\~{n}o},  \textsc{P.~Mu\~{n}oz},  and  \textsc{J.~Capmany},
in: {Proceedings} of the {2012 European Conference on Integrated Optics
  (ECIO)}, {Sitges}, {Spain}, {} (2012).


\bibitem{ZhangPTL2007}
 \textsc{J.~Zhang},  \textsc{A.~Hone},  and  \textsc{T.~Darcie},
 \jr{{IEEE} Photon. Technol. Lett.} \textbf{19}(14), 1033 --1035 (2007).


\bibitem{DarciePTL2007}
 \textsc{T.~Darcie} and  \textsc{P.~Driessen},
 \jr{{IEEE} Photon. Technol. Lett.} \textbf{18}(8), 929 --931 (2006).


\othercit
\bibitem{MarpaungMWP2006}
 \textsc{D.~Marpaung},  \textsc{C.~Roeloffzen},  and  \textsc{W.~van Etten},
in: {Proceedings} of the {IEEE Topical Meeting on Microwave Photonics (MWP
  2006)}, {Grenoble}, {France}, {} (2006).


\bibitem{WyrwasJLT2009}
 \textsc{J.~Wyrwas} and  \textsc{M.~Wu},
 \jr{J. Lightw. Technol.} \textbf{27}(24), 5552 --5562 (2009).


\bibitem{UrickMTT2007}
 \textsc{V.~Urick},  \textsc{F.~Bucholtz},  \textsc{P.~Devgan},
  \textsc{J.~McKinney},  and  \textsc{K.~Williams},
 \jr{{IEEE} Trans. Microw. Theory Tech.} \textbf{55}(9), 1978 --1985 (2007).


\bibitem{McKinneyJLT2009}
 \textsc{J.~McKinney},  \textsc{K.~Colladay},  and
  \textsc{K.~Williams},
 \jr{J. Lightw. Technol.} \textbf{27}(9), 1212 --1220 (2009).


\othercit
\bibitem{DarcieOFC2006}
 \textsc{T.\,E. Darcie},  \textsc{J.~Zhang},  \textsc{P.\,F. Driessen},  and
  \textsc{J.\,J. Eun},
in: {Proceedings} of the {Optical Fiber Communication and the National Fiber
  Optic Engineers Conference (OFC/NFOEC)}, {Anaheim,\,CA}, {USA}, {} (2006),
  p.\,PDP38.


\bibitem{DriessenJLT2008}
 \textsc{P.\,F. Driessen},  \textsc{T.\,E. Darcie},  and
  \textsc{J.~Zhang},
 \jr{J. Lightwave Technol.} \textbf{26}(15), 2740--2747 (2008).


\bibitem{XiePTL2002_grating}
 \textsc{X.~Xie},  \textsc{J.~Khurgin},  \textsc{J.~Kang},  and
  \textsc{F.~Choa},
 \jr{{IEEE} Photon. Technol. Lett.} \textbf{14}(3), 384 --386 (2002).


\bibitem{XiePTL2002_ring}
 \textsc{X.~Xie},  \textsc{J.~Khurgin},  \textsc{J.~Kang},  and  \textsc{F.\,S.
  Choa},
 \jr{{IEEE} Photon. Technol. Lett.} \textbf{14}(8), 1136 --1138 (2002).


\othercit
\bibitem{WyrwasOFC2010}
 \textsc{J.~Wyrwas} and  \textsc{M.~Wu},
in: {Proceedings} of the {Optical Fiber Communication and the National Fiber
  Optic Engineers Conference (OFC/NFOEC)}, {San Diego,\,CA}, {USA}, {} (2010).


\bibitem{LiJLT2009}
 \textsc{Y.~Li} and  \textsc{P.~Herczfeld},
 \jr{J. Lightw. Technol.} \textbf{27}(9), 1086 --1094 (2009).


\bibitem{ClarkPTL2007}
 \textsc{T.~Clark} and  \textsc{M.~Dennis},
 \jr{{IEEE} Photon. Technol. Lett.} \textbf{19}(16), 1206 --1208 (2007).


\bibitem{ClarkMTT2010}
 \textsc{T.~Clark},  \textsc{S.~O'Connor},  and  \textsc{M.~Dennis},
 \jr{{IEEE} Trans. Microw. Theory Tech.} \textbf{58}(11), 3039 --3058 (2010).


\bibitem{MinasianMTT2006}
 \textsc{R.~Minasian},
 \jr{{IEEE Trans. Microw. Theory Tech.}} \textbf{{54}}({2, Part 2}), {832--846}
  ({2006}).


\bibitem{CapmanyJLT2005}
 \textsc{J.~Capmany},  \textsc{B.~Ortega},  \textsc{D.~Pastor},  and
  \textsc{S.~Sales},
 \jr{J. Lightwave Technol.} \textbf{23}(2), 702 (2005).


\bibitem{CapmanyJLT2006}
 \textsc{J.~Capmany},  \textsc{B.~Ortega},  and  \textsc{D.~Pastor},
 \jr{J. Lightwave Technol.} \textbf{24}(1), 201 (2006).


\bibitem{MoraOL2003}
 \textsc{J.~Mora},  \textsc{M.\,V. Andr\'{e}s},  \textsc{J.\,L. Cruz},
  \textsc{B.~Ortega},  \textsc{J.~Capmany},  \textsc{D.~Pastor},  and
  \textsc{S.~Sales},
 \jr{Opt. Lett.} \textbf{28}(15), 1308--1310 (2003).


\bibitem{CapmanyOL2003}
 \textsc{J.~Capmany},  \textsc{D.~Pastor},  \textsc{A.~Martinez},
  \textsc{B.~Ortega},  and  \textsc{S.~Sales},
 \jr{Opt. Lett.} \textbf{28}(16), 1415--1417 (2003).


\bibitem{CapmanyOpex2005}
 \textsc{J.~Capmany},  \textsc{J.~Mora},  \textsc{B.~Ortega},  and
  \textsc{D.~Pastor},
 \jr{Opt. Express} \textbf{13}(5), 1412--1417 (2005).


\bibitem{SupradeepaNatPhotonics2012}
 \textsc{V.\,R. Supradeepa},  \textsc{C.\,M. Long},  \textsc{R.~Wu},
  \textsc{F.~Ferdous},  \textsc{E.~Hamidi},  \textsc{D.\,E. Leaird},  and
  \textsc{A.\,M. Weiner},
 \jr{{Nat. Photonics}} \textbf{{6}}({3}), {186--194} ({2012}).


\bibitem{TuJLT2010}
 \textsc{K.\,Y. Tu},  \textsc{M.\,S. Rasras},  \textsc{D.\,M. Gill},
  \textsc{S.\,S. Patel},  \textsc{Y.\,K. Chen},  \textsc{A.\,E. White},
  \textsc{A.~Pomerene},  \textsc{D.~Carothers},  \textsc{J.~Beattie},
  \textsc{M.~Beals},  \textsc{J.~Michel},  and  \textsc{L.\,C.
  Kimerling},
 \jr{{J. Lightwave Technol.}} \textbf{{28}}({20}), {3019--3028} ({2010}).


\bibitem{MunozJLT2002}
 \textsc{P.~Munoz},  \textsc{D.~Pastor},  and  \textsc{J.~Capmany},
 \jr{J. Lightwave Technol.} \textbf{20}(4), 661 (2002).


\bibitem{PastorOL2003}
 \textsc{D.~Pastor},  \textsc{B.~Ortega},  \textsc{J.~Capmany},
  \textsc{S.~Sales},  \textsc{A.~Martinez},  and  \textsc{P.\,M. {n}oz},
 \jr{Opt. Lett.} \textbf{28}(19), 1802--1804 (2003).


\bibitem{PoloPTL2003}
 \textsc{V.~Polo},  \textsc{B.~Vidal},  \textsc{J.~Corral},  and
  \textsc{J.~Marti},
 \jr{{IEEE} Photon. Technol. Lett.} \textbf{15}(4), 584 --586 (2003).


\bibitem{XuePTL2009}
 \textsc{W.~Xue},  \textsc{S.~Sales},  \textsc{J.~Mork},  and
  \textsc{J.~Capmany},
 \jr{{IEEE} Photon. Technol. Lett.} \textbf{21}(3), 167 --169 (2009).


\bibitem{LloretOpex2011}
 \textsc{J.~Lloret},  \textsc{J.~Sancho},  \textsc{M.~Pu},
  \textsc{I.~Gasulla},  \textsc{K.~Yvind},  \textsc{S.~Sales},  and
  \textsc{J.~Capmany},
 \jr{Opt. Express} \textbf{19}(13), 12402--12407 (2011).


\bibitem{MortonPTL2009}
 \textsc{P.\,A. Morton} and  \textsc{J.\,B. Khurgin},
 \jr{{IEEE Photonics Technol. Lett.}} \textbf{{21}}({22}), {1686--1688}
  ({2009}).


\bibitem{LenzJQE2006}
 \textsc{G.~Lenz},  \textsc{B.~Eggleton},  \textsc{C.~Madsen},  and
  \textsc{R.~Slusher},
 \jr{{IEEE} J. Quantum Electron.} \textbf{37}(4), 525 --532 (2001).


\bibitem{XueOL2009}
 \textsc{W.~Xue},  \textsc{S.~Sales},  \textsc{J.~Capmany},  and
  \textsc{J.~M{\o}rk},
 \jr{Opt. Lett.} \textbf{34}(7), 929--931 (2009).


\bibitem{XueOpex2010}
 \textsc{W.~Xue},  \textsc{S.~Sales},  \textsc{J.~Capmany},  and
  \textsc{J.~M{\o}rk},
 \jr{Opt. Express} \textbf{18}(6), 6156--6163 (2010).


\bibitem{ChinOpex2010}
 \textsc{S.~Chin},  \textsc{L.~Th\'{e}venaz},  \textsc{J.~Sancho},
  \textsc{S.~Sales},  \textsc{J.~Capmany},  \textsc{P.~Berger},
  \textsc{J.~Bourderionnet},  and  \textsc{D.~Dolfi},
 \jr{Opt. Express} \textbf{18}(21), 22599--22613 (2010).


\bibitem{NgJLT1991}
 \textsc{W.~Ng},  \textsc{A.~Walston},  \textsc{G.~Tangonan},  \textsc{J.~Lee},
   \textsc{I.~Newberg},  and  \textsc{N.~Bernstein},
 \jr{J. Lightw. Technol.} \textbf{9}(9), 1124 --1131 (1991).


\bibitem{FrigyesMTT1995}
 \textsc{I.~Frigyes} and  \textsc{A.~Seeds},
 \jr{{IEEE} Trans. Microw. Theory Tech.} \textbf{43}(9), 2378 --2386 (1995).


\bibitem{Frankel1995}
 \textsc{M.~Frankel} and  \textsc{R.~Esman},
 \jr{{IEEE} Trans. Microw. Theory Tech.} \textbf{43}(9), 2387--2394 (1995).


\bibitem{ZmudaPTL1997}
 \textsc{H.~Zmuda},  \textsc{R.~Soref},  \textsc{P.~Payson},
  \textsc{S.~Johns},  and  \textsc{E.~Toughlian},
 \jr{{IEEE} Photon. Technol. Lett.} \textbf{9}(2), 241 --243 (1997).


\othercit
\bibitem{AckermanIMS1992}
 \textsc{E.~Ackerman},  \textsc{S.~Wanuga},  \textsc{D.~Kasemset},
  \textsc{W.~Minford},  \textsc{N.~Thorsten},  and  \textsc{J.~Watson},
in: {Proceedings} of the {IEEE MTT-S International Microwave Symposium (IMS
  1992)}, {Albuquerque, NM}, {} (1992),  pp.\,681 --684 vol.2.


\bibitem{GrosskopfAntenna2003}
 \textsc{G.~Grosskopf},  \textsc{R.~Eggemann},  \textsc{H.~Ehlers},
  \textsc{A.~Kortke},  \textsc{B.~Kuhlow},  \textsc{G.~Przyrembel},
  \textsc{D.~Rohde},  and  \textsc{S.~Zinal},
 \jr{{IEEE} Trans. Antennas Propag.} \textbf{51}(11), 3040 -- 3046 (2003).


\bibitem{McKinneyOL2002}
 \textsc{J.\,D. McKinney},  \textsc{D.\,E. Leaird},  and  \textsc{A.\,M.
  Weiner},
 \jr{Opt. Lett.} \textbf{27}(15), 1345--1347 (2002).


\bibitem{ChouPTL2003}
 \textsc{J.~Chou},  \textsc{Y.~Han},  and  \textsc{B.~Jalali},
 \jr{{IEEE Photonics Technol. Lett.}} \textbf{{15}}({4}), {581--583} ({2003}).


\bibitem{LinUWB2005}
 \textsc{I.~Lin},  \textsc{J.~McKinney},  and  \textsc{A.~Weiner},
 \jr{{IEEE Microw. Wirel. Compon. Lett.}} \textbf{{15}}({4}), {226--228}
  ({2005}).


\bibitem{WangPTL2008}
 \textsc{C.~Wang} and  \textsc{J.~Yao},
 \jr{{IEEE} Photon. Technol. Lett.} \textbf{20}(11), 882 --884 (2008).


\bibitem{BoleaOpex2009}
 \textsc{M.~Bolea},  \textsc{J.~Mora},  \textsc{B.~Ortega},  and
  \textsc{J.~Capmany},
 \jr{Opt. Express} \textbf{17}(7), 5023--5032 (2009).


\bibitem{YaoJLT2007}
 \textsc{J.~Yao},  \textsc{F.~Zeng},  and  \textsc{Q.~Wang},
 \jr{J. Lightwave Technol.} \textbf{25}(11), 3219--3235 (2007).


\bibitem{YaoJOSAB1996}
 \textsc{X.\,S. Yao} and  \textsc{L.~Maleki},
 \jr{J. Opt. Soc. Am. B} \textbf{13}(8), 1725--1735 (1996).


\bibitem{MalekiNatPhotonics2011}
 \textsc{L.~Maleki},
 \jr{{Nat. Photonics}} \textbf{{5}}({12}), {728--730} ({2011}).


\bibitem{LiangOL2010}
 \textsc{W.~Liang},  \textsc{V.\,S. Ilchenko},  \textsc{A.\,A. Savchenkov},
  \textsc{A.\,B. Matsko},  \textsc{D.~Seidel},  and  \textsc{L.~Maleki},
 \jr{Opt. Lett.} \textbf{35}(16), 2822--2824 (2010).


\bibitem{DevganOEOPTL2010}
 \textsc{P.~Devgan},  \textsc{M.~Pruessner},  \textsc{V.~Urick},  and
  \textsc{K.~Williams},
 \jr{{IEEE} Photon. Technol. Lett.} \textbf{22}(3), 152 --154 (2010).


\bibitem{VolyanskiyOpex2010}
 \textsc{K.~Volyanskiy},  \textsc{P.~Salzenstein},  \textsc{H.~Tavernier},
  \textsc{M.~Pogurmirskiy},  \textsc{Y.\,K. Chembo},  and
  \textsc{L.~Larger},
 \jr{Opt. Express} \textbf{18}(21), 22358--22363 (2010).


\bibitem{KippenbergScience2011}
 \textsc{T.\,J. Kippenberg},  \textsc{R.~Holzwarth},  and  \textsc{S.\,A.
  Diddams},
 \jr{Science} \textbf{332}(6029), 555--559 (2011).


\bibitem{FosterOpex2011}
 \textsc{M.\,A. Foster},  \textsc{J.\,S. Levy},  \textsc{O.~Kuzucu},
  \textsc{K.~Saha},  \textsc{M.~Lipson},  and  \textsc{A.\,L. Gaeta},
 \jr{Opt. Express} \textbf{19}(15), 14233--14239 (2011).


\bibitem{FerdousNatPhotonics2011}
 \textsc{F.~Ferdous},  \textsc{H.~Miao},  \textsc{D.\,E. Leaird},
  \textsc{K.~Srinivasan},  \textsc{J.~Wang},  \textsc{L.~Chen},  \textsc{L.\,T.
  Varghese},  and  \textsc{A.\,M. Weiner},
 \jr{{Nat. Photonics}} \textbf{{5}}({12}), {770--776} ({2011}).


\bibitem{RutkowskaOpex2011}
 \textsc{K.~Rutkowska},  \textsc{D.~Duchesne},  \textsc{M.~Strain},
  \textsc{R.~Morandotti},  \textsc{M.~Sorel},  and  \textsc{J.~Azana},
 \jr{Opt. Express} \textbf{19}(20), 19514--19522 (2011).


\bibitem{FerreraOpex2011}
 \textsc{M.~Ferrera},  \textsc{Y.~Park},  \textsc{L.~Razzari},  \textsc{B.\,E.
  Little},  \textsc{S.\,T. Chu},  \textsc{R.~Morandotti},  \textsc{D.\,J.
  Moss},  and  \textsc{J.~Azana},
 \jr{Opt. Express} \textbf{19}(23), 23153--23161 (2011).


\bibitem{GutierrezOL2012}
 \textsc{A.\,M. Guti\'{e}rrez},  \textsc{A.~Brimont},  \textsc{J.~Herrera},
  \textsc{M.~Aamer},  \textsc{J.~Mart\'{i}},  \textsc{D.\,J. Thomson},
  \textsc{F.\,Y. Gardes},  \textsc{G.\,T. Reed},  \textsc{J.\,M. Fedeli},  and
  \textsc{P.~Sanchis},
 \jr{Opt. Lett.} \textbf{37}(10), 1721--1723 (2012).


\bibitem{JinPTL2012}
 \textsc{S.~Jin},  \textsc{A.~Bhardwaj},  \textsc{P.~Herczfeld},  and
  \textsc{Y.~Li},
 \jr{{IEEE} Photon. Technol. Lett.} \textbf{PP}(99), 1 (2012).


\bibitem{ValleyOpex2007}
 \textsc{G.\,C. Valley},
 \jr{Opt. Express} \textbf{15}(5), 1955--1982 (2007).


\bibitem{CorcoranOpex2010}
 \textsc{B.~Corcoran},  \textsc{T.\,D. Vo},  \textsc{M.\,D. Pelusi},
  \textsc{C.~Monat},  \textsc{D.\,X. Xu},  \textsc{A.~Densmore},
  \textsc{R.~Ma},  \textsc{S.~Janz},  \textsc{D.\,J. Moss},  and
  \textsc{B.\,J. Eggleton},
 \jr{Opt. Express} \textbf{18}(19), 20190--20200 (2010).


\othercit
\bibitem{WallCLEO2012}
 \textsc{W.~Wall} and  \textsc{M.~Foster},
in: {Proceedings} of the {2012 Conference on Lasers and Electro-Optics (CLEO)
  and Quantum Electronics and Laser Science Conference (QELS)}, {San Jose,
  \,CA}, {USA}, {} (2012),  p.\,CTu3A.4.


\bibitem{VidalPJ2012}
 \textsc{B.~Vidal},  \textsc{J.~Palaci},  and  \textsc{J.~Capmany},
 \jr{IEEE Photonics Journal} \textbf{4}(3), 759 --764 (2012).


\bibitem{GaetaNatPhoton2012}
 \textsc{Y.~Okawachi} and  \textsc{A.~Gaeta},
 \jr{{Nat. Photonics}} \textbf{{6}}({}), {274–276} ({2012}).


\bibitem{XueOptComm2011}
 \textsc{X.~Xue},  \textsc{X.~Zheng},  \textsc{H.~Zhang},  and
  \textsc{B.~Zhou},
 \jr{Opt. Commun.} \textbf{284}(12), 2695 -- 2699 (2011).


\bibitem{GasullaOpex2012}
 \textsc{I.~Gasulla} and  \textsc{J.~Capmany},
 \jr{Opt. Express} \textbf{20}(11), 11710--11717 (2012).


\othercit
\bibitem{UrickAVFOP2011}
 \textsc{V.~Urick},  \textsc{J.~McKinney},  \textsc{J.~Diehl},  and
  \textsc{J.~Singley},
in: {Proceedings} of the {IEEE Avionics, Fiber- Optics and Photonics Technology
  Conference (AVFOP 2011)}, {San Diego,\,CA}, {USA}, {} (2011),  pp.\,37 --38.


\bibitem{DongOpex2010_power}
 \textsc{P.~Dong},  \textsc{W.~Qian},  \textsc{H.~Liang},
  \textsc{R.~Shafiiha},  \textsc{D.~Feng},  \textsc{G.~Li},  \textsc{J.\,E.
  Cunningham},  \textsc{A.\,V. Krishnamoorthy},  and
  \textsc{M.~Asghari},
 \jr{Opt. Express} \textbf{18}(19), 20298--20304 (2010).


\bibitem{DeCortOL2011}
 \textsc{W.\,D. Cort},  \textsc{J.~Beeckman},  \textsc{T.~Claes},
  \textsc{K.~Neyts},  and  \textsc{R.~Baets},
 \jr{Opt. Lett.} \textbf{36}(19), 3876--3878 (2011).


\othercit
\bibitem{MelloniSPIE2012}
 \textsc{A.~Melloni},  \textsc{S.~Grillanda},  \textsc{A.~Canciamilla},
  \textsc{C.~Ferrari},  \textsc{F.~Morichetti},  \textsc{M.~Strain},
  \textsc{M.~Sorel},  \textsc{V.~Singh},  \textsc{A.~Agarwal},  and
  \textsc{L.\,C. Kimerling},
in: {SPIE Proceedings} of the {Silicon Photonics VII} (SPIE, ),  p.\,82660A.


\bibitem{SteedJSTQE2011}
 \textsc{R.~Steed},  \textsc{L.~Ponnampalam},  \textsc{M.~Fice},
  \textsc{C.~Renaud},  \textsc{D.~Rogers},  \textsc{D.~Moodie},
  \textsc{G.~Maxwell},  \textsc{I.~Lealman},  \textsc{M.~Robertson},
  \textsc{L.~Pavlovic},  \textsc{L.~Naglic},  \textsc{M.~Vidmar},  and
  \textsc{A.~Seeds},
 \jr{{IEEE} J. Sel. Topics Quantum Electron.} \textbf{17}(1), 210 --217 (2011).


\bibitem{BenYooTerahertz2012}
 \textsc{S.~Yoo},  \textsc{R.~Scott},  \textsc{D.~Geisler},
  \textsc{N.~Fontaine},  and  \textsc{F.~Soares},
 \jr{IEEE Trans. on Terahertz Science and Technology} \textbf{2}(2), 167 --176
  (2012).


\end{thebibliography}


\end{document}